\documentclass[a4paper,12pt]{article}
\pdfoutput=1
\usepackage{epsfig,graphicx,xcolor,amsbsy,amssymb,latexsym,amsfonts,amsmath,setspace}
\usepackage{pstricks}
\usepackage{color}
\usepackage{soul}
\usepackage[numbers,square,comma, compress]{natbib}
\usepackage{placeins}
\usepackage{wrapfig}

\usepackage{eurosym}
\usepackage[a4paper]{geometry}
\usepackage{subfig}
\usepackage{graphicx}
\usepackage{tikz}
\usepackage{xcolor}
\usepackage{amsmath,amssymb,amsfonts,pstricks,setspace}
\usepackage[enableskew]{youngtab}

\numberwithin{equation}{section}

\newcommand{\bea}{\begin{eqnarray}\displaystyle}
\newcommand{\eea}{\end{eqnarray}}

\newcommand{\figref}[1]{Fig.~\protect\ref{#1}}

\newcommand{\pf}[2]{\mathcal{Z}_{#1,#2}}

\title{
\vspace{-3cm}
\begin{flushright}
{\small LYCEN 2018-07}\\[40pt]
\end{flushright}
\bf{Beyond Triality: Dual Quiver Gauge Theories and Little String Theories}\\[15pt]}
\author{\large \textsc{Brice Bastian\footnote{\tt b.bastian@ipnl.in2p3.fr}},~~\textsc{Stefan~Hohenegger\footnote{\tt s.hohenegger@ipnl.in2p3.fr},~ Amer Iqbal\footnote{\tt  amer@alum.mit.edu},~ Soo-Jong Rey\,\footnote{\tt rey.soojong@gmail.com}}}
\date{}

\begin{document}

\maketitle

\begin{center}
\renewcommand{\thefootnote}{\fnsymbol{footnote}}\vspace{-0.5cm}
${}^{\footnotemark[1]\footnotemark[2]}$ Universit\'e de Lyon\\
UMR 5822, CNRS/IN2P3, Institut de Physique Nucl\'eaire de Lyon\\ 4 rue Enrico Fermi, 69622 Villeurbanne Cedex, \rm FRANCE\\[0.4cm]
${}^{\footnotemark[3]}$ Abdus Salam School of Mathematical Sciences \\ Government College University, Lahore, PAKISTAN\\[0.4cm]
${}^{\footnotemark[3]}$ Center for Theoretical Physics, Lahore, PAKISTAN\\[0.4cm]
${}^{\footnotemark[4]}$ School of Physics and Astronomy\\
Seoul National University, Seoul 08826 \rm KOREA\\[1cm]
\end{center}

\begin{abstract}
The web of dual gauge theories engineered from a class of toric Calabi-Yau threefolds is explored. In previous work, we have argued for a triality structure by compiling evidence for the fact that every such manifold $X_{N,M}$ (for given $(N,M)$) engineers three a priori different, weakly coupled quiver gauge theories in five dimensions. The strong coupling regime of the latter is in general described by Little String Theories. Furthermore, we also conjectured that the manifold $X_{N,M}$ is dual to $X_{N',M'}$ if $NM=N'M'$ and $\text{gcd}(N,M)=\text{gcd}(N',M')$. Combining this result with the triality structure, we currently argue for a large number of dual quiver gauge theories, whose instanton partition functions can be computed explicitly as specific expansions of the topological partition function $\mathcal{Z}_{N,M}$ of $X_{N,M}$. We illustrate this web of dual theories by studying explicit examples in detail. We also undertake first steps in further analysing the extended moduli space of $X_{N,M}$ with the goal of finding other dual gauge theories.
\end{abstract}

\newpage

\tableofcontents

\onehalfspacing

\vskip1cm

%%%%%%%%%%%%%%%%%%%%%%%%%%%%%%%%%%%%%%
\section{Introduction}
The study of gauge theories in dimensions larger than four has over the years brought to light a variety of unexpected and interesting dualities. In order to explore the latter in a systematic fashion, the connection to string theory (or M- and F-theory as its higher-dimensional avatars) has proven to be a powerful tool. Besides (in many cases) providing a physical explanation for many a priori surprising dualities, the connection also allows to analyze aspects (\emph{e.g.} the non-perturbative regime) that would be inaccessible by more traditional, purely field theoretic methods alone. An example is the zoo of six-dimensional superconformal theories which are engineered using string theory and which do not have a Lagrangian description \cite{Witten:1995ex,Witten:1995zh,Strominger:1995ac,Seiberg:1996qx}. It was also observed that various properties and dualities of lower-dimensional theories follow from compactification of these six-dimensional theories \cite{Gaiotto:2009we}.

A class of theories \cite{Haghighat:2013gba,Haghighat:2013tka,Hohenegger:2013ala,Hohenegger:2015btj} that has recently attracted attention concerns the theories that are engineered from F-theory compactifications \cite{Vafa:1996xn,Morrison:1996na,Morrison:1996pp,Bershadsky:1996nu,Aspinwall:1996vc,Bershadsky:1997sb,Esole:2015xfa} on a two-parameter family of toric, non-compact Calabi-Yau threefolds $X_{N,M}$. These manifolds have a double elliptic fibration structure, where the two fibrations have singularities of type $I_{N-1}$ and $I_{M-1}$ (with $N,M\in\mathbb{N}$), respectively, and can be realised as the $\mathbb{Z}_N \times \mathbb{Z}_M$ orbifold of $X_{1,1}$ (see \cite{Kanazawa:2016tnt,Hohenegger:2015btj}). What makes the theories constructed in this fashion interesting is that they are in fact Little String Theories (LSTs) \cite{Seiberg:1996vs,Berkooz:1997cq,Blum:1997mm,Seiberg:1997zk,Losev:1997hx,Intriligator:1997dh,Aharony:1998ub}: at a low-energy scale, they correspond to conventional gauge theories with a specific gauge group. However, these theories contain an effective scale beyond which they can no longer be described as simple point-particle field theories, but require the inclusion of noncritical string degrees of freedom (see also \cite{Aharony:1999ks,Kutasov:2001uf}). They retain (part of) their fundamental string theoretic origin, although all gravitational degrees of freedom are decoupled. The supersymmetric partition function $\mathcal{Z}_{N,M}$ of these theories is captured by the (refined) topological string partition function of the Calabi-Yau manifold $X_{N,M}$, which (given the toric nature of the latter) can be computed in an efficient manner \cite{Haghighat:2013gba,Haghighat:2013tka,Hohenegger:2013ala,Hohenegger:2015btj} with the help of the refined topological vertex formalism \cite{Aganagic:2003db, Hollowood:2003cv, Iqbal:2007ii}. Indeed, in \cite{Bastian:2017ing}, a generic building block was constructed which allows to compute different expansions of $\mathcal{Z}_{N,M}$ in terms of certain K\"ahler parameters for generic $(N,M)$.

In the recent papers \cite{Hohenegger:2016yuv} and \cite{Bastian:2017ary}, two intriguing observations have been made which imply that a low-energy theory constructed from a given $X_{N,M}$ permits a large number of dual descriptions. First of all, in \cite{Hohenegger:2016yuv}, the extended K\"ahler moduli space of $X_{N,M}$ was analysed: The K\"ahler moduli space of $X_{N,M}$ has the shape of a cone, whose codimension-one faces (referred to as walls in the following) include loci where one of the K\"ahler parameters vanishes. Crossing such a wall corresponds to a so-called \emph{flop transition}\,\cite{TianYau,Kollar}: it leads to a neighbouring cone that describes again a Calabi-Yau manifold and lies in the extended K\"ahler moduli space \cite{Aspinwall:1993yb,Morrison} of $X_{N,M}$. By using a particular series of flop transitions, it was conjectured in \cite{Hohenegger:2016yuv} that 
\begin{align}
&X_{N,M}\sim X_{N',M'}&&\text{if} &&\begin{array}{l} NM=N'M' \,,\,\,\text{and}\\ \text{gcd}(N,M)=\text{gcd}(N',M')\,,\end{array}\,\label{RelDuality}
\end{align}
\emph{i.e.} under the given conditions, the K\"ahler cones of the Calabi-Yau manifolds $X_{N,M}$ and $X_{N',M'}$ are part of a common extended moduli space. The relation (\ref{RelDuality}) was explicitly shown for a large class of examples and passed highly non-trivial consistency checks for generic $(N,M)$. Moreover, it is expected that the partition functions $\mathcal{Z}_{N,M}$ and $\mathcal{Z}_{N',M'}$ agree upon taking into account the duality map implicit in (\ref{RelDuality}). This fact was explicitly checked for $\text{gcd}(N,M)=1$ in \cite{Bastian:2017ing}.

Secondly, in \cite{Bastian:2017ary} the question was analysed what type of (low-energy) gauge theories can be engineered from a given Calabi-Yau manifold $X_{N,M}$. By studying different series expansions of $\mathcal{Z}_{N,M}$ (in terms of different sets of the K\"ahler parameters of $X_{N,M}$) it was found that the K\"ahler cone of $X_{N,M}$ contains three different regions that engineer five-dimensional quiver gauge theories with gauge groups   
\begin{align}
&G_{\text{hor}}=[U(M)]^N\,, &&G_{\text{vert}}=[U(N)]^M\,, &&G_{\text{diag}}=[U(\tfrac{MN}{k})]^k\,. \label{TrialityGroups}
\end{align} 
respectively, where $k=\text{gcd}(N,M)$.  It is expected that the UV completions of (at least some of) these gauge theories are LSTs. The relation among the three low-energy theories with gauge groups (\ref{TrialityGroups}) was dubbed \emph{triality} in \cite{Bastian:2017ary}, reflecting the fact that all three are engineered by the same Calabi-Yau threefold. It is important to realise that the mapping of the various gauge theory parameters is highly non-trivial among the members of the triality: indeed, the coupling constants, mass parameters and other Coulomb branch parameters are mixed in a highly non-trivial fashion when going from one theory to another.

Combining the triality discussed in \cite{Bastian:2017ary} with the observations of \cite{Hohenegger:2016yuv} that the Calabi-Yau manifold $X_{N,M}$ itself can be dualised to other manifolds in the sense of eq.~(\ref{RelDuality}) suggests an even more elaborate picture at the level of the low-energy theories: indeed it implies that a (circular) quiver gauge theory with $N$ gauge nodes of type $U(M)$ (that is engineered by $X_{N,M}$ for arbitrary $(N,M)$) is dual to a whole web of other quiver gauge theories that have $N'$ gauge nodes of type $U(M')$ such that $NM=N'M'$ and $\text{gcd}(N,M)=\text{gcd}(N',M')$. In this paper, we elaborate on this idea and perform a first step towards extending the web of dualities even further to include yet new regions in the extended moduli space of the Calabi-Yau threefolds $X_{N,M}$. Indeed, the latter also contains cones that do not represent manifolds of the type $X_{N',M'}$ for some $N',M'\in\mathbb{N}$, \emph{i.e.} that do not fall into the class of doubly elliptically fibered threefolds introduced above. Among these, we focus on a class of manifolds that can be represented by so-called \emph{twisted toric web diagrams} (see section~\ref{Sect:IntermediateCones} for a detailed definition) and show (by using some elucidating examples) that they can also engineer low-energy gauge theories. The details of analysing the structure of the underlying gauge group (as well as their affine extension, possibly by non-perturbative effects) by geometric means is relegated to a separate publication~\cite{ToAppear}.

This paper is organised as follows: In section~\ref{Sect:DualLSTs}, we review in more detail the structure of the extended moduli space of $X_{N,M}$ as well as the triality relation proposed in \cite{Bastian:2017ary}. We then combine these two results to argue for a large web of dualities for the theories constructed from the Calabi-Yau manifolds of type $X_{N,M}$. As a non-trivial example, we present the configuration $(N,M)=(6,5)$. In section~\ref{Sect:IntermediateCones}, we propose to extend the web of dualities to include theories engineered by other regions in the extended moduli space of $X_{N,M}$. We focus on a particular class of manifolds, which are described by so-called shifted we diagrams and discuss explicitly the cases $(N,M)=(6,1)$ and $(4,1)$. This work is accompanied by 3 appendices: they contain technical details on the geometric realisation of the gauge algebra from the perspective of the web diagram in the case $(N,M)=(2,2)$, as well as definition of the Nekrasov subfunctions and further details on a particular class of duality transformations, respectively.
%%%%%%%%
%%%%%%%%
\section{Symmetry Transformations of $X_{N,M}$ and Dual Little String Theories}\label{Sect:DualLSTs}
In this section, we discuss various degenerations of $X_{N,M}$ and the corresponding walls in the extended K\"ahler moduli space. We also discuss the gauge theories which appear in various cones of the latter.

\subsection{(Extended) K\"ahler Moduli Space of $X_{N,M}$}
\begin{figure}[htb]
\begin{center}
\scalebox{0.6}{\parbox{16.5cm}{\begin{tikzpicture}[scale = 1.5]
%horizontal lines%%%%%%%%%%
%first layer
\draw[ultra thick] (-1,0) -- (0,0);
\draw[ultra thick] (1,1) -- (2,1);
\draw[ultra thick] (3,2) -- (4,2);
\node at (4.5,2) {\Large $\cdots$};
\draw[ultra thick] (5,2) -- (6,2);
\draw[ultra thick] (7,3) -- (8,3);
%second layer
\draw[ultra thick] (0,2) -- (1,2);
\draw[ultra thick] (2,3) -- (3,3);
\draw[ultra thick] (4,4) -- (5,4);
\node at (5.5,4) {\Large $\cdots$};
\draw[ultra thick] (6,4) -- (7,4);
\draw[ultra thick] (8,5) -- (9,5);
%third layer
\draw[ultra thick] (1,6) -- (2,6);
\draw[ultra thick] (3,7) -- (4,7);
\draw[ultra thick] (5,8) -- (6,8);
\node at (6.5,8) {\Large $\cdots$};
\draw[ultra thick] (7,8) -- (8,8);
\draw[ultra thick] (9,9) -- (10,9);
%vertical lines%%%%%%%%%%
%first layer
\draw[ultra thick] (0,0) -- (0,-1);
\draw[ultra thick] (2,1) -- (2,0);
\draw[ultra thick] (6,2) -- (6,1);
%second layer
\draw[ultra thick] (1,1) -- (1,2);
\draw[ultra thick] (3,2) -- (3,3);
\draw[ultra thick] (7,3) -- (7,4);
%third layer
\draw[ultra thick] (2,3) -- (2,4);
\draw[ultra thick] (4,4) -- (4,5);
\draw[ultra thick] (8,5) -- (8,6);
%dots
\node[rotate=90] at (2,4.5) {\Large $\cdots$};
\node[rotate=90] at (4,5.5) {\Large $\cdots$};
\node[rotate=90] at (8,6.5) {\Large $\cdots$};
%fourth layer
\draw[ultra thick] (2,5) -- (2,6);
\draw[ultra thick] (4,6) -- (4,7);
\draw[ultra thick] (8,7) -- (8,8);
%fifth layer
\draw[ultra thick] (3,7) -- (3,8);
\draw[ultra thick] (5,8) -- (5,9);
\draw[ultra thick] (9,9) -- (9,10);
%diagonal lines%%%%%%%%%%
%first layer
\draw[ultra thick] (0,0) -- (1,1);
\draw[ultra thick] (2,1) -- (3,2);
\draw[ultra thick] (6,2) -- (7,3);
%second layer
\draw[ultra thick] (1,2) -- (2,3);
\draw[ultra thick] (3,3) -- (4,4);
\draw[ultra thick] (7,4) -- (8,5);
%third layer
\draw[ultra thick] (2,6) -- (3,7);
\draw[ultra thick] (4,7) -- (5,8);
\draw[ultra thick] (8,8) -- (9,9);
%%%%%%%%%%%%%%%%%%%%%%
%connectors
\node[rotate=90] at (-0.75,0) {$=$}; 
\node at (-0.65,-0.2) {{\small $1$}};
\node[rotate=90] at (0.25,2) {$=$}; 
\node at (0.35,1.8) {{\small $2$}};
\node[rotate=90] at (1.25,6) {$=$}; 
\node at (1.35,5.8) {{\small $M$}};
\node[rotate=90] at (7.75,3) {$=$}; 
\node at (7.85,2.8) {{\small $1$}};
\node[rotate=90] at (8.75,5) {$=$};
\node at (8.85,4.8) {{\small $2$}};
\node[rotate=90] at (9.75,9) {$=$};
\node at (9.85,8.8) {{\small $M$}}; 
\node at (0,-0.75) {--}; 
\node at (0.2,-0.9) {{\small $1$}};
\node at (2,0.25) {--}; 
\node at (2.2,0.2) {{\small $2$}};
\node at (6,1.25) {--}; 
\node at (6.2,1.2) {{\small $N$}};
\node at (3,7.75) {--};
\node at (3.2,7.65) {{\small $1$}}; 
\node at (5,8.75) {--}; 
\node at (5.2,8.65) {{\small $2$}}; 
\node at (9,9.75) {--}; 
\node at (9.2,9.65) {{\small $N$}}; 
\end{tikzpicture}}}
\end{center}
\caption{\sl Web Diagram of the toric Calabi-Yau manifold $X_{N,M}$. The vertical and horizontal lines are glued together as indicated by the labels of the outermost legs.}
\label{Fig:WebDiagramGeneric}
\end{figure}
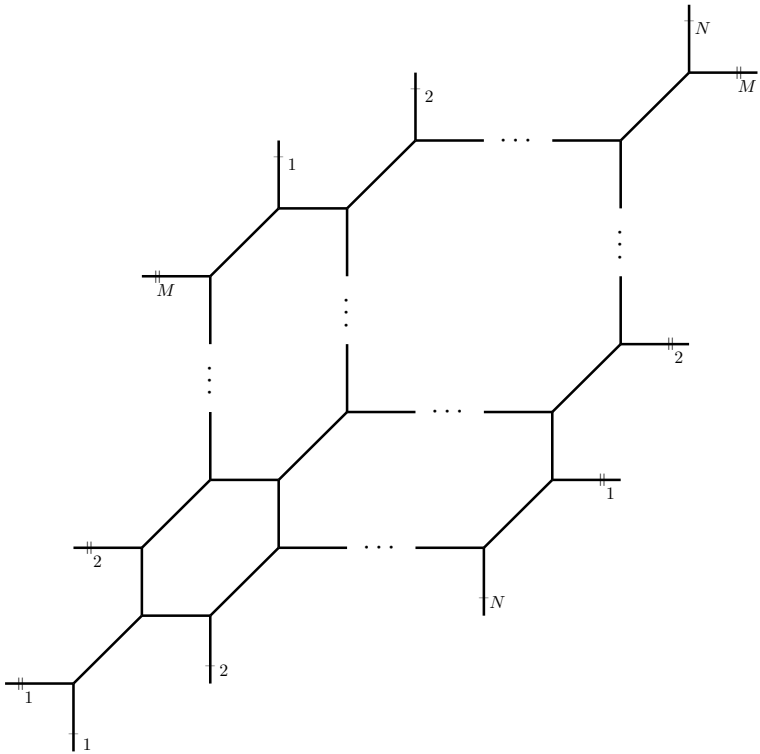

The main subject of study of this paper is relations among certain $\mathcal{N}=(1,0)$ LSTs (\emph{i.e.} with 8 supercharges), focusing in particular on their low-energy descriptions in terms of quiver gauge theories \cite{Nekrasov:2012xe}. These theories are defined by F-theory compactified on a class of toric, non-compact Calabi-Yau manifolds $X_{N,M}$, whose associated web diagram is shown in~\figref{Fig:WebDiagramGeneric}.
Each line $\Sigma_i$ of this diagram represents a holomorphic curve in $X_{N,M}$, whose area is given by
\begin{align}
\int_{\Sigma_i}\omega\,,
\end{align}
where $\omega$ is the K\"ahler form of $X_{N,M}$.\footnote{In the following, unless otherwise indicated, by abuse of notation, we use the same symbol for the holomorphic curves and their areas.}  There are a total $NM$ horizontal $\{{\bf h}\}$, $NM$ vertical $\{{\bf v}\}$ and $NM$ diagonal $\{{\bf m}\}$ curves, respectively. However, the corresponding $3NM$ areas are not all independent of one another. Indeed, as discussed in~\cite{Haghighat:2013tka,Hohenegger:2013ala}, consistency conditions of the web diagram enforce linear relations among the latter, leaving $(NM+2)$ independent parameters which parametrise the K\"ahler moduli space of $X_{N,M}$. More precisely, denoting the two-dimensional complex submanifolds of $X_{N,M}$ as $P_a$, the K\"ahler cone of $X_{N,M}$ is given by
\begin{align}
&\int_{X_{N,M}}\omega\wedge\omega\wedge\omega\geq 0\,,&&\text{and} &&\int _{P_a}\omega\wedge\omega\geq 0\,,&&\text{and} &&\int_{\Sigma_i}\omega\geq 0\,.\label{DefKahlerCone}
\end{align}
The walls of the cone correspond to the regions in which any of these inequalities turns into an actual equality. In particular, they include the regions in which one or more of the areas $\{\mathbf{h},\mathbf{v},\mathbf{m}\}$ vanish, such that the web diagram in~\figref{Fig:WebDiagramGeneric} degenerates. 

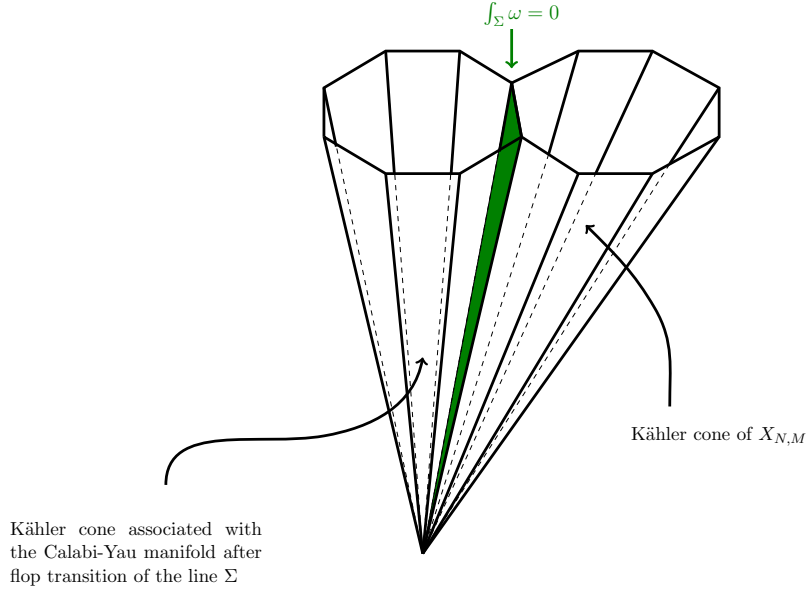
\begin{figure}[htb]
\begin{center}
\scalebox{0.65}{\parbox{16.5cm}{\begin{tikzpicture}[scale = 1]
%center cone 
\draw[fill=green!50!black] (0,-4) -- (1.8,5.6) -- (2,4.5) -- (0,-4);
\node[green!50!black,thick] at (2,7) {$\int_\Sigma\omega=0$};
\draw[ultra thick,->,green!50!black] (1.8,6.7) -- (1.8,5.9);
\draw[ultra thick] (-2,4.5) -- (-0.75,3.75) -- (0.75,3.75)-- (2,4.5) -- (1.8,5.6) -- (0.75,6.25) -- (-0.75,6.25) -- (-2,5.5) -- (-2,4.5);
\draw[ultra thick] (0,-4) -- (-2,4.5);
\draw[ultra thick] (0,-4) -- (-0.75,3.75);
\draw[ultra thick] (0,-4) -- (0.75,3.75);
\draw[ultra thick] (0,-4) -- (2,4.5);
\draw[ultra thick] (-2,5.5) -- (-1.75,4.3);
\draw[dashed] (-2,5.5) -- (0,-4);
\draw[ultra thick] (-0.75,6.25) -- (-0.575,3.75);
\draw[dashed] (-0.75,6.25) -- (0,-4);
\draw[ultra thick] (1.8,5.6) -- (1.55,4.2);
\draw[dashed] (1.8,5.6) -- (0,-4);
\draw[ultra thick] (0.75,6.25) -- (0.575,3.75);
\draw[dashed] (0.75,6.25) -- (0,-4);
\draw[ultra thick,->] (-5.2,-2.6) to [out=90,in=190] (-2,-1.6) to [out=10,in=260] (0,0);
\node at (-5.8,-4) {\small \parbox{5.1cm}{K\"ahler cone associated with the Calabi-Yau manifold after flop transition of the line $\Sigma$}};
%%%%%%%%%%%
%cone right
\draw[ultra thick] (2,4.5) -- (3.15,3.75) -- (4.65,3.75)-- (6,4.5) -- (6,5.5) -- (4.65,6.25) -- (3.15,6.25) -- (1.8,5.6) -- (2,4.5);
\draw[ultra thick] (0,-4) -- (3.15,3.75);
\draw[ultra thick] (0,-4) -- (4.65,3.75);
\draw[ultra thick] (0,-4) -- (6,4.5);
\draw[dashed] (0,-4) -- (3.5,3.75);
\draw[ultra thick] (3.5,3.75) -- (4.65,6.25);
\draw[dashed] (0,-4) -- (2.55,4.15);
\draw[ultra thick] (2.55,4.15) -- (3.15,6.25);
\draw[dashed] (0,-4) -- (5.0,3.95);
\draw[ultra thick] (5.0,3.95) -- (6,5.5);
\draw[ultra thick,->] (5,-1) to [out=90,in=300] (4.7,1) to [out=120,in=315] (3.3,2.7);
\node at (6,-1.6) {\small K\"ahler cone of $X_{N,M}$};
%
%%%%%%%
\end{tikzpicture}}}
\caption{\sl K\"ahler cones of two Calabi-Yau manifolds connected through a flop transition of the curve $\Sigma$.The corresponding wall, along which the cones are glued together, is characterised by $\int_\Sigma\omega=0$ and is shown in green.}
\label{Fig:TwoCones}
\end{center}
\end{figure}

The cone described by (\ref{DefKahlerCone}), however, can be embedded in a larger, so-called \emph{extended moduli space} \cite{Aspinwall:1993yb}. As explained above, while the shrinking of a rational curve of $X_{N,M}$ with normal bundle $\mathcal{O}(-1)\oplus\mathcal{O}(-1)$ results in a singular geometry, the latter can again be resolved. This leads to a geometry which lies outside of the original K\"ahler cone (\ref{DefKahlerCone}) since, from the point of view of the latter, the curve that resolves the degeneracies has negative area. This process is called a \emph{flop transition} and connects the K\"ahler cones of two Calabi-Yau manifolds with the same Hodge numbers (but possibly different intersection numbers) \cite{TianYau, Kollar,Morrison}, as shown in \figref{Fig:TwoCones}.
In this way, we can characterise the extended K\"ahler moduli space as the collection of K\"ahler cones of Calabi-Yau threefolds that are connected through (sequences of) flop transitions.
%%%%%%%%%
\subsection{Web of Dual Calabi-Yau Manifolds and Associated LSTs}
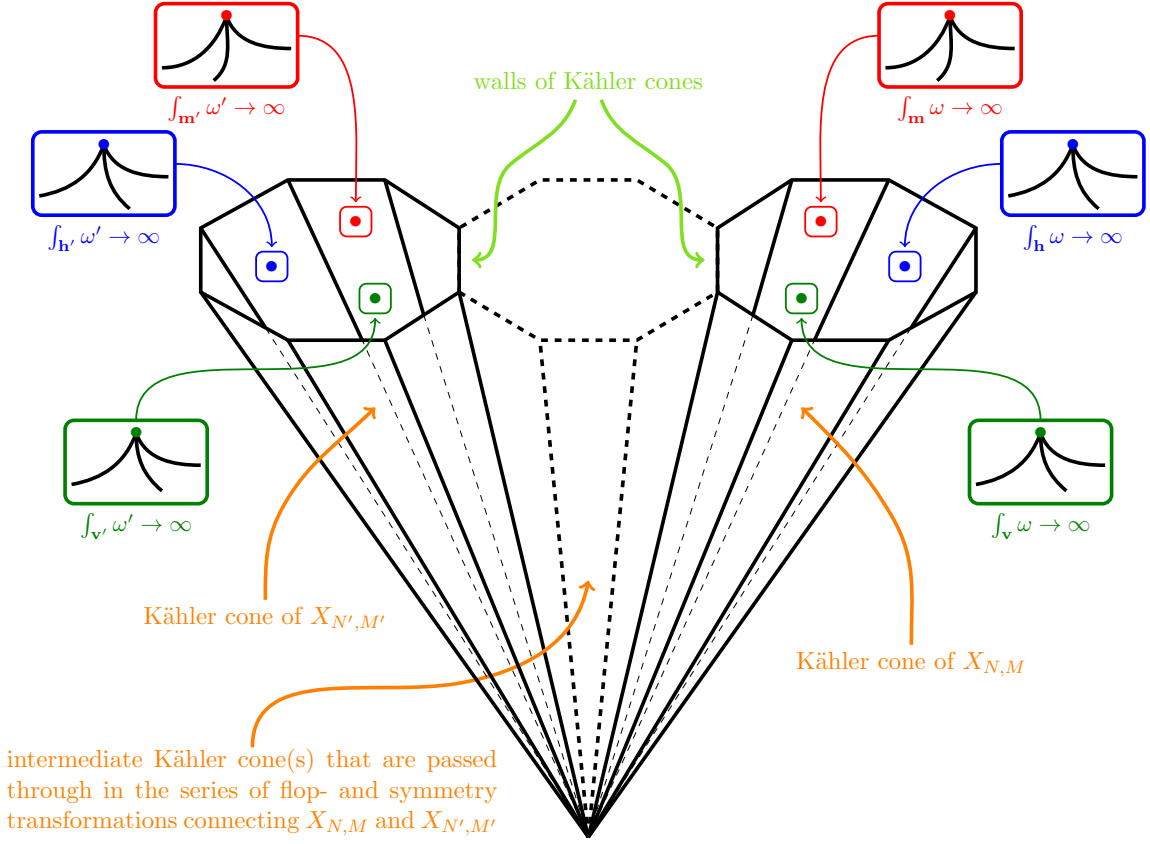
\begin{figure}[htb]
\begin{center}
\scalebox{0.85}{\parbox{17.8cm}{\begin{tikzpicture}[scale = 1]
%ghost cone 
\draw[ultra thick,dashed] (-2,4.5) -- (-0.75,3.75) -- (0.75,3.75)-- (2,4.5) -- (2,5.5) -- (0.75,6.25) -- (-0.75,6.25) -- (-2,5.5) -- (-2,4.5);
\draw[ultra thick] (0,-4) -- (-2,4.5);
\draw[dashed, ultra thick] (0,-4) -- (-0.75,3.75);
\draw[dashed, ultra thick] (0,-4) -- (0.75,3.75);
\draw[ultra thick] (0,-4) -- (2,4.5);
\draw[ultra thick, red!50!yellow,->] (-5.2,-2.6) to [out=90,in=190] (-2,-1.6) to [out=10,in=260] (0,0);
\node[red!50!yellow] at (-5.2,-3.3) {\small \parbox{7.6cm}{intermediate K\"ahler cone(s) that are passed through in the series of flop- and symmetry transformations connecting $X_{N,M}$ and $X_{N',M'}$}};
%%%%%%%%%%%
%cone right
\draw[ultra thick] (2,4.5) -- (3.15,3.75) -- (4.65,3.75)-- (6,4.5) -- (6,5.5) -- (4.65,6.25) -- (3.15,6.25) -- (2,5.5) -- (2,4.5);
\draw[ultra thick] (0,-4) -- (3.15,3.75);
\draw[ultra thick] (0,-4) -- (4.65,3.75);
\draw[ultra thick] (0,-4) -- (6,4.5);
\draw[dashed] (0,-4) -- (3.5,3.75);
\draw[ultra thick] (3.5,3.75) -- (4.65,6.25);
\draw[dashed] (0,-4) -- (2.55,4.15);
\draw[ultra thick] (2.55,4.15) -- (3.15,6.25);
\draw[dashed] (0,-4) -- (5.0,3.95);
\draw[ultra thick] (5.0,3.95) -- (6,5.5);
\draw[ultra thick, red!50!yellow,->] (5,-1) to [out=90,in=300] (4.7,1) to [out=120,in=315] (3.3,2.7);
\node[red!50!yellow] at (5,-1.3) {\small K\"ahler cone of $X_{N,M}$};
%cone left
\draw[ultra thick] (-2,4.5) -- (-3.15,3.75) -- (-4.65,3.75)-- (-6,4.5) -- (-6,5.5) -- (-4.65,6.25) -- (-3.15,6.25) -- (-2,5.5) -- (-2,4.5);
\draw[ultra thick] (0,-4) -- (-3.15,3.75);
\draw[ultra thick] (0,-4) -- (-4.65,3.75);
\draw[ultra thick] (0,-4) -- (-6,4.5);
\draw[dashed] (0,-4) -- (-3.5,3.75);
\draw[ultra thick] (-3.5,3.75) -- (-4.65,6.25);
\draw[dashed] (0,-4) -- (-2.55,4.15);
\draw[ultra thick] (-2.55,4.15) -- (-3.15,6.25);
\draw[dashed] (0,-4) -- (-5.0,3.95);
\draw[ultra thick] (-5.0,3.95) -- (-6,5.5);
\draw[ultra thick, red!50!yellow,->] (-5,-0.3) to [out=90,in=240] (-4.7,1.2) to [out=60,in=225] (-3.3,2.7);
\node[red!50!yellow] at (-5,-0.6) {\small K\"ahler cone of $X_{N',M'}$};
%%%%%%%
%labelling walls
\draw[ultra thick,blue!25!green!50!yellow,->] (0.2,7.5) to [out=290,in=140] (1.2,6.5) to [out=320,in=180] (1.8,5);
\draw[ultra thick,blue!25!green!50!yellow,->] (-0.2,7.5) to [out=250,in=40] (-1.2,6.5) to [out=220,in=0] (-1.8,5);
\node[blue!25!green!50!yellow] at (0,7.8) {\small walls of K\"ahler cones};
%%%%%%%
%theories right cone
\node[red, thick, rounded corners=3pt, draw] at (3.6,5.6) {$\bullet$}; 
\node[blue, thick, rounded corners=3pt, draw] at (4.9,4.9) {$\bullet$};
\node[green!50!black, thick, rounded corners=3pt, draw] at (3.3,4.4) {$\bullet$}; 
%%%%%%%%%%%%%%%%%%%%%%%%%%%%%%%
%%%%%%%
\draw[thick,red,<-] (3.6,5.9) to [out=90,in=180] (4.5,8.5);
\begin{scope}[xshift=6cm,yshift=2cm]
\draw[ultra thick,rounded corners,red] (-1.5,5.7) -- (0.7,5.7) -- (0.7,7) -- (-1.5,7) -- cycle;
\draw[ultra thick] (-1.4,6) to [out =0,in=250] (-0.4,6.8) to [out=290,in=180] (0.6,6.3); 
\draw[ultra thick] (-0.6,5.8) to [out=40,in=270] (-0.4,6.8);
\node[red] at (-0.4,6.8) {$\bullet$};
\node[red] at (-0.4,5.35) {{\footnotesize $\int_{\mathbf{m}}\omega\to \infty$}};
\end{scope}
%%%%%%%%%%%%%%%%%%%%
\draw[thick,blue,<-] (4.9,5.2) to [out=90,in=180] (6.4,6.5);
\begin{scope}[xshift=8cm]
\draw[ultra thick,rounded corners,blue] (-1.6,5.7) -- (0.6,5.7) -- (0.6,7) -- (-1.6,7) -- cycle;
\draw[ultra thick] (-1.5,6) to [out =10,in=250] (-0.5,6.8) to [out=290,in=180] (0.5,6.3); 
\draw[ultra thick] (-0.1,5.8) to [out=140,in=270] (-0.5,6.8);
\node[blue] at (-0.5,6.8) {$\bullet$};
\node[blue] at (-0.5,5.35) {{\footnotesize $\int_{\mathbf{h}}\omega\to \infty$}};
\end{scope} 
%%%%%%%%%%%%%%%%%%%%
%%%%%%%%%%%%%%%%%%%%
\draw[thick,green!50!black,<-] (3.3,4.1) to [out=270,in=90] (7,2.5);
\begin{scope}[xshift=7.5cm,yshift=-4.5cm]
\draw[ultra thick,rounded corners,green!50!black] (-1.6,5.7) -- (0.6,5.7) -- (0.6,7) -- (-1.6,7) -- cycle;
\draw[ultra thick] (-1.5,6) to [out =10,in=250] (-0.5,6.8) to [out=290,in=180] (0.5,6.3); 
\draw[ultra thick] (-0.1,5.9) to [out=140,in=270] (-0.5,6.8);
\node[green!50!black] at (-0.5,6.8) {$\bullet$};
\node[green!50!black] at (-0.5,5.35) {{\footnotesize $\int_{\mathbf{v}}\omega\to \infty$}};
\end{scope} 
%%%%%%%%%%%%%%%%%%%%%%%%%%%%%%%%%%%%%%%%
%%%%%%%
%theories left cone
\node[red, thick, rounded corners=3pt, draw] at (-3.6,5.6) {$\bullet$}; 
\node[blue, thick, rounded corners=3pt, draw] at (-4.9,4.9) {$\bullet$};
\node[green!50!black, thick, rounded corners=3pt, draw] at (-3.3,4.4) {$\bullet$}; 
%%%%%%%%%%%%%%%%%%%%%%%%%%%%%%%
%%%%%%%
\draw[thick,red,<-] (-3.6,5.9) to [out=90,in=0] (-4.5,8.5);
\begin{scope}[xshift=-5.2cm,yshift=2cm]
\draw[ultra thick,rounded corners,red] (-1.5,5.7) -- (0.7,5.7) -- (0.7,7) -- (-1.5,7) -- cycle;
\draw[ultra thick] (-1.4,6) to [out =0,in=250] (-0.4,6.8) to [out=290,in=180] (0.6,6.3); 
\draw[ultra thick] (-0.6,5.8) to [out=40,in=270] (-0.4,6.8);
\node[red] at (-0.4,6.8) {$\bullet$};
\node[red] at (-0.4,5.35) {{\footnotesize $\int_{\mathbf{m}'}\omega'\to \infty$}};
\end{scope}
%%%%%%%%%%%%%%%%%%%%
\draw[thick,blue,<-] (-4.9,5.2) to [out=90,in=0] (-6.4,6.5);
\begin{scope}[xshift=-7cm]
\draw[ultra thick,rounded corners,blue] (-1.6,5.7) -- (0.6,5.7) -- (0.6,7) -- (-1.6,7) -- cycle;
\draw[ultra thick] (-1.5,6) to [out =10,in=250] (-0.5,6.8) to [out=290,in=180] (0.5,6.3); 
\draw[ultra thick] (-0.1,5.8) to [out=140,in=270] (-0.5,6.8);
\node[blue] at (-0.5,6.8) {$\bullet$};
\node[blue] at (-0.5,5.35) {{\footnotesize $\int_{\mathbf{h}'}\omega'\to \infty$}};
\end{scope} 
%%%%%%%%%%%%%%%%%%%%
%%%%%%%%%%%%%%%%%%%%
\draw[thick,green!50!black,<-] (-3.3,4.1) to [out=270,in=90] (-7,2.5);
\begin{scope}[xshift=-6.5cm,yshift=-4.5cm]
\draw[ultra thick,rounded corners,green!50!black] (-1.6,5.7) -- (0.6,5.7) -- (0.6,7) -- (-1.6,7) -- cycle;
\draw[ultra thick] (-1.5,6) to [out =10,in=250] (-0.5,6.8) to [out=290,in=180] (0.5,6.3); 
\draw[ultra thick] (-0.1,5.9) to [out=140,in=270] (-0.5,6.8);
\node[green!50!black] at (-0.5,6.8) {$\bullet$};
\node[green!50!black] at (-0.5,5.35) {{\footnotesize $\int_{\mathbf{v}'}\omega'\to \infty$}};
\end{scope} 
\end{tikzpicture}}}
\caption{\sl Weak coupling regions in the extended moduli space of $X_{N,M}$.}
\label{Fig:FlopTransformations}
\end{center}
\end{figure}

Recent studies \cite{Hohenegger:2013ala,Bastian:2017ing,Bastian:2017ary} of the extended K\"ahler moduli space of the toric Calabi-Yau threefolds $X_{N,M}$ have suggested the existence of a large number of new dualities of the type
\begin{align}
&X_{N,M}\sim X_{N',M'}\,,&&\text{for} &&\begin{array}{l}  NM=N'M'\text{ and} \\ \text{gcd}(N,M)=k = \text{gcd}(N',M') \,.\end{array}\label{DualityCalabiYauFlop}
\end{align}
The relation $\sim$ means that the two Calabi-Yau threefolds are related by a certain combination of flop transitions (and other symmetry transformations). This process is shown schematically in \figref{Fig:FlopTransformations}: starting at a point in the K\"ahler cone of $X_{N,M}$, one can reach the K\"ahler cone of $X_{N',M'}$.
The K\"ahler form of $X_{N',M'}$ is denoted $\omega'$ and the corresponding web configuration consists of $M'N'$ many line segments $(\mathbf{h}',\mathbf{v}',\mathbf{m}')$. The explicit duality map relating $(\mathbf{h},\mathbf{v},\mathbf{m})$ to $(\mathbf{h}',\mathbf{v}',\mathbf{m}')$ was proposed in \cite{Hohenegger:2016yuv}, where it was argued (and proven explicitly in \cite{Bastian:2017ing} for $k=1$) that the topological string partition function of $X_{N,M}$ is invariant under the duality transformation, \emph{i.e.}
\begin{align}
\pf{N}{M}(\mathbf{h},\mathbf{v},\mathbf{m},\epsilon_{1,2})=\pf{N'}{M'}(\mathbf{h}',\mathbf{v}',\mathbf{m}',\epsilon_{1,2})\,.\label{InvariancePartitionFunction}
\end{align}
Furthermore, in \cite{Bastian:2017ary}, we argued that within the K\"ahler cone of $X_{N,M}$ (for given $N,M\in\mathbb{N}$) there exist three regions that represent the weak-coupling regime of three (in general different) quiver gauge theories with gauge groups
\begin{align}
&G_{\text{hor}}=[U(M)]^N\,, &&G_{\text{vert}}=[U(N)]^M\,, &&G_{\text{diag}}=[U(MN/k)]^k\,. \label{DualGaugeTheories}
\end{align} 
These regions are schematically shown in \figref{Fig:FlopTransformations} and afford three different series expansions of the topological string partition function of $X_{N,M}$\cite{Bastian:2017ary}\footnote{$Z_{p}$ being the perturbative contribution depending on the parameters other than the gauge couplings.}
{\allowdisplaybreaks
\begin{align}
\pf{N}{M}({\bf h},{\bf v},{\bf m},\epsilon_{1,2})&=Z_p({\bf v},{\bf m})\sum_{\vec{k}}\,e^{-\vec{k}\cdot \mathbf{h}}\,Z_{\vec{k}}({\bf v},{\bf m})=Z^{(N,M)}_{\text{hor}}\nonumber\\
&=Z_p({\bf h},{\bf m})\sum_{\vec{k}}\,e^{-\vec{k}\cdot \mathbf{v}}\,Z_{\vec{k}}({\bf h},{\bf m})=Z^{(N,M)}_{\text{vert}}\nonumber\\
&=Z_p({\bf h}, {\bf v})\sum_{\vec{k}}\,e^{-\vec{k}\cdot \mathbf{m}}\,Z_{\vec{k}}({\bf h},{\bf v})=Z^{(N,M)}_{\text{diag}}\,,\label{TrialitySchemat}
\end{align}}
which can be interpreted as the instanton partition functions of the three gauge theories (\ref{DualGaugeTheories}).\footnote{For an explicit and efficient manner to compute the series expansions in eq.~(\ref{TrialitySchemat}) by means of a universal building block, we refer the reader to \cite{Bastian:2017ing}. This construction is also briefly reviewed section~\ref{Sect:ReviewPartitionFunction}. The explicit choice of (independent) K\"ahler parameters underlying the three series expansions in eq.~(\ref{TrialitySchemat}) is discussed in detail in section~\ref{Sect:BaseChoice} below.} As shown in \figref{Fig:FlopTransformations}, since $X_{N',M'}$ has a similar web diagram as the one shown in \figref{Fig:WebDiagramGeneric}, the K\"ahler cone of $X_{N',M'}$ also contains three regions which allow series expansions of the partition function of the following form
\begin{align}
\pf{N}{M}({\bf h},{\bf v},{\bf m},\epsilon_{1,2})&=Z_p({\bf v}',{\bf m}')\sum_{\vec{k}}\,e^{-\vec{k}\cdot \mathbf{h}'}\,Z_{\vec{k}}({\bf v}',{\bf m}')=Z^{(N',M')}_{\text{hor}}\nonumber\\
&=Z_p({\bf h}',{\bf m}')\sum_{\vec{k}}\,e^{-\vec{k}\cdot \mathbf{v}'}\,Z_{\vec{k}}({\bf h}',{\bf m}')=Z^{(N',M')}_{\text{vert}}\nonumber\\
&=Z_p({\bf h}',{\bf v}')\sum_{\vec{k}}\,e^{-\vec{k}\cdot \mathbf{m}'}\,Z_{\vec{k}}({\bf h}',\ {\bf v}')=Z^{(N',M')}_{\text{diag}}\,,\label{TrialitySchemat2}
\end{align}
where the invariance (\ref{InvariancePartitionFunction}) was used. The series expansions $Z^{(N',M')}_{\text{hor}}$, $Z^{(N',M')}_{\text{vert}}$ and $Z^{(N',M')}_{\text{diag}}$ can be interpreted as the partition functions of quiver gauge theories with gauge groups
\begin{align}
&G'_{\text{hor}} = [U(M')]^{N'}\,,&&\text{and} && G'_{\text{vert}} = [U(N')]^{M'}\,, &&\text{and} && G'_{\text{diag}} = [U(M'N'/k)]^k\,, 
\label{trialityMN}
\end{align}
respectively. As their partition functions are equal, these three gauge theories are not only dual to one another, but are also dual to the theories with the gauge groups (\ref{DualGaugeTheories}).\footnote{Note that, even though $G_{\text{diag}}=G'_{\text{diag}}$, the coupling constants, Coulomb branch parameters and hypermultiplet masses may in general be different for the two theories because of the non-trivial duality map $(\mathbf{h},\mathbf{v},\mathbf{m}) \to(\mathbf{h}',\mathbf{v}',\mathbf{m}')$. In particular, the coupling constants may be shifted by functions of the remaining parameters of the theory.} From the perspective of the gauge theories, depending on the line segments of the web diagram that undergo flop transitions, there are two possibilities for the nature of these dualities:
\begin{itemize}
\item If none of the line segments that are associated with the coupling constants of the theory are flopped (\emph{i.e.} $\mathbf{h}$ for the horizontal, $\mathbf{v}$ for the vertical and $\mathbf{m}$ for the diagonal theories), the duality map corresponds to a finite shift of the coupling constants and a symmetry transformation of the Coulomb branch moduli. This can be interpreted as a finite reparametrisation of the Coulomb branch: both the theory before and after the combined symmetry and flop-transformation are in the weak coupling regime and their perturbative descriptions can directly be compared. 
\item If line segments associated with the coupling constants undergo flop transitions, the duality requires to pass through a region in which the instanton counting parameter becomes of order 1, \emph{i.e.} a regime in which the gauge coupling constants blow up. In this case, the duality cannot be understood from the perspective of the weakly coupled gauge theories alone: indeed, in the corresponding K\"ahler cone in \figref{Fig:FlopTransformations}, even before hitting the actual wall that signals the flop transition, the theory enters into a strong coupling phase, in which the description in terms of a (quiver) gauge theory breaks down and needs to be replaced (in general) by an LST.
\end{itemize} 
We discuss an explicit example (namely the duality between theories engineerd by $X_{6,5}$ and $X_{10,3}$) in section~\ref{Sect:Example65}.
%%%%%%%%%%%%%%%%%
\subsection{Independent K\"ahler Parameters and Gauge Group Structure}\label{Sect:BaseChoice}
In writing the series expansions (\ref{TrialitySchemat}) and (\ref{TrialitySchemat2}), we have not yet addressed an important technical aspect, namely the choice of independent K\"ahler parameters, which is closely related to the gauge groups (\ref{DualGaugeTheories}) and (\ref{trialityMN}), respectively \cite{Bastian:2017ary}. As alluded to briefly above, not all of the $3NM$ areas of the curves in the web diagram of $X_{N,M}$ are independent. Instead, there are consistency conditions among the former, which ultimately leave $NM+2$ independent K\"ahler parameters. These can be chosen in various different manners, but in \cite{Bastian:2017ary} three different proposals for a choice were made, each of which suitable for one of the weak coupling regions in the K\"ahler cone of $X_{N,M}$. Indeed, considering the low-energy theory with gauge group
\begin{align}
&G=[U(A)]^B\,,&&\text{with} &&(A,B)=\left\{\begin{array}{lcl}(M,N) & \ldots & \text{horizontal theory,} \\ (N,M) & \ldots & \text{vertical theory,} \\ (\tfrac{MN}{k},k) & \ldots & \text{diagonal theory,}   \end{array}\right.\label{GeneralGroups}
\end{align}
such that $AB=NM$ and $\text{gcd}(A,B)=k=\text{gcd}(N,M)$, the $NM+2$ independent parameters are organised as follows:
\begin{itemize}
\item $B$ parameters related to the coupling constants $g_{1,\ldots,B}$ of the $B$ different $U(A)$ gauge factors
\item $B$ sets of $A-1$ parameters representing the roots of $B$ copies of the Lie algebra $\mathfrak{a}_{A-1}$
\item a single parameter that is related to an affine root thereby extending each of the $B$ copies of $\mathfrak{a}_{A-1}$ to affine algebras 
\item a single parameter which is related to the mass scale in the hypermultiplet sector
\end{itemize}
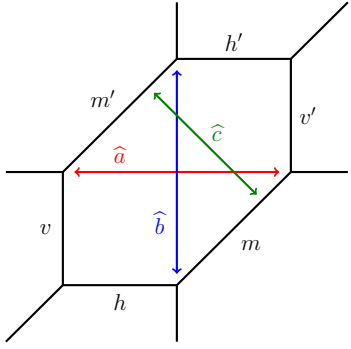
\begin{wrapfigure}{L}{0.3\textwidth}
\centering
\scalebox{0.5}{\parbox{9cm}{\begin{tikzpicture}[scale = 1.5]
\draw[ultra thick] (0,0) -- (2,0) -- (4,2) -- (4,4) -- (2,4) -- (0,2) -- (0,0);
\draw[ultra thick] (-1,-1) -- (0,0);
\draw[ultra thick] (4,4) -- (5,5);
\draw[ultra thick] (2,0) -- (2,-1);
\draw[ultra thick] (4,2) -- (5,2);
\draw[ultra thick] (2,4) -- (2,5); 
\draw[ultra thick] (0,2) -- (-1,2); 
%lines
\node at (1,-0.3) {\Large $h$};
\node at (3,4.3) {\Large$h'$};
\node at (-0.3,1) {\Large$v$};
\node at (4.3,3) {\Large$v'$};
\node at (0.7,3.3) {\Large$m'$};
\node at (3.3,0.7) {\Large$m$};
%roots
\draw[red, ultra thick,<->] (0.2,2) -- (3.8,2);
\node[red] at (1,2.3) {\Large $\widehat{a}$};
\draw[blue, ultra thick,<->] (2,0.2) -- (2,3.8);
\node[blue] at (1.7,1.1) {\Large $\widehat{b}$};
\draw[green!50!black, ultra thick,<->] (1.6,3.4) -- (3.4,1.6);
\node[green!50!black] at (2.7,2.7) {\Large $\widehat{c}$};
\end{tikzpicture}}}
\caption{\sl Three classes of curves within a given hexagon $S_i$ whose areas are identified with roots in the gauge theory perspective.}
\label{Fig:22Roots}
\end{wrapfigure}
These are indeed a total of $B+B(A-1)+1+1=NM+2$ parameters, which have been conjectured in \cite{Haghighat:2013tka, Bastian:2017ary} to be independent of one another\footnote{This has explicitly been verified for a large number of examples.} and thus represent a viable parametrisation of the K\"ahler cone of $X_{N,M}$. 

The curves whose areas make up the roots of the gauge algebras $\mathfrak{a}_{A-1}$, can be identified directly in the web diagram in~\figref{Fig:WebDiagramGeneric}. Indeed, focusing on a particular hexagon in the latter, as shown in~\figref{Fig:22Roots}, we can identify the roots $\widehat{a}$ associated with the gauge theory stemming from the horizontal expansion, $\widehat{b}$ associated with the gauge theory stemming from the vertical expansion and $\widehat{c}$ associated with the gauge theory stemming from the diagonal expansion. The distinction between the finite (positive simple) roots and the affine root of each $\widehat{\mathfrak{a}}_{A-1}$ is a choice that fixes an automorphism of $\widehat{\mathfrak{a}}_{A-1}$. The definition of the remaining parameters mentioned above (particularly the one related to the mass scale of the hypermultiplet sector of the gauge theory) is somewhat more involved and is related to the fact that the web diagram is defined on a torus (\emph{i.e.} the fact that all external legs in~\figref{Fig:WebDiagramGeneric} are glued together). Since these precise definitions are not important for the current work, we refer the reader to \cite{Bastian:2017ary} for more information and explicit examples. 
%%%%%%%%%%%%%%%%%%%%%%%%%%%
%%%%%%%%%%%%%%%%%%%%%%%%%%%%%%%%%%%%%%%%%%%
%%%%%%%%%%%%%%%%%%%%%%%%%%%%%%%%%%%%%%%%%%%%%%
\subsection{Example: $(N,M)=(6,5)$}\label{Sect:Example65}
\begin{figure}[htbp]
\begin{center}
\rotatebox{90}{\scalebox{0.65}{\parbox{29cm}{\begin{tikzpicture}[scale = 1.50]
\draw[ultra thick] (-1,0) -- (0,0) -- (1,1) -- (2,1) -- (3,2) -- (4,2) -- (5,3) -- (6,3) -- (7,4) -- (8,4) -- (9,5) -- (10,5) -- (11,6) -- (12,6);
\draw[ultra thick] (0,2) -- (1,2) -- (2,3) -- (3,3) -- (4,4) -- (5,4) -- (6,5) -- (7,5) -- (8,6) -- (9,6) -- (10,7) -- (11,7) -- (12,8) -- (13,8);
\draw[ultra thick] (1,4) -- (2,4) -- (3,5) -- (4,5) -- (5,6) -- (6,6) -- (7,7) -- (8,7) -- (9,8) -- (10,8) -- (11,9) -- (12,9) -- (13,10) -- (14,10);
\draw[ultra thick] (2,6) -- (3,6) -- (4,7) -- (5,7) -- (6,8) -- (7,8) -- (8,9) -- (9,9) -- (10,10) -- (11,10) -- (12,11) -- (13,11) -- (14,12) -- (15,12);
\draw[ultra thick] (3,8) -- (4,8) -- (5,9) -- (6,9) -- (7,10) -- (8,10) -- (9,11) -- (10,11) -- (11,12) -- (12,12) -- (13,13) -- (14,13) -- (15,14) -- (16,14);
%bottom vertical
\draw[ultra thick] (0,-1) -- (0,0);
\draw[ultra thick] (2,0) -- (2,1);
\draw[ultra thick] (4,1) -- (4,2);
\draw[ultra thick] (6,2) -- (6,3);
\draw[ultra thick] (8,3) -- (8,4);
\draw[ultra thick] (10,4) -- (10,5);
%first vertical
\draw[ultra thick] (1,1) -- (1,2);
\draw[ultra thick] (3,2) -- (3,3);
\draw[ultra thick] (5,3) -- (5,4);
\draw[ultra thick] (7,4) -- (7,5);
\draw[ultra thick] (9,5) -- (9,6);
\draw[ultra thick] (11,6) -- (11,7);
%second vertical
\draw[ultra thick] (2,3) -- (2,4);
\draw[ultra thick] (4,4) -- (4,5);
\draw[ultra thick] (6,5) -- (6,6);
\draw[ultra thick] (8,6) -- (8,7);
\draw[ultra thick] (10,7) -- (10,8);
\draw[ultra thick] (12,8) -- (12,9);
%third vertical
\draw[ultra thick] (3,5) -- (3,6);
\draw[ultra thick] (5,6) -- (5,7);
\draw[ultra thick] (7,7) -- (7,8);
\draw[ultra thick] (9,8) -- (9,9);
\draw[ultra thick] (11,9) -- (11,10);
\draw[ultra thick] (13,10) -- (13,11);
%fourth vertical
\draw[ultra thick] (4,7) -- (4,8);
\draw[ultra thick] (6,8) -- (6,9);
\draw[ultra thick] (8,9) -- (8,10);
\draw[ultra thick] (10,10) -- (10,11);
\draw[ultra thick] (12,11) -- (12,12);
\draw[ultra thick] (14,12) -- (14,13);
%top vertical
\draw[ultra thick] (5,9) -- (5,10);
\draw[ultra thick] (7,10) -- (7,11);
\draw[ultra thick] (9,11) -- (9,12);
\draw[ultra thick] (11,12) -- (11,13);
\draw[ultra thick] (13,13) -- (13,14);
\draw[ultra thick] (15,14) -- (15,15);
%ID labels bottom
\node at (0,-1.2) {{\small \bf $1$}};
\node at (2,-0.2) {{\small \bf $2$}};
\node at (4,0.8) {{\small \bf $3$}};
\node at (6,1.8) {{\small \bf $4$}};
\node at (8,2.8) {{\small \bf $5$}};
\node at (10,3.8) {{\small \bf $6$}};
%ID labels top
\node at (5,10.2) {{\small \bf $1$}};
\node at (7,11.2) {{\small \bf $2$}};
\node at (9,12.2) {{\small \bf $3$}};
\node at (11,13.2) {{\small \bf $4$}};
\node at (13,14.2) {{\small \bf $5$}};
\node at (15,15.2) {{\small \bf $6$}};
%ID labels left
\node at (-1.2,0) {{\small \bf $a$}};
\node at (-0.2,2) {{\small \bf $b$}};
\node at (0.8,4) {{\small \bf $c$}};
\node at (1.8,6) {{\small \bf $d$}};
\node at (2.8,8) {{\small \bf $e$}};
%ID labels right
\node at (12.2,6) {{\small \bf $a$}};
\node at (13.2,8) {{\small \bf $b$}};
\node at (14.2,10) {{\small \bf $c$}};
\node at (15.2,12) {{\small \bf $d$}};
\node at (16.2,14) {{\small \bf $e$}};
%vertical-labels bottom layer
\node at (-0.2,-0.5) {{\small \bf $v_1$}};
\node at (1.8,0.5) {{\small \bf $v_2$}};
\node at (3.8,1.5) {{\small \bf $v_3$}};
\node at (5.8,2.5) {{\small \bf $v_4$}};
\node at (7.8,3.5) {{\small \bf $v_5$}};
\node at (9.8,4.5) {{\small \bf $v_6$}};
%vertical-labels first layer
\node at (0.75,1.5) {{\small \bf $v_{7}$}};
\node at (2.75,2.5) {{\small \bf $v_{8}$}};
\node at (4.75,3.5) {{\small \bf $v_{9}$}};
\node at (6.75,4.5) {{\small \bf $v_{10}$}};
\node at (8.75,5.5) {{\small \bf $v_{11}$}};
\node at (10.75,6.5) {{\small \bf $v_{12}$}};
%vertical-labels second layer
\node at (1.75,3.5) {{\small \bf $v_{13}$}};
\node at (3.75,4.5) {{\small \bf $v_{14}$}};
\node at (5.75,5.5) {{\small \bf $v_{15}$}};
\node at (7.75,6.5) {{\small \bf $v_{16}$}};
\node at (9.75,7.5) {{\small \bf $v_{17}$}};
\node at (11.75,8.5) {{\small \bf $v_{18}$}};
%vertical-labels third layer
\node at (2.75,5.5) {{\small \bf $v_{19}$}};
\node at (4.75,6.5) {{\small \bf $v_{20}$}};
\node at (6.75,7.5) {{\small \bf $v_{21}$}};
\node at (8.75,8.5) {{\small \bf $v_{22}$}};
\node at (10.75,9.5) {{\small \bf $v_{23}$}};
\node at (12.75,10.5) {{\small \bf $v_{24}$}};
%vertical-labels fourth layer
\node at (3.75,7.5) {{\small \bf $v_{25}$}};
\node at (5.75,8.5) {{\small \bf $v_{26}$}};
\node at (7.75,9.5) {{\small \bf $v_{27}$}};
\node at (9.75,10.5) {{\small \bf $v_{28}$}};
\node at (11.75,11.5) {{\small \bf $v_{29}$}};
\node at (13.75,12.5) {{\small \bf $v_{30}$}};
%vertical-labels top layer
\node at (4.8,9.5) {{\small \bf $v_{1}$}};
\node at (6.8,10.5) {{\small \bf $v_{2}$}};
\node at (8.8,11.5) {{\small \bf $v_{3}$}};
\node at (10.8,12.5) {{\small \bf $v_{4}$}};
\node at (12.8,13.5) {{\small \bf $v_{5}$}};
\node at (14.8,14.5) {{\small \bf $v_{6}$}};
%horizontal-labels first layer
\node at (-0.5,0.2) {{\small \bf $h_{6}$}};
\node at (1.5,1.2) {{\small \bf $h_{1}$}};
\node at (3.5,2.2) {{\small \bf $h_{2}$}};
\node at (5.5,3.2) {{\small \bf $h_{3}$}};
\node at (7.5,4.2) {{\small \bf $h_{4}$}};
\node at (9.5,5.2) {{\small \bf $h_{5}$}};
\node at (11.5,6.2) {{\small \bf $h_{6}$}};
%horizontal-labels second layer
\node at (0.5,2.2) {{\small \bf $h_{12}$}};
\node at (2.5,3.2) {{\small \bf $h_{7}$}};
\node at (4.5,4.2) {{\small \bf $h_{8}$}};
\node at (6.5,5.2) {{\small \bf $h_{9}$}};
\node at (8.5,6.2) {{\small \bf $h_{10}$}};
\node at (10.5,7.2) {{\small \bf $h_{11}$}};
\node at (12.5,8.2) {{\small \bf $h_{12}$}};
%horizontal-labels third layer
\node at (1.5,4.2) {{\small \bf $h_{18}$}};
\node at (3.5,5.2) {{\small \bf $h_{13}$}};
\node at (5.5,6.2) {{\small \bf $h_{14}$}};
\node at (7.5,7.2) {{\small \bf $h_{15}$}};
\node at (9.5,8.2) {{\small \bf $h_{16}$}};
\node at (11.5,9.2) {{\small \bf $h_{17}$}};
\node at (13.5,10.2) {{\small \bf $h_{18}$}};
%horizontal-labels fourth layer
\node at (2.5,6.2) {{\small \bf $h_{24}$}};
\node at (4.5,7.2) {{\small \bf $h_{19}$}};
\node at (6.5,8.2) {{\small \bf $h_{20}$}};
\node at (8.5,9.2) {{\small \bf $h_{21}$}};
\node at (10.5,10.2) {{\small \bf $h_{22}$}};
\node at (12.5,11.2) {{\small \bf $h_{23}$}};
\node at (14.5,12.2) {{\small \bf $h_{24}$}};
%horizontal-labels top layer
\node at (3.5,8.2) {{\small \bf $h_{30}$}};
\node at (5.5,9.2) {{\small \bf $h_{25}$}};
\node at (7.5,10.2) {{\small \bf $h_{26}$}};
\node at (9.5,11.2) {{\small \bf $h_{27}$}};
\node at (11.5,12.2) {{\small \bf $h_{28}$}};
\node at (13.5,13.2) {{\small \bf $h_{29}$}};
\node at (15.5,14.2) {{\small \bf $h_{30}$}};
%diagonal-labels bottom layer
\node at (0.65,0.3) {{\small \bf $m_1$}};
\node at (2.65,1.3) {{\small \bf $m_{2}$}};
\node at (4.65,2.3) {{\small \bf $m_{3}$}};
\node at (6.65,3.3) {{\small \bf $m_{4}$}};
\node at (8.65,4.3) {{\small \bf $m_{5}$}};
\node at (10.65,5.3) {{\small \bf $m_{6}$}};
%diagonal-labels first layer
\node at (1.7,2.3) {{\small \bf $m_{7}$}};
\node at (3.7,3.3) {{\small \bf $m_{8}$}};
\node at (5.7,4.3) {{\small \bf $m_{9}$}};
\node at (7.7,5.3) {{\small \bf $m_{10}$}};
\node at (9.7,6.3) {{\small \bf $m_{11}$}};
\node at (11.7,7.3) {{\small \bf $m_{12}$}};
%diagonal-labels second layer
\node at (2.7,4.3) {{\small \bf $m_{13}$}};
\node at (4.7,5.3) {{\small \bf $m_{14}$}};
\node at (6.7,6.3) {{\small \bf $m_{15}$}};
\node at (8.7,7.3) {{\small \bf $m_{16}$}};
\node at (10.7,8.3) {{\small \bf $m_{17}$}};
\node at (12.7,9.3) {{\small \bf $m_{18}$}};
%diagonal-labels third layer
\node at (3.7,6.3) {{\small \bf $m_{19}$}};
\node at (5.7,7.3) {{\small \bf $m_{20}$}};
\node at (7.7,8.3) {{\small \bf $m_{21}$}};
\node at (9.7,9.3) {{\small \bf $m_{22}$}};
\node at (11.7,10.3) {{\small \bf $m_{23}$}};
\node at (13.7,11.3) {{\small \bf $m_{24}$}};
%diagonal-labels top layer
\node at (4.7,8.3) {{\small \bf $m_{25}$}};
\node at (6.7,9.3) {{\small \bf $m_{26}$}};
\node at (8.7,10.3) {{\small \bf $m_{27}$}};
\node at (10.7,11.3) {{\small \bf $m_{28}$}};
\node at (12.7,12.3) {{\small \bf $m_{29}$}};
\node at (14.7,13.3) {{\small \bf $m_{30}$}};
%%%%%%%%%%%%%%%%%%%%%%%%%%%%%%%%%%%%%%%%%%%%%%%
%roots bottom layer
\draw[thick,<->,red] (1.7,-0.2) -- (0.8,0.7);
\node[red] at (1.25,-0.1) {{\small $\widehat{a}_{26}$}};
\draw[thick,<->,red] (3.7,0.8) -- (2.8,1.7);
\node[red] at (3.15,1) {{\small $\widehat{a}_{21}$}};
\draw[thick,<->,red] (5.7,1.8) -- (4.8,2.7);
\node[red] at (5.1,1.95) {{\small $\widehat{a}_{16}$}};
\draw[thick,<->,red] (7.7,2.8) -- (6.8,3.7);
\node[red] at (7.1,3) {{\small $\widehat{a}_{11}$}};
\draw[thick,<->,red] (9.7,3.8) -- (8.8,4.7);
\node[red] at (9.2,4) {{\small $\widehat{a}_{6}$}};
\draw[thick,<->,red] (11.7,4.8) -- (10.8,5.7);
\node[red] at (11.2,5) {{\small $\widehat{a}_1$}};
%roots second layer
\draw[thick,<->,red] (0.7,0.8) -- (-0.2,1.7);
\node[red] at (0.15,1) {{\small $\widehat{a}_{25}$}};
\draw[thick,<->,red] (2.7,1.8) -- (1.8,2.7);
\node[red] at (2.1,2) {{\small $\widehat{a}_{20}$}};
\draw[thick,<->,red] (4.7,2.8) -- (3.8,3.7);
\node[red] at (4.2,2.9) {{\small $\widehat{a}_{15}$}};
\draw[thick,<->,red] (6.7,3.8) -- (5.8,4.7);
\node[red] at (6.1,4) {{\small $\widehat{a}_{10}$}};
\draw[thick,<->,red] (8.7,4.8) -- (7.8,5.7);
\node[red] at (8.2,5) {{\small $\widehat{a}_{5}$}};
\draw[thick,<->,red] (10.7,5.8) -- (9.8,6.7);
\node[red] at (10.15,5.95) {{\small $\widehat{a}_{30}$}};
%roots third layer
\draw[thick,<->,red] (1.7,2.8) -- (0.8,3.7);
\node[red] at (1.1,3) {{\small $\widehat{a}_{19}$}};
\draw[thick,<->,red] (3.7,3.8) -- (2.8,4.7);
\node[red] at (3.15,3.95) {{\small $\widehat{a}_{14}$}};
\draw[thick,<->,red] (5.7,4.8) -- (4.8,5.7);
\node[red] at (5.2,5) {{\small $\widehat{a}_{9}$}};
\draw[thick,<->,red] (7.7,5.8) -- (6.8,6.7);
\node[red] at (7.25,5.9) {{\small $\widehat{a}_{4}$}};
\draw[thick,<->,red] (9.7,6.8) -- (8.8,7.7);
\node[red] at (9.15,7) {{\small $\widehat{a}_{29}$}};
\draw[thick,<->,red] (11.7,7.8) -- (10.8,8.7);
\node[red] at (11.1,8) {{\small $\widehat{a}_{24}$}};
%roots fourth layer
\draw[thick,<->,red] (2.7,4.8) -- (1.8,5.7);
\node[red] at (2.1,5) {{\small $\widehat{a}_{13}$}};
\draw[thick,<->,red] (4.7,5.8) -- (3.8,6.7);
\node[red] at (4.2,5.95) {{\small $\widehat{a}_8$}};
\draw[thick,<->,red] (6.7,6.8) -- (5.8,7.7);
\node[red] at (6.2,7) {{\small $\widehat{a}_{3}$}};
\draw[thick,<->,red] (8.7,7.8) -- (7.8,8.7);
\node[red] at (8.1,8) {{\small $\widehat{a}_{28}$}};
\draw[thick,<->,red] (10.7,8.8) -- (9.8,9.7);
\node[red] at (10.2,8.9) {{\small $\widehat{a}_{23}$}};
\draw[thick,<->,red] (12.7,9.8) -- (11.8,10.7);
\node[red] at (12.1,10) {{\small $\widehat{a}_{18}$}};
%roots fifth layer
\draw[thick,<->,red] (3.7,6.8) -- (2.8,7.7);
\node[red] at (3.1,7) {{\small $\widehat{a}_{7}$}};
\draw[thick,<->,red] (5.7,7.8) -- (4.8,8.7);
\node[red] at (5.15,7.95) {{\small $\widehat{a}_2$}};
\draw[thick,<->,red] (7.7,8.8) -- (6.8,9.7);
\node[red] at (7.1,9) {{\small $\widehat{a}_{27}$}};
\draw[thick,<->,red] (9.7,9.8) -- (8.8,10.7);
\node[red] at (9.2,10) {{\small $\widehat{a}_{22}$}};
\draw[thick,<->,red] (11.7,10.8) -- (10.8,11.7);
\node[red] at (11.15,11) {{\small $\widehat{a}_{17}$}};
\draw[thick,<->,red] (13.7,11.8) -- (12.8,12.7);
\node[red] at (13.2,11.9) {{\small $\widehat{a}_{12}$}};
%%%%M-line
\draw[thick,<->,red] (-0.6,-1.4) -- (-2.9,0.9);
\node[red] at (-2,-0.5) {{\small $M$}};
\draw[dashed] (4,8) -- (-3,1);
\draw[dashed] (14,13) -- (-0.5,-1.5);
%%%%rho line
\draw[thick,red,<->] (0,-1.55) -- (12,-1.55);
\node[red] at (6,-1.9) {$6V-54M+5\sum_{i=1}^{30}\widehat{a}_i$};
\end{tikzpicture}}}}
\caption{\sl Parametrisation of the $(6,5)$ web diagram: out of the 90 curves $(h_i,v_i,m_i)$ (for $i=1,\ldots,30$) only 32 parameters are independent. The red curves constitute a maximal set of independent parameters which makes a $U(30)$ symmetry visible.}
\label{Fig:WebDiag65}
\end{center}
\end{figure}
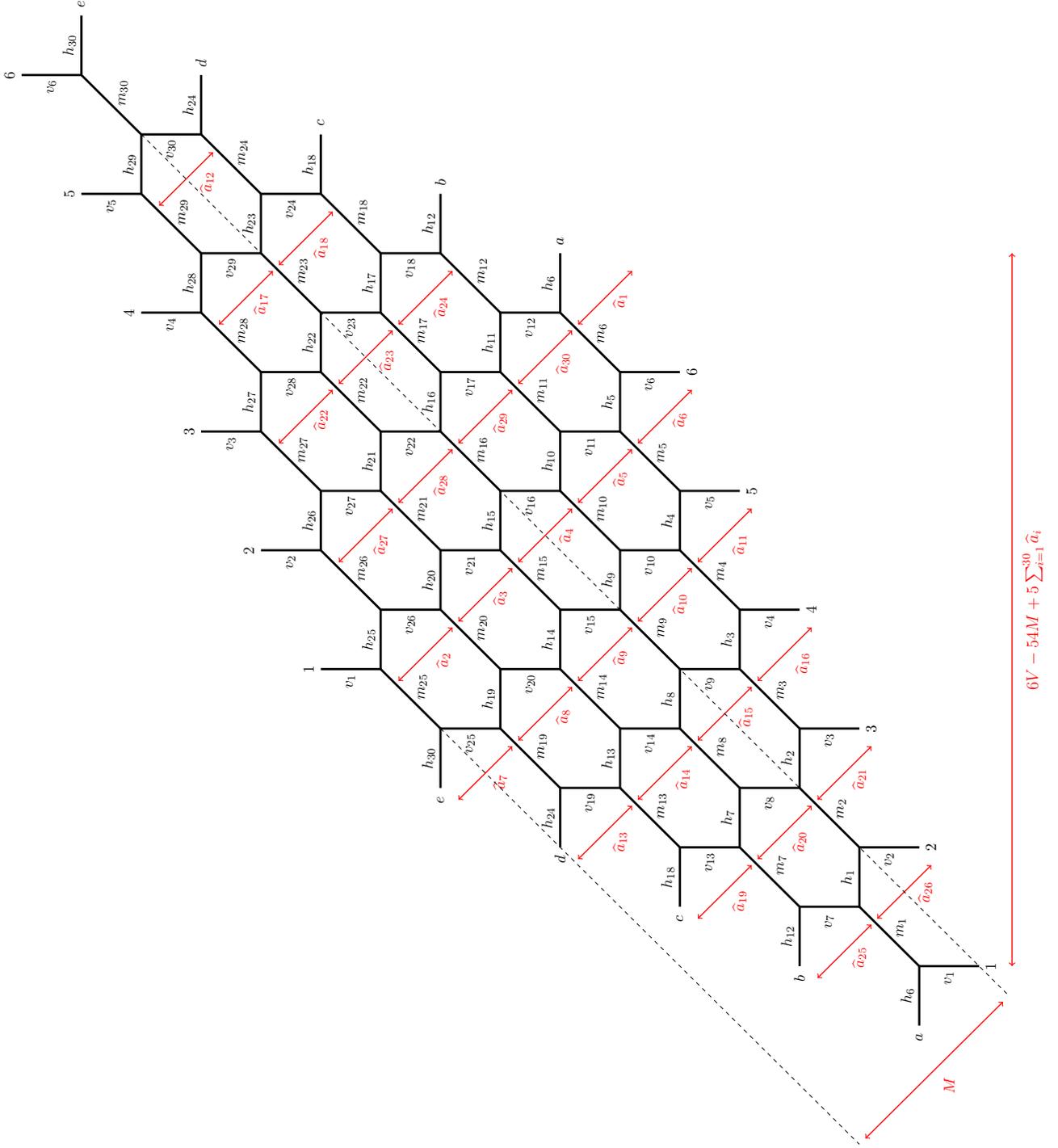

To illustrate the fact that the combination of eq.(\ref{DualityCalabiYauFlop}) with the triality in the sense of eq.(\ref{TrialitySchemat}) leads to a large web of dual theories, we discuss in more detail a non-trivial example, namely $(N,M)=(6,5)$ (with $k=\text{gcd}(6,5)=1$). The corresponding web diagram is shown in \figref{Fig:WebDiag65}.
%%%%%%%%%%%%%%%%%%%%%%%%%%%%%%%%%%%%%%%%%%%%%%
%%%%%%%%%%%%%%%%%%%%%%%%%%%%
%%%%%%%%%%%%%%%%%%%%%%%%%%%%%%%%%%%%%%%%%%%%%%
%%%%%%%%%%%%%%%%%%%%%%%%%%%%%%%%%%%%%%%%%%%%%%
According to \cite{Hohenegger:2013ala,Haghighat:2013tka},  the horizontal and vertical expansions can be associated with gauge theories of gauge group $[U(5)]^6$ and $[U(6)]^5$, respectively, while according to \cite{Bastian:2017ary}, the diagonal expansion gives rise to a gauge theory with gauge group $U(30)$. To obtain the latter theory, in particular, to make the structure of $U(30)$ manifest, we need to expand $\pf{6}{5}$ in terms of a specific set of variables, a procedure which was proposed in full generality in \cite{Bastian:2017ary}. Concretely, in the present case, these variables are depicted in red in \figref{Fig:WebDiag65} and consist of $(M,V,\widehat{a}_{1,\ldots,30})$, where $M$ and $V$ are given explicitly as 
{\allowdisplaybreaks
\begin{align}
M&=h_6 + v_1 + h_{25} + v_{26} + h_{20} + v_{21} + h_{15} + v_{16} + h_{10} + v_{11} + h_5\,,\nonumber\\
V&=m_{30} + (6 - 1) (h_{29} + h_{30}) + (6 - 2) (v_5 + h_4 + v_{25} + h_{19})\nonumber\\
&\hspace{1.25cm} + (6 -
    3) (v_{10} + h_9 + v_{20} + h_{14}) + (6 - 4) (v_{15} + h_{14} + v_{15} + h_9) \nonumber\\
&\hspace{1.25cm}    + (6 -
    5) (v_{20} + h_{19} + v_{10} + h_4)\,.
\end{align}}
Here, the last relation follows from the general duality map relating $X_{6,5}\sim X_{30,1}$ that was conjectured previously in \cite{Hohenegger:2016yuv}. In this basis, we have\footnote{We stress that all 90 parameters $(h_i,v_i,m_i)$ can be expressed as linear combinations of the 32 elements $(M,V,\widehat{a}_{1,\ldots,30})$. However, we refrain to write down these relations explicitly, since they will not be needed for the following discussions. }
\begin{align}
&m_i=V+p_i(M,\widehat{a}_{1,\ldots,30})\,,&&\forall \ i=1,\ldots,30\,,\label{DualFrame1V}
\end{align}
where $p_i$ are multi-linear functions in the 31 variables $(M,\widehat{a}_{1,\ldots,30})$, while $h_{1,\ldots,30}$ and $v_{1,\ldots,30}$ are independent of $V$. Thus, formulated in a different manner, the diagonal expansion written schematically in (\ref{TrialitySchemat}), can be understood as a power series expansion in $Q_V=e^{-V}$. Furthermore, the $\widehat{a}_i$ play the role of the roots of $\widehat{\mathfrak{a}}_{29}$, \emph{i.e.} the affine extension of the Lie algebra associated with the gauge group $U(30)$ that is associated with the diagonal expansion: indeed, in the weak coupling limit $V\to \infty$, the diagonal lines in \figref{Fig:WebDiag65} are cut (as the area of the corresponding curves in the toric Calabi-Yau threefold becomes infinite) and the remaining diagram can be presented as a single strip of length 30.

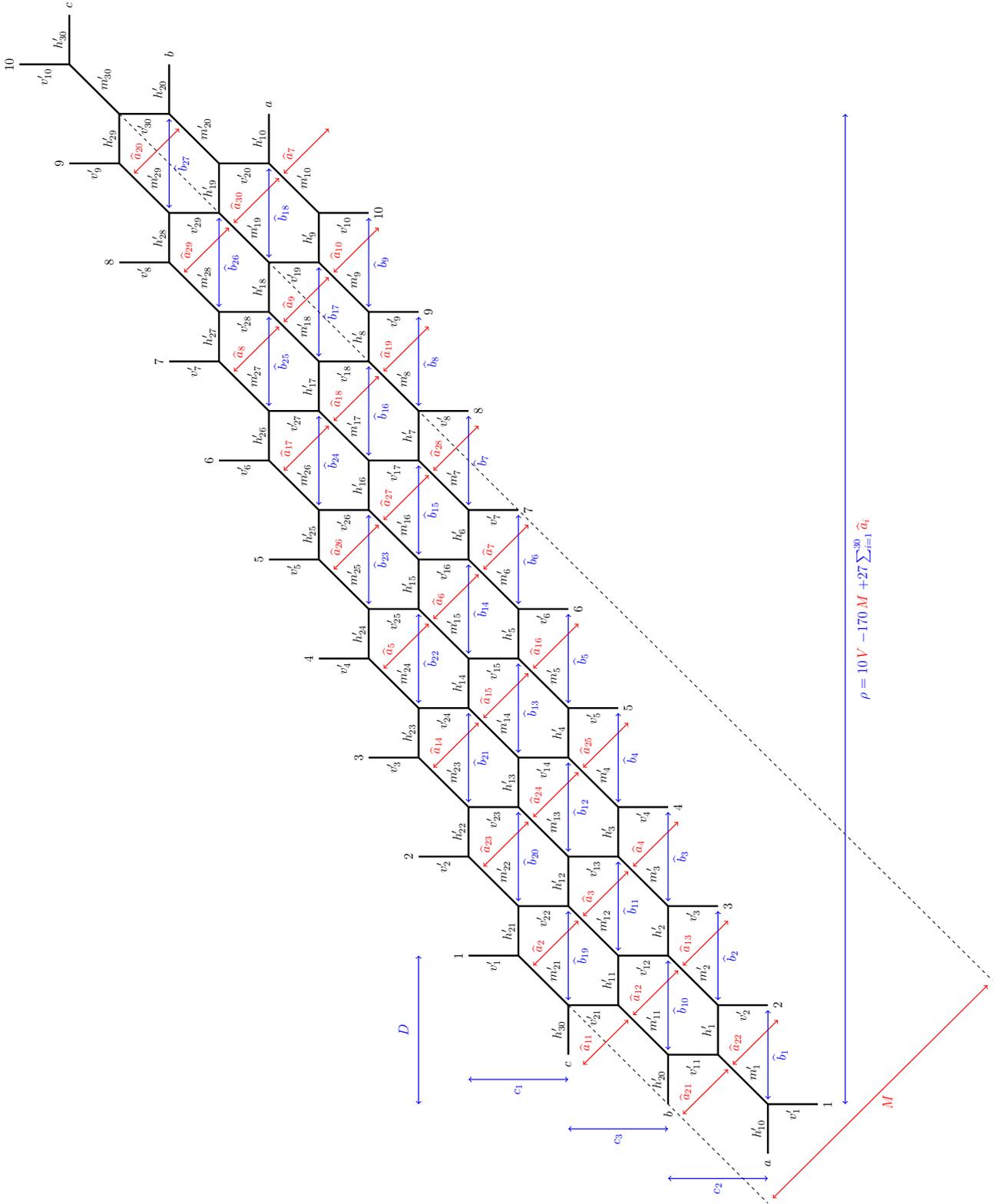
\begin{figure}[htbp]
\rotatebox{90}{\scalebox{0.57}{\parbox{36.5cm}{\begin{tikzpicture}[scale = 1.50]
\draw[ultra thick] (-1,0) -- (0,0) -- (1,1) -- (2,1) -- (3,2) -- (4,2) -- (5,3) -- (6,3) -- (7,4) -- (8,4) -- (9,5) -- (10,5) -- (11,6) -- (12,6) -- (13,7) -- (14,7) -- (15,8) -- (16,8) -- (17,9) -- (18,9) -- (19,10) -- (20,10);
\draw[ultra thick] (0,2) -- (1,2) -- (2,3) -- (3,3) -- (4,4) -- (5,4) -- (6,5) -- (7,5) -- (8,6) -- (9,6) -- (10,7) -- (11,7) -- (12,8) -- (13,8) -- (14,9) -- (15,9) -- (16,10) -- (17,10) -- (18,11) -- (19,11) -- (20,12) -- (21,12);
\draw[ultra thick] (1,4) -- (2,4) -- (3,5) -- (4,5) -- (5,6) -- (6,6) -- (7,7) -- (8,7) -- (9,8) -- (10,8) -- (11,9) -- (12,9) -- (13,10) -- (14,10) -- (15,11) -- (16,11) -- (17,12) -- (18,12) -- (19,13) -- (20,13) -- (21,14) -- (22,14);
%bottom vertical
\draw[ultra thick] (0,-1) -- (0,0);
\draw[ultra thick] (2,0) -- (2,1);
\draw[ultra thick] (4,1) -- (4,2);
\draw[ultra thick] (6,2) -- (6,3);
\draw[ultra thick] (8,3) -- (8,4);
\draw[ultra thick] (10,4) -- (10,5);
\draw[ultra thick] (12,5) -- (12,6);
\draw[ultra thick] (14,6) -- (14,7);
\draw[ultra thick] (16,7) -- (16,8);
\draw[ultra thick] (18,8) -- (18,9);
%first vertical
\draw[ultra thick] (1,1) -- (1,2);
\draw[ultra thick] (3,2) -- (3,3);
\draw[ultra thick] (5,3) -- (5,4);
\draw[ultra thick] (7,4) -- (7,5);
\draw[ultra thick] (9,5) -- (9,6);
\draw[ultra thick] (11,6) -- (11,7);
\draw[ultra thick] (13,7) -- (13,8);
\draw[ultra thick] (15,8) -- (15,9);
\draw[ultra thick] (17,9) -- (17,10);
\draw[ultra thick] (19,10) -- (19,11);
%second vertical
\draw[ultra thick] (2,3) -- (2,4);
\draw[ultra thick] (4,4) -- (4,5);
\draw[ultra thick] (6,5) -- (6,6);
\draw[ultra thick] (8,6) -- (8,7);
\draw[ultra thick] (10,7) -- (10,8);
\draw[ultra thick] (12,8) -- (12,9);
\draw[ultra thick] (14,9) -- (14,10);
\draw[ultra thick] (16,10) -- (16,11);
\draw[ultra thick] (18,11) -- (18,12);
\draw[ultra thick] (20,12) -- (20,13);
%top vertical
\draw[ultra thick] (3,5) -- (3,6);
\draw[ultra thick] (5,6) -- (5,7);
\draw[ultra thick] (7,7) -- (7,8);
\draw[ultra thick] (9,8) -- (9,9);
\draw[ultra thick] (11,9) -- (11,10);
\draw[ultra thick] (13,10) -- (13,11);
\draw[ultra thick] (15,11) -- (15,12);
\draw[ultra thick] (17,12) -- (17,13);
\draw[ultra thick] (19,13) -- (19,14);
\draw[ultra thick] (21,14) -- (21,15);
%ID labels bottom
\node at (0,-1.2) {{\small \bf $1$}};
\node at (2,-0.2) {{\small \bf $2$}};
\node at (4,0.8) {{\small \bf $3$}};
\node at (6,1.8) {{\small \bf $4$}};
\node at (8,2.8) {{\small \bf $5$}};
\node at (10,3.8) {{\small \bf $6$}};
\node at (12,4.8) {{\small \bf $7$}};
\node at (14,5.8) {{\small \bf $8$}};
\node at (16,6.8) {{\small \bf $9$}};
\node at (18,7.8) {{\small \bf $10$}};
%ID labels top
\node at (3,6.2) {{\small \bf $1$}};
\node at (5,7.2) {{\small \bf $2$}};
\node at (7,8.2) {{\small \bf $3$}};
\node at (9,9.2) {{\small \bf $4$}};
\node at (11,10.2) {{\small \bf $5$}};
\node at (13,11.2) {{\small \bf $6$}};
\node at (15,12.2) {{\small \bf $7$}};
\node at (17,13.2) {{\small \bf $8$}};
\node at (19,14.2) {{\small \bf $9$}};
\node at (21,15.2) {{\small \bf $10$}};
%ID labels left
\node at (-1.2,0) {{\small \bf $a$}};
\node at (-0.2,2) {{\small \bf $b$}};
\node at (0.8,4) {{\small \bf $c$}};
%ID labels right
\node at (20.2,10) {{\small \bf $a$}};
\node at (21.2,12) {{\small \bf $b$}};
\node at (22.2,14) {{\small \bf $c$}};
%vertical-labels bottom layer
\node at (-0.2,-0.5) {{\small \bf $v'_1$}};
\node at (1.8,0.5) {{\small \bf $v'_2$}};
\node at (3.8,1.5) {{\small \bf $v'_3$}};
\node at (5.8,2.5) {{\small \bf $v'_4$}};
\node at (7.8,3.5) {{\small \bf $v'_5$}};
\node at (9.8,4.5) {{\small \bf $v'_6$}};
\node at (11.8,5.5) {{\small \bf $v'_7$}};
\node at (13.8,6.5) {{\small \bf $v'_8$}};
\node at (15.8,7.5) {{\small \bf $v'_9$}};
\node at (17.75,8.5) {{\small \bf $v'_{10}$}};
%vertical-labels first layer
\node at (0.75,1.5) {{\small \bf $v'_{11}$}};
\node at (2.75,2.5) {{\small \bf $v'_{12}$}};
\node at (4.75,3.5) {{\small \bf $v'_{13}$}};
\node at (6.75,4.5) {{\small \bf $v'_{14}$}};
\node at (8.75,5.5) {{\small \bf $v'_{15}$}};
\node at (10.75,6.5) {{\small \bf $v'_{16}$}};
\node at (12.75,7.5) {{\small \bf $v'_{17}$}};
\node at (14.75,8.5) {{\small \bf $v'_{18}$}};
\node at (16.75,9.5) {{\small \bf $v'_{19}$}};
\node at (18.75,10.5) {{\small \bf $v'_{20}$}};
%vertical-labels second layer
\node at (1.75,3.5) {{\small \bf $v'_{21}$}};
\node at (3.75,4.5) {{\small \bf $v'_{22}$}};
\node at (5.75,5.5) {{\small \bf $v'_{23}$}};
\node at (7.75,6.5) {{\small \bf $v'_{24}$}};
\node at (9.75,7.5) {{\small \bf $v'_{25}$}};
\node at (11.75,8.5) {{\small \bf $v'_{26}$}};
\node at (13.75,9.5) {{\small \bf $v'_{27}$}};
\node at (15.75,10.5) {{\small \bf $v'_{28}$}};
\node at (17.75,11.5) {{\small \bf $v'_{29}$}};
\node at (19.75,12.5) {{\small \bf $v'_{30}$}};
%vertical-labels top layer
\node at (2.8,5.5) {{\small \bf $v'_{1}$}};
\node at (4.8,6.5) {{\small \bf $v'_{2}$}};
\node at (6.8,7.5) {{\small \bf $v'_{3}$}};
\node at (8.8,8.5) {{\small \bf $v'_{4}$}};
\node at (10.8,9.5) {{\small \bf $v'_{5}$}};
\node at (12.8,10.5) {{\small \bf $v'_{6}$}};
\node at (14.8,11.5) {{\small \bf $v'_{7}$}};
\node at (16.8,12.5) {{\small \bf $v'_{8}$}};
\node at (18.8,13.5) {{\small \bf $v'_{9}$}};
\node at (20.75,14.5) {{\small \bf $v'_{10}$}};
%horizontal-labels first layer
\node at (-0.5,0.2) {{\small \bf $h'_{10}$}};
\node at (1.5,1.2) {{\small \bf $h'_{1}$}};
\node at (3.5,2.2) {{\small \bf $h'_{2}$}};
\node at (5.5,3.2) {{\small \bf $h'_{3}$}};
\node at (7.5,4.2) {{\small \bf $h'_{4}$}};
\node at (9.5,5.2) {{\small \bf $h'_{5}$}};
\node at (11.5,6.2) {{\small \bf $h'_{6}$}};
\node at (13.5,7.2) {{\small \bf $h'_{7}$}};
\node at (15.5,8.2) {{\small \bf $h'_{8}$}};
\node at (17.5,9.2) {{\small \bf $h'_{9}$}};
\node at (19.5,10.2) {{\small \bf $h'_{10}$}};
%horizontal-labels second layer
\node at (0.5,2.2) {{\small \bf $h'_{20}$}};
\node at (2.5,3.2) {{\small \bf $h'_{11}$}};
\node at (4.5,4.2) {{\small \bf $h'_{12}$}};
\node at (6.5,5.2) {{\small \bf $h'_{13}$}};
\node at (8.5,6.2) {{\small \bf $h'_{14}$}};
\node at (10.5,7.2) {{\small \bf $h'_{15}$}};
\node at (12.5,8.2) {{\small \bf $h'_{16}$}};
\node at (14.5,9.2) {{\small \bf $h'_{17}$}};
\node at (16.5,10.2) {{\small \bf $h'_{18}$}};
\node at (18.5,11.2) {{\small \bf $h'_{19}$}};
\node at (20.5,12.2) {{\small \bf $h'_{20}$}};
%horizontal-labels top layer
\node at (1.5,4.2) {{\small \bf $h'_{30}$}};
\node at (3.5,5.2) {{\small \bf $h'_{21}$}};
\node at (5.5,6.2) {{\small \bf $h'_{22}$}};
\node at (7.5,7.2) {{\small \bf $h'_{23}$}};
\node at (9.5,8.2) {{\small \bf $h'_{24}$}};
\node at (11.5,9.2) {{\small \bf $h'_{25}$}};
\node at (13.5,10.2) {{\small \bf $h'_{26}$}};
\node at (15.5,11.2) {{\small \bf $h'_{27}$}};
\node at (17.5,12.2) {{\small \bf $h'_{28}$}};
\node at (19.5,13.2) {{\small \bf $h'_{29}$}};
\node at (21.5,14.2) {{\small \bf $h'_{30}$}};
%diagonal-labels bottom layer
\node at (0.65,0.3) {{\small \bf $m'_1$}};
\node at (2.65,1.3) {{\small \bf $m'_{2}$}};
\node at (4.65,2.3) {{\small \bf $m'_{3}$}};
\node at (6.65,3.3) {{\small \bf $m'_{4}$}};
\node at (8.65,4.3) {{\small \bf $m'_{5}$}};
\node at (10.65,5.3) {{\small \bf $m'_{6}$}};
\node at (12.65,6.3) {{\small \bf $m'_{7}$}};
\node at (14.65,7.3) {{\small \bf $m'_{8}$}};
\node at (16.65,8.3) {{\small \bf $m'_{9}$}};
\node at (18.7,9.3) {{\small \bf $m'_{10}$}};
%diagonal-labels first layer
\node at (1.7,2.3) {{\small \bf $m'_{11}$}};
\node at (3.7,3.3) {{\small \bf $m'_{12}$}};
\node at (5.7,4.3) {{\small \bf $m'_{13}$}};
\node at (7.7,5.3) {{\small \bf $m'_{14}$}};
\node at (9.7,6.3) {{\small \bf $m'_{15}$}};
\node at (11.7,7.3) {{\small \bf $m'_{16}$}};
\node at (13.7,8.3) {{\small \bf $m'_{17}$}};
\node at (15.7,9.3) {{\small \bf $m'_{18}$}};
\node at (17.7,10.3) {{\small \bf $m'_{19}$}};
\node at (19.7,11.3) {{\small \bf $m'_{20}$}};
%diagonal-labels second layer
\node at (2.7,4.3) {{\small \bf $m'_{21}$}};
\node at (4.7,5.3) {{\small \bf $m'_{22}$}};
\node at (6.7,6.3) {{\small \bf $m'_{23}$}};
\node at (8.7,7.3) {{\small \bf $m'_{24}$}};
\node at (10.7,8.3) {{\small \bf $m'_{25}$}};
\node at (12.7,9.3) {{\small \bf $m'_{26}$}};
\node at (14.7,10.3) {{\small \bf $m'_{27}$}};
\node at (16.7,11.3) {{\small \bf $m'_{28}$}};
\node at (18.7,12.3) {{\small \bf $m'_{29}$}};
\node at (20.7,13.3) {{\small \bf $m'_{30}$}};
%%%
%%%%%basis (30,1)%%%%
%%%
%roots bottom layer
\draw[thick,<->,red] (1.7,-0.2) -- (0.8,0.7);
\node[red] at (1.25,0.65) {{\small $\widehat{a}_{22}$}};
\draw[thick,<->,red] (3.7,0.8) -- (2.8,1.7);
\node[red] at (3.25,1.65) {{\small $\widehat{a}_{13}$}};
\draw[thick,<->,red] (5.7,1.8) -- (4.8,2.7);
\node[red] at (5.2,2.6) {{\small $\widehat{a}_4$}};
\draw[thick,<->,red] (7.7,2.8) -- (6.8,3.7);
\node[red] at (7.25,3.65) {{\small $\widehat{a}_{25}$}};
\draw[thick,<->,red] (9.7,3.8) -- (8.8,4.7);
\node[red] at (9.25,4.65) {{\small $\widehat{a}_{16}$}};
\draw[thick,<->,red] (11.7,4.8) -- (10.8,5.7);
\node[red] at (11.2,5.6) {{\small $\widehat{a}_7$}};
\draw[thick,<->,red] (13.7,5.8) -- (12.8,6.7);
\node[red] at (13.25,6.65) {{\small $\widehat{a}_{28}$}};
\draw[thick,<->,red] (15.7,6.8) -- (14.8,7.7);
\node[red] at (15.25,7.65) {{\small $\widehat{a}_{19}$}};
\draw[thick,<->,red] (17.7,7.8) -- (16.8,8.7);
\node[red] at (17.25,8.65) {{\small $\widehat{a}_{10}$}};
\draw[thick,<->,red] (19.7,8.8) -- (18.8,9.7);
\node[red] at (19.2,9.6) {{\small $\widehat{a}_7$}};
%roots second layer
\draw[thick,<->,red] (0.7,0.8) -- (-0.2,1.7);
\node[red] at (0.25,1.65) {{\small $\widehat{a}_{21}$}};
\draw[thick,<->,red] (2.7,1.8) -- (1.8,2.7);
\node[red] at (2.25,2.65) {{\small $\widehat{a}_{12}$}};
\draw[thick,<->,red] (4.7,2.8) -- (3.8,3.7);
\node[red] at (4.2,3.6) {{\small $\widehat{a}_3$}};
\draw[thick,<->,red] (6.7,3.8) -- (5.8,4.7);
\node[red] at (6.25,4.65) {{\small $\widehat{a}_{24}$}};
\draw[thick,<->,red] (8.7,4.8) -- (7.8,5.7);
\node[red] at (8.25,5.65) {{\small $\widehat{a}_{15}$}};
\draw[thick,<->,red] (10.7,5.8) -- (9.8,6.7);
\node[red] at (10.2,6.6) {{\small $\widehat{a}_6$}};
\draw[thick,<->,red] (12.7,6.8) -- (11.8,7.7);
\node[red] at (12.25,7.65) {{\small $\widehat{a}_{27}$}};
\draw[thick,<->,red] (14.7,7.8) -- (13.8,8.7);
\node[red] at (14.25,8.65) {{\small $\widehat{a}_{18}$}};
\draw[thick,<->,red] (16.7,8.8) -- (15.8,9.7);
\node[red] at (16.2,9.6) {{\small $\widehat{a}_9$}};
\draw[thick,<->,red] (18.7,9.8) -- (17.8,10.7);
\node[red] at (18.25,10.65) {{\small $\widehat{a}_{30}$}};
%roots third layer
\draw[thick,<->,red] (1.7,2.8) -- (0.8,3.7);
\node[red] at (1.25,3.65) {{\small $\widehat{a}_{11}$}};
\draw[thick,<->,red] (3.7,3.8) -- (2.8,4.7);
\node[red] at (3.2,4.6) {{\small $\widehat{a}_2$}};
\draw[thick,<->,red] (5.7,4.8) -- (4.8,5.7);
\node[red] at (5.25,5.65) {{\small $\widehat{a}_{23}$}};
\draw[thick,<->,red] (7.7,5.8) -- (6.8,6.7);
\node[red] at (7.25,6.65) {{\small $\widehat{a}_{14}$}};
\draw[thick,<->,red] (9.7,6.8) -- (8.8,7.7);
\node[red] at (9.2,7.6) {{\small $\widehat{a}_5$}};
\draw[thick,<->,red] (11.7,7.8) -- (10.8,8.7);
\node[red] at (11.25,8.65) {{\small $\widehat{a}_{26}$}};
\draw[thick,<->,red] (13.7,8.8) -- (12.8,9.7);
\node[red] at (13.25,9.65) {{\small $\widehat{a}_{17}$}};
\draw[thick,<->,red] (15.7,9.8) -- (14.8,10.7);
\node[red] at (15.2,10.6) {{\small $\widehat{a}_8$}};
\draw[thick,<->,red] (17.7,10.8) -- (16.8,11.7);
\node[red] at (17.25,11.65) {{\small $\widehat{a}_{29}$}};
\draw[thick,<->,red] (19.7,11.8) -- (18.8,12.7);
\node[red] at (19.25,12.65) {{\small $\widehat{a}_{20}$}};
\draw[dashed] (20,13) -- (2.5,-4.5);
\draw[dashed] (2,4) -- (-2,0);
\draw[thick,red,<->] (-1.9,-0.1) -- (2.4,-4.4);
\node[red] at (0.05,-2.4) {$M$};
\draw[thick,blue,<->] (0,-1.55) -- (20,-1.55);
\node[blue] at (10,-1.9) {$\rho=10$\,{\color{red}{$V$} }$-170$\,{\color{red}{$M$ }}$+27\sum_{i=1}^{30}\,${\color{red}{$\widehat{a}_i$}}};
%%%
%%%%%basis (10,3)%%%%
%%%
%roots first layer
\draw[thick,blue,<->] (0.1,0) -- (1.9,0);
\node[blue] at (1,-0.25) {{\small $\widehat{b}_1$}};
\draw[thick,blue,<->] (2.1,1) -- (3.9,1);
\node[blue] at (3,0.75) {{\small $\widehat{b}_2$}};
\draw[thick,blue,<->] (4.1,2) -- (5.9,2);
\node[blue] at (5,1.75) {{\small $\widehat{b}_3$}};
\draw[thick,blue,<->] (6.1,3) -- (7.9,3);
\node[blue] at (7,2.75) {{\small $\widehat{b}_4$}};
\draw[thick,blue,<->] (8.1,4) -- (9.9,4);
\node[blue] at (9,3.75) {{\small $\widehat{b}_5$}};
\draw[thick,blue,<->] (10.1,5) -- (11.9,5);
\node[blue] at (11,4.75) {{\small $\widehat{b}_6$}};
\draw[thick,blue,<->] (12.1,6) -- (13.9,6);
\node[blue] at (13,5.75) {{\small $\widehat{b}_7$}};
\draw[thick,blue,<->] (14.1,7) -- (15.9,7);
\node[blue] at (15,6.75) {{\small $\widehat{b}_8$}};
\draw[thick,blue,<->] (16.1,8) -- (17.9,8);
\node[blue] at (17,7.75) {{\small $\widehat{b}_9$}};
%roots second layer
\draw[thick,blue,<->] (1.1,2) -- (2.9,2);
\node[blue] at (2,1.75) {{\small $\widehat{b}_{10}$}};
\draw[thick,blue,<->] (3.1,3) -- (4.9,3);
\node[blue] at (4,2.75) {{\small $\widehat{b}_{11}$}};
\draw[thick,blue,<->] (5.1,4) -- (6.9,4);
\node[blue] at (6,3.75) {{\small $\widehat{b}_{12}$}};
\draw[thick,blue,<->] (7.1,5) -- (8.9,5);
\node[blue] at (8,4.75) {{\small $\widehat{b}_{13}$}};
\draw[thick,blue,<->] (9.1,6) -- (10.9,6);
\node[blue] at (10,5.75) {{\small $\widehat{b}_{14}$}};
\draw[thick,blue,<->] (11.1,7) -- (12.9,7);
\node[blue] at (12,6.75) {{\small $\widehat{b}_{15}$}};
\draw[thick,blue,<->] (13.1,8) -- (14.9,8);
\node[blue] at (14,7.75) {{\small $\widehat{b}_{16}$}};
\draw[thick,blue,<->] (15.1,9) -- (16.9,9);
\node[blue] at (16,8.75) {{\small $\widehat{b}_{17}$}};
\draw[thick,blue,<->] (17.1,10) -- (18.9,10);
\node[blue] at (18,9.75) {{\small $\widehat{b}_{18}$}};
%roots third layer
\draw[thick,blue,<->] (2.1,4) -- (3.9,4);
\node[blue] at (3,3.75) {{\small $\widehat{b}_{19}$}};
\draw[thick,blue,<->] (4.1,5) -- (5.9,5);
\node[blue] at (5,4.75) {{\small $\widehat{b}_{20}$}};
\draw[thick,blue,<->] (6.1,6) -- (7.9,6);
\node[blue] at (7,5.75) {{\small $\widehat{b}_{21}$}};
\draw[thick,blue,<->] (8.1,7) -- (9.9,7);
\node[blue] at (9,6.75) {{\small $\widehat{b}_{22}$}};
\draw[thick,blue,<->] (10.1,8) -- (11.9,8);
\node[blue] at (11,7.75) {{\small $\widehat{b}_{23}$}};
\draw[thick,blue,<->] (12.1,9) -- (13.9,9);
\node[blue] at (13,8.75) {{\small $\widehat{b}_{24}$}};
\draw[thick,blue,<->] (14.1,10) -- (15.9,10);
\node[blue] at (15,9.75) {{\small $\widehat{b}_{25}$}};
\draw[thick,blue,<->] (16.1,11) -- (17.9,11);
\node[blue] at (17,10.75) {{\small $\widehat{b}_{26}$}};
\draw[thick,blue,<->] (18.1,12) -- (19.9,12);
\node[blue] at (19,11.75) {{\small $\widehat{b}_{27}$}};
%vertical distances
\draw[thick,blue,<->] (-1.5,0) -- (-1.5,2);
\node[blue] at (-1.7,1) {{\small $c_2$}};
\draw[thick,blue,<->] (-0.5,2) -- (-0.5,4);
\node[blue] at (-0.7,3) {{\small $c_3$}};
\draw[thick,blue,<->] (0.5,4) -- (0.5,6);
\node[blue] at (0.3,5) {{\small $c_1$}};
%D distance
\draw[thick,blue,<->] (0,7) -- (3,7);
\node[blue] at (1.5,7.3) {$D$};
\end{tikzpicture}}}}
\caption{\sl Two different maximal sets of independent K\"ahler parameters in the $(10,3)$ web diagram. After a series of suitable flop- and symmetry transformations, the red parametrisation agrees with the maximal set of independent parameters $(M,V,\widehat{a}_{1,\ldots,30})$ used in the $(6,5)$ web diagram in \figref{Fig:WebDiag65}.}
\label{Fig:Web103}
\end{figure}

As discussed in \cite{Hohenegger:2016yuv}, the $(6, 5)$ web diagram in \figref{Fig:WebDiag65} can be dualised to other webs of the type  $(N',M')$ with $NM=N'M'$ and $\text{gcd}(N,M)=\text{gcd}(N',M')$. In the case at hand, one such configuration is $(N',M')=(10,3)$, whose web-diagram is drawn in \figref{Fig:Web103}.
%%%%%%%%%%%%%%%%%%%%%%%%%%%%%%%%%%%%%%%%%%%%%%
%%%%%%%%%%%%%%%%%%%%%%%%%%%%%%%%%%%%%%%%%%%%%%
%%%%%%%%%%%%%%%%%%%%%%%%%%%%
%%%%%%%%%%%%%%%%%%%%%%%%%%%%
%%%%%%%%%%%%%%%%%%%%%%%%%%%%
%%%%%%%%%%%%%%%%%%%%%%%%%%%%
An explicit form for the duality map relating $(\mathbf{h},\mathbf{v},\mathbf{m})$ to $(\mathbf{h}',\mathbf{v}',\mathbf{m}')$ has been conjectured in \cite{Hohenegger:2016yuv}, which allows us to recover the same set of parameters $(M,V,\widehat{a}_{1,\ldots,30})$ (drawn in red) also in \figref{Fig:Web103}. In terms of $(\mathbf{h}',\mathbf{v}',\mathbf{m}')$,  we have explicit relations
{\allowdisplaybreaks
\begin{align}
M&=h'_{10} + v'_1 + h'_{21} + v'_{22} + h'_{12} + v'_{13} + h'_{3} + v'_4 + h'_{24} + v'_{25} + h'_{15} + v'_{16} + h'_{6} + v'_7\nonumber\\
&\hspace{1.25cm} + h'_{27} + v'_{28} + h'_{18} + v'_{19} + h'_9\,,\nonumber\\
V&=m'_{30} + (10 - 1) (h'_{29} + h'_{30}) + (10 - 2) (v'_9 + h'_8 + v'_{21} + h'_{11})\nonumber\\
&\hspace{1.25cm} + (10 - 3) (v'_{18} + h'_{17} + v'_{12} + h'_{2}) + (10 - 4) (v'_{27} + h'_{26} + v'_3 + h'_{23})\nonumber\\
&\hspace{1.25cm} + (10 - 5) (v'_6 + h'_5 + v'_{24} + h'_{14}) + (10 - 6) (v'_{15} + h'_{14} + v'_{15} + h'_5)\nonumber\\
&\hspace{1.25cm} + (10 - 7) (v'_{24} + h'_{23} + v'_6 +  h'_{26}) + (10 - 8) (v'_3 + h'_2 + v'_{27} + h'_{17})\nonumber\\
&\hspace{1.25cm} + (10 - 9) (v'_{12} + h'_{11} + v'_{18} + h'_8)\,.
\end{align}}
Note that, analogous to (\ref{DualFrame1V}), we also have in the dual web diagram
\begin{align}
&m'_i=V+p'_i(M,\widehat{a}_{1,\ldots,30})\,,&&\forall \,  i=1,\ldots,30\,,
\end{align}
for some multi-linear functions  $p'_i$, while $h'_{1,\ldots,30}$ and $v'_{1,\ldots,30}$ are independent of $V$. Therefore, the diagonal expansions (in the sense of (\ref{TrialitySchemat})) of $\pf{6}{5}$ and $\pf{10}{3}$ both give rise to gauge theories with gauge group $U(30)$, as implied by \cite{Hohenegger:2016yuv} and as explained  above.

In \figref{Fig:Web103}, however, we have also shown (in blue) a different set of maximally independent K\"ahler parameters $(D,\rho,c_{1,2,3},\widehat{b}_{1,\ldots,27})$. In terms of these variables, we have
\begin{align}
v'_i=\left\{\begin{array}{lcl}c_1+p_i^{(1)}(D,\rho,\widehat{b}_{1,\ldots,27}) & \text{if} & 1\leq i\leq 10\\[2pt] c_2+p_i^{(2)}(D,\rho,\widehat{b}_{1,\ldots,27}) & \text{if} & 11\leq i\leq 20\\[2pt]c_3+p_i^{(3)}(D,\rho,\widehat{b}_{1,\ldots,27}) & \text{if} & 21\leq i\leq 30\\[2pt] \end{array}\right.
\end{align}
for some polynomials $p_i^{(1,2,3)}$, while $h'_{1,\ldots,30}$ and $v'_{1,\ldots,30}$ are independent of $c_{1,2,3}$. Thus, in the limit $c_i\to \infty$ for $i=1,\ldots,3$, the vertical lines in \figref{Fig:Web103} are cut and the diagram decomposes into three strips of length 10 (similarly to the examples in the previous section). We can interpret this as the weak coupling limit of a gauge theory with gauge group $[U(10)]^{3}$. This indicates that, upon expanding $\pf{6}{5}$ as a power series in $e^{-c_1}$, $e^{-c_2}$ and $e^{-c_3}$ (which is equivalent to the expansion in terms of $e^{-v'_i}$ for $i=1,\ldots,10$) the latter can be interpreted as an instanton expansion of a gauge theory with gauge group $[U(10)]^{3}$ (which via $SL(2,\mathbb{Z})$ transforms is in turn dual to a theory with gauge group $[U(3)]^{10}$). It is worth noticing that, in this manner, the $\widehat{b}_{1,\ldots,27}$ play the role of roots of Lie algebras associated with this group. 

Exploiting further dualities of $X_{6,5}$ we can in the same fashion engineer a large set of dual quiver gauge theories whose gauge groups include
\begin{align}
&U(30)\,,&&[U(15)]^2\,,&&[U(10)]^3\,,&&[U(6)]^5\,,&&[U(5)]^6\,,&&[U(3)]^{10}\,,&&[U(2)]^{15}\,,&&[U(1)]^{30}\,, 
\end{align}
all of which are compatible with the condition in eq.~(\ref{DualityCalabiYauFlop}).
%%%%%%%%%%%%%%%%%%%%%%%%%%%%%%%%%%%%%%%%%%%%%%%%%%%%%%%%%%%%%%%
\section{Intermediate K\"ahler Cones and Other Dual Theories}\label{Sect:IntermediateCones}
As schematically indicated in \figref{Fig:FlopTransformations}, the duality transformation that relates $X_{N,M}$ and $X_{N',M'}$ does in general not relate regions in directly adjacent cones of the extended K\"ahler moduli space. Instead, the series of flop transitions and other symmetry  transformations discussed in~\cite{Hohenegger:2016yuv} (and reviewed in appendix \ref{App:ShiftDuality}), generically passes through several other regions (labelled `intermediate K\"ahler cones' in \figref{Fig:FlopTransformations}). An interesting question is whether any of these cones also contains regions which engineer a (weak coupling) description of a gauge theory of some type and/or a Little String Theory. While we are not able to give an answer for a generic K\"ahler cone in the extended moduli space of $X_{N,M}$, in this section, we discuss a particular type among the former, namely, those cones where the toric web takes a similar form as in \figref{Fig:WebDiagramGeneric}, except that the external legs are identified after a cyclic rotation with a shift $\delta\in [0,N-1]$, as shown in \figref{Fig:WebDiagramGenericShift}.\footnote{In the following we refer to web diagrams of this type as \emph{shifted} with shift parameter $\delta$.}
\begin{figure}[htb]
\begin{center}
\scalebox{0.6}{\parbox{16.5cm}{\begin{tikzpicture}[scale = 1.5]
%horizontal lines%%%%%%%%%%
%first layer
\draw[ultra thick] (-1,0) -- (0,0);
\draw[ultra thick] (1,1) -- (2,1);
\draw[ultra thick] (3,2) -- (4,2);
\node at (4.5,2) {\Large $\cdots$};
\draw[ultra thick] (5,2) -- (6,2);
\draw[ultra thick] (7,3) -- (8,3);
%second layer
\draw[ultra thick] (0,2) -- (1,2);
\draw[ultra thick] (2,3) -- (3,3);
\draw[ultra thick] (4,4) -- (5,4);
\node at (5.5,4) {\Large $\cdots$};
\draw[ultra thick] (6,4) -- (7,4);
\draw[ultra thick] (8,5) -- (9,5);
%third layer
\draw[ultra thick] (1,6) -- (2,6);
\draw[ultra thick] (3,7) -- (4,7);
\draw[ultra thick] (5,8) -- (6,8);
\node at (6.5,8) {\Large $\cdots$};
\draw[ultra thick] (7,8) -- (8,8);
\draw[ultra thick] (9,9) -- (10,9);
%vertical lines%%%%%%%%%%
%first layer
\draw[ultra thick] (0,0) -- (0,-1);
\draw[ultra thick] (2,1) -- (2,0);
\draw[ultra thick] (6,2) -- (6,1);
%second layer
\draw[ultra thick] (1,1) -- (1,2);
\draw[ultra thick] (3,2) -- (3,3);
\draw[ultra thick] (7,3) -- (7,4);
%third layer
\draw[ultra thick] (2,3) -- (2,4);
\draw[ultra thick] (4,4) -- (4,5);
\draw[ultra thick] (8,5) -- (8,6);
%dots
\node[rotate=90] at (2,4.5) {\Large $\cdots$};
\node[rotate=90] at (4,5.5) {\Large $\cdots$};
\node[rotate=90] at (8,6.5) {\Large $\cdots$};
%fourth layer
\draw[ultra thick] (2,5) -- (2,6);
\draw[ultra thick] (4,6) -- (4,7);
\draw[ultra thick] (8,7) -- (8,8);
%fifth layer
\draw[ultra thick] (3,7) -- (3,8);
\draw[ultra thick] (5,8) -- (5,9);
\draw[ultra thick] (9,9) -- (9,10);
%diagonal lines%%%%%%%%%%
%first layer
\draw[ultra thick] (0,0) -- (1,1);
\draw[ultra thick] (2,1) -- (3,2);
\draw[ultra thick] (6,2) -- (7,3);
%second layer
\draw[ultra thick] (1,2) -- (2,3);
\draw[ultra thick] (3,3) -- (4,4);
\draw[ultra thick] (7,4) -- (8,5);
%third layer
\draw[ultra thick] (2,6) -- (3,7);
\draw[ultra thick] (4,7) -- (5,8);
\draw[ultra thick] (8,8) -- (9,9);
%%%%%%%%%%%%%%%%%%%%%%
%connectors
\node[rotate=90] at (-0.75,0) {$=$}; 
\node at (-0.65,-0.2) {{\small $1$}};
\node[rotate=90] at (0.25,2) {$=$}; 
\node at (0.35,1.8) {{\small $2$}};
\node[rotate=90] at (1.25,6) {$=$}; 
\node at (1.35,5.8) {{\small $M$}};
\node[rotate=90] at (7.75,3) {$=$}; 
\node at (7.85,2.8) {{\small $1$}};
\node[rotate=90] at (8.75,5) {$=$};
\node at (8.85,4.8) {{\small $2$}};
\node[rotate=90] at (9.75,9) {$=$};
\node at (9.85,8.8) {{\small $M$}}; 
\node at (0,-0.75) {--}; 
\node at (0.2,-0.9) {{\small $1$}};
\node at (2,0.25) {--}; 
\node at (2.2,0.2) {{\small $2$}};
\node at (6,1.25) {--}; 
\node at (6.2,1.2) {{\small $N$}};
\node at (3,7.85) {--};
\node at (3.85,7.75) {{\footnotesize $1+\delta+ \text{mod}[N]$}}; 
\node at (5,8.85) {--}; 
\node at (5.85,8.75) {{\footnotesize $2+\delta+\text{mod}[N]$}}; 
\node at (9,9.85) {--}; 
\node at (10,9.7) {{\small $N+\delta+\text{mod}[N]$}}; 
\end{tikzpicture}}}
\caption{\sl Toric web diagram with a shifted identification of the external legs for $\delta\in[0,N-1]$.}
\label{Fig:WebDiagramGenericShift}
\end{center}
\end{figure}
Indeed, as explained in appendix~\ref{App:ShiftDuality}, starting from the web-diagram of $X_{N,M}$ as shown in \figref{Fig:WebDiagramGeneric}, there exists a duality transformation $\mathcal{F}$ (introduced in eq.~(\ref{FlopShift})) to a shifted web of the type shown in  \figref{Fig:WebDiagramGenericShift} with $\delta=M-N\text{ mod }N$. This duality transformation is based on a series of flop and symmetry transformations first discussed in \cite{Hohenegger:2016yuv}, which for $M=1$ was also reviewed in \cite{Bastian:2017ing}.

The transformation described in appendix~\ref{App:ShiftDuality} involves no flop transformations on the vertical lines. Since, for example, in the vertical gauge theory, the latter are related to the gauge coupling constants, this transformation therefore relates two points in the weak coupling regime of this theory. In other words, even after the transformation, $Z_{\text{vert}}^{(N,M)}$ in (\ref{TrialitySchemat}) is still a valid series expansion that can be identified with an instanton series of a (weakly coupled) gauge theory. This strongly suggests that even a Calabi-Yau manifold with a shifted web diagram engineers at least one weakly coupled gauge theory, if it can be related to an unshifted web diagram with a transformation of the type described in appendix~\ref{App:ShiftDuality}. Below we will analyse possible weak coupling theories engineered from such webs in more detail by focusing on the examples $(N,M)=(6,1)$ and $(N,M)=(4,1)$. To analyse the latter, we will have to introduce as a further notion a purely geometric realisation of the gauge algebra, that can directly be read off from the web diagram.
%%%%%%%%%%%%%%%%%%%%%%%%%%%
%%%%%%%%%%%%%%%%%%%%%%%%%%%
\subsection{Example: $(N,M)=(6,1)$}
Our first example is the diagram $X_{6,1}$ which is schematically drawn in \figref{Fig:61NoFlop}, along with a parametrisation of the K\"ahler parameters that is compatible with all consistency conditions.
\begin{figure}
\begin{center}
\scalebox{0.68}{\parbox{20.5cm}{\begin{tikzpicture}[scale = 1.5]
\draw[ultra thick] (-1,0) -- (0,0) -- (1,1) -- (2,1) -- (3,2) -- (4,2) -- (5,3) -- (6,3) -- (7,4) -- (8,4) -- (9,5) -- (10,5) -- (11,6) -- (12,6);
%above
\draw[ultra thick] (1,1) -- (1,2);
\draw[ultra thick] (3,2) -- (3,3);
\draw[ultra thick] (5,3) -- (5,4);
\draw[ultra thick] (7,4) -- (7,5);
\draw[ultra thick] (9,5) -- (9,6);
\draw[ultra thick] (11,6) -- (11,7);
%below
\draw[ultra thick] (0,-1) -- (0,0);
\draw[ultra thick] (2,0) -- (2,1);
\draw[ultra thick] (4,1) -- (4,2);
\draw[ultra thick] (6,2) -- (6,3);
\draw[ultra thick] (8,3) -- (8,4);
\draw[ultra thick] (10,4) -- (10,5);
%labels
\node at (-1.2,0) {{$a$}};
\node at (12.2,6) {{$a$}};
\node at (1,2.2) {{$1$}};
\node at (3,3.2) {{$2$}};
\node at (5,4.2) {{$3$}};
\node at (7,5.2) {{$4$}};
\node at (9,6.2) {{$5$}};
\node at (11,7.2) {{$6$}};
\node at (0,-1.2) {{$1$}};
\node at (2,-0.2) {{$2$}};
\node at (4,0.8) {{$3$}};
\node at (6,1.8) {{$4$}};
\node at (8,2.8) {{$5$}};
\node at (10,3.8) {{$6$}};
%horizontal
\node at (-0.5,-0.25) {{\bf $h_{1}$}};
\node at (1.5,0.75) {{\bf $h_{2}$}};
\node at (3.5,1.75) {{\bf $h_{3}$}};
\node at (5.5,2.75) {{\bf $h_{4}$}};
\node at (7.5,3.75) {{\bf $h_{5}$}};
\node at (9.5,4.75) {{\bf $h_{6}$}};
\node at (11.5,5.75) {{\bf $h_{1}$}};
%vertical
\node at (0.2,-0.5) {{\bf $v$}};
\node at (2.2,0.5) {{\bf $v$}};
\node at (4.2,1.5) {{\bf $v$}};
\node at (6.2,2.5) {{\bf $v$}};
\node at (8.2,3.5) {{\bf $v$}};
\node at (10.2,4.5) {{\bf $v$}};
\node at (0.8,1.5) {{\bf $v$}};
\node at (2.8,2.5) {{\bf $v$}};
\node at (4.8,3.5) {{\bf $v$}};
\node at (6.8,4.5) {{\bf $v$}};
\node at (8.8,5.5) {{\bf $v$}};
\node at (10.8,6.5) {{\bf $v$}};
%diagonal
\node at (0.65,0.35) {{\bf $m$}};
\node at (2.65,1.35) {{\bf $m$}};
\node at (4.65,2.35) {{\bf $m$}};
\node at (6.65,3.35) {{\bf $m$}};
\node at (8.65,4.35) {{\bf $m$}};
\node at (10.65,5.35) {{\bf $m$}};
%rootsa
\draw[<->,red, thick] (0.1,0) -- (1.9,0);
\node[red] at (1,-0.25) {{\bf $\widehat{b}_2$}};
\draw[<->,red, thick] (2.1,1) -- (3.9,1);
\node[red] at (3,0.75) {{\bf $\widehat{b}_3$}};
\draw[<->,red, thick] (4.1,2) -- (5.9,2);
\node[red] at (5,1.75) {{\bf $\widehat{b}_4$}};
\draw[<->,red, thick] (6.1,3) -- (7.9,3);
\node[red] at (7,2.75) {{\bf $\widehat{b}_5$}};
\draw[<->,red, thick] (10.1,5) -- (11.9,5);
\node[red] at (11,4.75) {{\bf $\widehat{b}_1$}};
%rootsc
\draw[<->,green!50!black, thick] (1.95,1.1) -- (1.1,1.95);
\node[green!50!black] at (1.7,1.65) {{\bf $\widehat{a}_1$}};
\draw[<->,green!50!black, thick] (3.95,2.1) -- (3.1,2.95);
\node[green!50!black] at (3.7,2.65) {{\bf $\widehat{a}_2$}};
\draw[<->,green!50!black, thick] (5.95,3.1) -- (5.1,3.95);
\node[green!50!black] at (5.7,3.65) {{\bf $\widehat{a}_3$}};
\draw[<->,green!50!black, thick] (7.95,4.1) -- (7.1,4.95);
\node[green!50!black] at (7.7,4.65) {{\bf $\widehat{a}_4$}};
\draw[<->,green!50!black, thick] (9.95,5.1) -- (9.1,5.95);
\node[green!50!black] at (9.7,5.65) {{\bf $\widehat{a}_5$}};
%tau
\draw[blue, thick, <->] (0,0.1) -- (0,2);
\node[blue] at (-0.25,1) {{\bf $\tau$}};
\draw[blue, thick, <->] (-1,7.5) -- (11,7.5);
\node[blue] at (5,7.2) {{\bf $\rho$}};
\draw[dashed] (-1,7.4) -- (-1,0.1);
\end{tikzpicture}}}
\caption{\sl Toric web diagram of $X_{6,1}$ (with shift parameter $\delta=0$) with a consistent labelling of the areas of all curves.}
\label{Fig:61NoFlop}
\end{center}
\end{figure}
We can arrange the latter in the following form:
{\allowdisplaybreaks
\begin{align}
&\widehat{a}_i=h_{i+1}+v\,,&&\widehat{b}_i=h_i+m\,,&&\forall i\in\{1,2,3,4,5\}\,,\nonumber\\
&L=\sum_{i=1}^5\widehat{a}_i+h_1+v\,,&&\rho=\sum_{i=1}^5\widehat{b}_i+h_6+m\,,&&\tau= m+v\,,&&D=E/6=m\,.
\end{align}}
which is more adapted to the description of three different gauge theories engineered by $X_{N,M}$. Indeed, as explained in \cite{Bastian:2017ary} we can engineer the following theories
\begin{itemize}
\item horizontal theory of $X_{6,1}^{(\delta=0)}$\\
The horizontal expansion of $\mathcal{Z}_{6,1}$ can be interpreted as the instanton partition function of a gauge theory with gauge group $[U(1)]^6$. This theory is parametrised in the following fashion:
\begin{itemize}
\item coupling constants: the parameters $\widehat{b}_{1,2,3,4,5}$ and $\rho-\sum_{i=1}^5\widehat{b}_i$ are related to the coupling constants
\item roots: there is no finite Lie algebra associated with $U(1)$, however, the parameter $\tau$ can be interpreted as the affine root for 6 copies of the Heisenberg algebra $\widehat{\mathfrak{a}}_0$
\item mass scale: the hypermultiplet mass scale of the theory is set by the parameter $E$
\end{itemize}
\item vertical theory of $X_{6,1}^{(\delta=0)}$\\
The vertical expansion of $\mathcal{Z}_{6,1}$ can be interpreted as the instanton partition function of a gauge theory with gauge group $U(6)$, which is parametrised in the following fashion:
\begin{itemize}
\item coupling constant: the parameter $v$ is related to the coupling constant
\item roots: the parameters $\widehat{b}_{1,2,3,4,5}$ play the role of the simple positive roots of $\mathfrak{a}_5$, which is extended to $\widehat{\mathfrak{a}}_5$ by $\rho$
\item mass scale: the hypermultiplet mass scale of the theory is set by the parameter $D$
\end{itemize}
\item diagonal theory of $X_{6,1}^{(\delta=0)}$\\
The diagonal expansion of $\mathcal{Z}_{6,1}$ can be interpreted as the instanton partition function of a gauge theory with gauge group $U(6)$, which is parametrised in the following fashion:
\begin{itemize}
\item coupling constant: the parameter $m$ is related to the coupling constant
\item roots: the parameters $\widehat{a}_{1,2,3,4,5}$ play the role of the simple positive roots of $\mathfrak{a}_5$, which is extended to $\widehat{\mathfrak{a}}_5$ by $L$
\item mass scale: the hypermultiplet mass scale of the theory is set by the parameter $v$
\end{itemize}
\end{itemize}

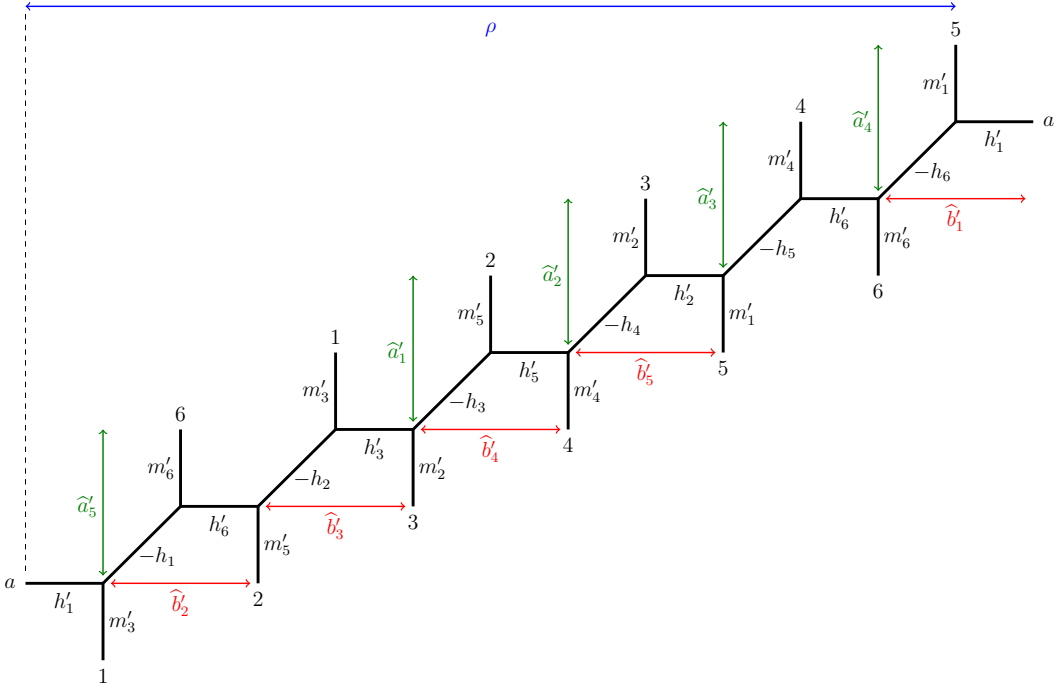
\begin{figure}
\begin{center}
\scalebox{0.68}{\parbox{20.5cm}{\begin{tikzpicture}[scale = 1.5]
\draw[ultra thick] (-1,0) -- (0,0) -- (1,1) -- (2,1) -- (3,2) -- (4,2) -- (5,3) -- (6,3) -- (7,4) -- (8,4) -- (9,5) -- (10,5) -- (11,6) -- (12,6);
%above
\draw[ultra thick] (1,1) -- (1,2);
\draw[ultra thick] (3,2) -- (3,3);
\draw[ultra thick] (5,3) -- (5,4);
\draw[ultra thick] (7,4) -- (7,5);
\draw[ultra thick] (9,5) -- (9,6);
\draw[ultra thick] (11,6) -- (11,7);
%below
\draw[ultra thick] (0,-1) -- (0,0);
\draw[ultra thick] (2,0) -- (2,1);
\draw[ultra thick] (4,1) -- (4,2);
\draw[ultra thick] (6,2) -- (6,3);
\draw[ultra thick] (8,3) -- (8,4);
\draw[ultra thick] (10,4) -- (10,5);
%labels
\node at (-1.2,0) {{$a$}};
\node at (12.2,6) {{$a$}};
\node at (1,2.2) {{$6$}};
\node at (3,3.2) {{$1$}};
\node at (5,4.2) {{$2$}};
\node at (7,5.2) {{$3$}};
\node at (9,6.2) {{$4$}};
\node at (11,7.2) {{$5$}};
\node at (0,-1.2) {{$1$}};
\node at (2,-0.2) {{$2$}};
\node at (4,0.8) {{$3$}};
\node at (6,1.8) {{$4$}};
\node at (8,2.8) {{$5$}};
\node at (10,3.8) {{$6$}};
%horizontal
\node at (-0.5,-0.25) {{\bf $h'_{1}$}};
\node at (1.5,0.75) {{\bf $h'_{6}$}};
\node at (3.5,1.75) {{\bf $h'_{3}$}};
\node at (5.5,2.75) {{\bf $h'_{5}$}};
\node at (7.5,3.75) {{\bf $h'_{2}$}};
\node at (9.5,4.75) {{\bf $h'_{6}$}};
\node at (11.5,5.75) {{\bf $h'_{1}$}};
%vertical
\node at (0.25,-0.5) {{\bf $m'_3$}};
\node at (2.25,0.5) {{\bf $m'_5$}};
\node at (4.25,1.5) {{\bf $m'_2$}};
\node at (6.25,2.5) {{\bf $m'_4$}};
\node at (8.25,3.5) {{\bf $m'_1$}};
\node at (10.25,4.5) {{\bf $m'_6$}};
\node at (0.75,1.5) {{\bf $m'_6$}};
\node at (2.75,2.5) {{\bf $m'_3$}};
\node at (4.75,3.5) {{\bf $m'_5$}};
\node at (6.75,4.5) {{\bf $m'_2$}};
\node at (8.75,5.5) {{\bf $m'_4$}};
\node at (10.75,6.5) {{\bf $m'_1$}};
%diagonal
\node at (0.7,0.35) {{\bf $-h_1$}};
\node at (2.7,1.35) {{\bf $-h_2$}};
\node at (4.7,2.35) {{\bf $-h_3$}};
\node at (6.7,3.35) {{\bf $-h_4$}};
\node at (8.7,4.35) {{\bf $-h_5$}};
\node at (10.7,5.35) {{\bf $-h_6$}};
%rootsa
\draw[<->,red, thick] (0.1,0) -- (1.9,0);
\node[red] at (1,-0.25) {{\bf $\widehat{b}'_2$}};
\draw[<->,red, thick] (2.1,1) -- (3.9,1);
\node[red] at (3,0.75) {{\bf $\widehat{b}'_3$}};
\draw[<->,red, thick] (4.1,2) -- (5.9,2);
\node[red] at (5,1.75) {{\bf $\widehat{b}'_4$}};
\draw[<->,red, thick] (6.1,3) -- (7.9,3);
\node[red] at (7,2.75) {{\bf $\widehat{b}'_5$}};
\draw[<->,red, thick] (10.1,5) -- (11.9,5);
\node[red] at (11,4.75) {{\bf $\widehat{b}'_1$}};
%rootsc
\draw[<->,green!50!black, thick] (0,0.1) -- (0,2);
\node[green!50!black] at (-0.2,1) {{\bf $\widehat{a}'_5$}};
\draw[<->,green!50!black, thick] (4,2.1) -- (4,4);
\node[green!50!black] at (3.8,3) {{\bf $\widehat{a}'_1$}};
\draw[<->,green!50!black, thick] (6,3.1) -- (6,5);
\node[green!50!black] at (5.8,4) {{\bf $\widehat{a}'_2$}};
\draw[<->,green!50!black, thick] (8,4.1) -- (8,6);
\node[green!50!black] at (7.8,5) {{\bf $\widehat{a}'_3$}};
\draw[<->,green!50!black, thick] (10,5.1) -- (10,7);
\node[green!50!black] at (9.8,6) {{\bf $\widehat{a}'_4$}};
\draw[blue, thick, <->] (-1,7.5) -- (11,7.5);
\node[blue] at (5,7.2) {{\bf $\rho$}};
\draw[dashed] (-1,7.4) -- (-1,0.1);
\end{tikzpicture}}}
\caption{\sl Toric web diagram of $X_{6,1}$ (with shift parameter $\delta=0$) with a consistent labelling of the areas of all curves.}
\label{Fig:61Flop}
\end{center}
\end{figure}

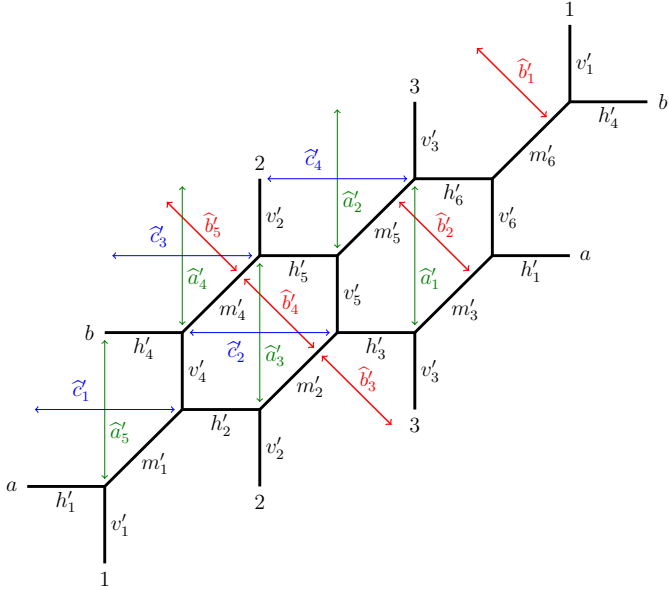
\begin{wrapfigure}{L}{0.52\textwidth}
\centering
\scalebox{0.68}{\parbox{13cm}{\begin{tikzpicture}[scale = 1.5]
\draw[ultra thick] (0,-1) -- (0,0);
\draw[ultra thick] (2,0) -- (2,1);
\draw[ultra thick] (4,1) -- (4,2);
\draw[ultra thick] (-1,0) -- (0,0) -- (1,1) -- (2,1) -- (3,2) -- (4,2) -- (5,3) -- (6,3);
\draw[ultra thick] (1,1) -- (1,2);
\draw[ultra thick] (3,2) -- (3,3);
\draw[ultra thick] (5,3) -- (5,4);
\draw[ultra thick] (0,2) -- (1,2) -- (2,3) -- (3,3) -- (4,4) -- (5,4) -- (6,5) -- (7,5);
\draw[ultra thick] (2,3) -- (2,4);
\draw[ultra thick] (4,4) -- (4,5);
\draw[ultra thick] (6,5) -- (6,6);
%labels
\node at (-1.2,0) {{$a$}};
\node at (-0.2,2) {{$b$}};
\node at (6.2,3) {{$a$}};
\node at (7.2,5) {{$b$}};
\node at (0,-1.2) {{$1$}};
\node at (2,-0.2) {{$2$}};
\node at (4,0.8) {{$3$}};
\node at (2,4.2) {{$2$}};
\node at (4,5.2) {{$3$}};
\node at (6,6.2) {{$1$}};
%verticals
\node at (0.2,-0.5) {{\bf $v'_1$}};
\node at (2.2,0.5) {{\bf $v'_2$}};
\node at (4.2,1.5) {{\bf $v'_3$}};
\node at (1.2,1.5) {{\bf $v'_4$}};
\node at (3.2,2.5) {{\bf $v'_5$}};
\node at (5.2,3.5) {{\bf $v'_6$}};
\node at (2.2,3.5) {{\bf $v'_2$}};
\node at (4.2,4.5) {{\bf $v'_3$}};
\node at (6.2,5.5) {{\bf $v'_1$}};
%horizontals
\node at (-0.5,-0.2) {{\bf $h'_1$}};
\node at (1.5,0.8) {{\bf $h'_2$}};
\node at (3.5,1.8) {{\bf $h'_3$}};
\node at (5.5,2.8) {{\bf $h'_1$}};
\node at (0.5,1.8) {{\bf $h'_4$}};
\node at (2.5,2.8) {{\bf $h'_5$}};
\node at (4.5,3.8) {{\bf $h'_6$}};
\node at (6.5,4.8) {{\bf $h'_4$}};
%diagonals
\node at (0.65,0.3) {{\bf $m'_1$}};
\node at (2.65,1.3) {{\bf $m'_2$}};
\node at (4.65,2.3) {{\bf $m'_3$}};
\node at (1.65,2.3) {{\bf $m'_4$}};
\node at (3.65,3.3) {{\bf $m'_5$}};
\node at (5.65,4.3) {{\bf $m'_6$}};
%roots diagonal
\draw[thick,<->,red] (5.7,4.8) -- (4.8,5.7);
\node[red] at (5.45,5.4) {{\small $\widehat{b}'_1$}};
\draw[thick,<->,red] (1.7,2.8) -- (0.8,3.7);
\node[red] at (1.4,3.4) {{\small $\widehat{b}'_5$}};
\draw[thick,<->,red] (2.7,1.8) -- (1.8,2.7);
\node[red] at (2.4,2.4) {{\small $\widehat{b}'_4$}};
\draw[thick,<->,red] (4.7,2.8) -- (3.8,3.7);
\node[red] at (4.4,3.4) {{\small $\widehat{b}'_2$}};
\draw[thick,<->,red] (3.7,0.8) -- (2.8,1.7);
\node[red] at (3.4,1.4) {{\small $\widehat{b}'_3$}};
%roots vertical
\draw[green!50!black,<->] (0,0.1) -- (0,1.9);
\node[green!50!black,<->] at (0.2,0.7) {$\widehat{a}'_5$};
\draw[green!50!black,<->] (2,1.1) -- (2,2.9);
\node[green!50!black,<->] at (2.2,1.7) {$\widehat{a}'_3$};
\draw[green!50!black,<->] (1,2.1) -- (1,3.9);
\node[green!50!black,<->] at (1.2,2.7) {$\widehat{a}'_4$};
\draw[green!50!black,<->] (3,3.1) -- (3,4.9);
\node[green!50!black,<->] at (3.2,3.7) {$\widehat{a}'_2$};
\draw[green!50!black,<->] (4,2.1) -- (4,3.9);
\node[green!50!black,<->] at (4.2,2.7) {$\widehat{a}'_1$};
%roots horizontal
\draw[blue,<->] (-0.9,1) -- (0.9,1);
\node[blue,<->] at (-0.3,1.25) {$\widehat{c}'_1$};
\draw[blue,<->] (1.1,2) -- (2.9,2);
\node[blue,<->] at (1.7,1.75) {$\widehat{c}'_2$};
\draw[blue,<->] (0.1,3) -- (1.9,3);
\node[blue,<->] at (0.7,3.25) {$\widehat{c}'_3$};
\draw[blue,<->] (2.1,4) -- (3.9,4);
\node[blue,<->] at (2.7,4.25) {$\widehat{c}'_4$};
\end{tikzpicture}}}
\caption{\sl Toric web diagram of $X_{3,2}^{(1)}$ with shift $\delta=1$, and a labelling of the K\"ahler parameters.}
\label{Fig:32delta}
\end{wrapfigure}

\noindent
After performing the transformation $\mathcal{F}$ in (\ref{FlopShift}) of appendix~\ref{App:ShiftDuality} (which consists of a series of flop transformations on the horizontal curves with areas $h_{1,\ldots, 6}$) along with other symmetry transformations), the web diagram can be brought into the form shown in \figref{Fig:61Flop}. This diagram is of the form of $X_{6,1}^{(\delta=5)}=\mathcal{F}(X_{6,1}^{(\delta=0)})$ (\emph{i.e.} a diagram with $(N,M)=(6,1)$, however, with shift parameter $\delta=5\sim -1$). It can also be presented in the form of $X_{3,2}^{(\delta=1)}$ as shown in \figref{Fig:32delta}, \emph{i.e.} a web with $(N,M)=(3,2)$ and shift parameter $\delta=1$. To arrive at this presentation only $SL(2,\mathbb{Z})$ transformations were used, but in particular no flop transformations. The K\"ahler parameters of the new web diagram in \figref{Fig:61Flop} (and equivalently \figref{Fig:32delta}) can be expressed in terms of the original K\"ahler parameters $\{h_{1,\ldots,6},v,m\}$. Explicitly, we have for the areas of all curves
{\allowdisplaybreaks
\begin{align}
&v'_1=-h_6\,,&&v'_2=-h_4\,,&&v'_3=-h_2\,,\nonumber\\
&v'_4=-h_5\,,&&v'_5=-h_3\,,&&v'_6=-h_1\,,\nonumber\\[4pt]
&h'_1=m+h_1+h_6\,,&&h'_2=m+h_4+h_5\,,&&h'_3=m+h_2+h_3\,,\nonumber\\
&h'_4=m+h_5+h_6\,,&&h'_5=m+h_3+h_4\,,&&h'_6=m+h_1+h_2\,,\nonumber\\[4pt]
&m'_1=v+h_5+h_6\,,&&m'_2=v+h_3+h_4\,,&&m'_3=v+h_1+h_2\,,\nonumber\\
&m'_4=v+h_4+h_5\,,&&m'_5=v+h_2+h_3\,,&&m'_6=v+h_1+h_6\,.\nonumber
\end{align}}
With these parameters we can furthermore define
{\allowdisplaybreaks
\begin{align}
&\widehat{b}'_1=h'_1+v'_1=\widehat{b}_1\,,&&\widehat{b}'_2=h'_6+v'_6=\widehat{b}_2\,,&&\widehat{b}'_3=h'_3+v'_3=\widehat{b}_3\,,\nonumber\\
&\widehat{b}'_4=h'_5+v'_5=\widehat{b}_4\,,&&\widehat{b}'_5=h'_2+v'_2=\widehat{b}_5\,,\\[10pt]
&\widehat{a}'_1=m'_3+v'_6=\widehat{a}_1\,,&&\widehat{a}'_2=m'_5+v'_3=\widehat{a}_2\,,&&\widehat{a}'_3=m'_2+v'_5=\widehat{a}_3\,,\nonumber\\
&\widehat{a}'_4=m'_4+v'_2=\widehat{a}_4\,,&&\widehat{a}'_5=m'_1+v'_4=\widehat{a}_5\,,\\[10pt]
&\widehat{c}'_1=h'_1+m'_1\,,&&\widehat{c}'_2=h'_2+m'_2\,,&&\widehat{c}'_3=h'_4+m'_4\,,&&\widehat{c}'_4=h'_5+m'_5\,,\nonumber\\[10pt]
&L'=m'_1+h'_1+m'_2+h_2'+m_3'+h_3'\,,&&M'=\sum_{i=1}^6v'_i\,,&&v'=v'_1+v'_6+m'_6\,,\\[10pt]
&E'=m'_4+m'_5+m'_6\,,&&D'=\sum_{i=1}^6m'_i\,, &&\tau'=\sum_{i=1}^6(v_i'+m_i')
\end{align}}
which is more appropriate for their interpretation in terms of gauge theories: Indeed, in the same way as above, there are three regions in the K\"ahler cone of $X_{3,2}^{(\delta=1)}$ which suggest an interpretation as weak coupling regions of three gauge theories: 
\begin{itemize}
\item horizontal theory of $X_{6,1}^{(\delta=5)}$\\
In the limit $L' \to \infty$, the diagram $X_{6,1}^{(\delta=5)}$ decomposes into a single strip of length $6$, which suggests an interpretation as the weak coupling limit of a gauge theory with gauge group $U(6)$. In analogy to the theories with $\delta=0$ we call this theory the \emph{horizontal} theory, which is parametrised as follows
\begin{itemize}
\item coupling constant: the parameter $L'$ is related to the coupling constant
\item roots: the parameters $\widehat{a}'_{1,\ldots,5}$ play the role of the simple positive roots of $\mathfrak{a}_5$, which is extended to $\widehat{\mathfrak{a}}_5$ by $\tau'$
\item mass scale: the hypermultiplet mass scale of the theory is set by the parameter $D'$.
\end{itemize}
\item vertical theory of $X_{6,1}^{(\delta=5)}$\\
In the limit $\widehat{a}'_1\to \infty$ and $ 3\tau'-\widehat{a}'_1\to \infty$, the diagram $X_{3,2}^{(\delta=1)}$ decomposes into two strips, each of length $3$, which suggests an interpretation as the weak coupling limit of a gauge theory with gauge group $U(3)\times U(3)$. In analogy to the theories with $\delta=0$ we call this theory the \emph{vertical} theory, which is parametrised as follows
\begin{itemize}
\item coupling constants: the parameters $\widehat{a}'_1$ and $3\tau'-\widehat{a}'_1$ are related to the coupling constants
\item roots: the parameters $\widehat{c}'_{1,\ldots,4}$ play the role of the simple positive roots of two copies of $\mathfrak{a}_2$, which are extended to $\widehat{\mathfrak{a}}_2$ by $L'$
\item mass scale: the hypermultiplet mass scale of the theory is set by the parameter $E'$
\end{itemize}
\item diagonal theory of $X_{6,1}^{(\delta=5)}$\\
In the limit $v'\to \infty$, the diagram $X_{3,2}^{(\delta=1)}$ decomposes into a single strip of length $6$, which suggests an interpretation as the weak coupling limit of a gauge theory with gauge group $U(6)$. In analogy to the theories with $\delta=0$ we call this theory the 
\begin{itemize}
\item coupling constant: the parameter $v'$ is related to the coupling constant
\item roots: the parameters $\widehat{b}_{1,\ldots,5}$ play the role of the simple positive roots of $\mathfrak{a}_5$, which is extended to $\widehat{\mathfrak{a}}_5$ by $\rho$
\item mass scale: the hypermultiplet mass scale of the theory is set by the parameter $M'$
\end{itemize}
\end{itemize}
In all three cases, in the limit when the designated coupling constants vanish, the web diagram decomposes into several strips that engineer the perturbative limit of the corresponding gauge theory. This argument is a direct generalisation of \cite{Bastian:2017ary} in the case of the 'unshifted' $X_{N,M}$.

We also remark that the duality transformation $\mathcal{F}$, which relates $X_{6,1}^{(\delta=5)}=\mathcal{F}(X_{6,1}^{(\delta=0)})$, acts as a flop transformation on the parameters $h_{1,\ldots,6}$. Therefore, from the perspective of the horizontal gauge theory of $X_{6,1}^{(\delta=0)}$, this transformation goes through a strong coupling regime and is not realised purely perturbatively. As a consequence, $X_{6,1}^{(\delta=5)}$ does not engineer a theory with gauge group $U(1)^6$, but rather with its `strong coupling dual', which we called the vertical theory of $X_{6,1}^{(\delta=5)}$ and which we conjecture to have gauge group $U(3)\times U(3)$ in the perturbative limit of vanishing coupling constant.\footnote{The precise aspects of the perturbative expansion and specifically the strong coupling regime of this theory might in fact be more delicate, as showcased in a different example in the following section.} From the perspective of the remaining two gauge theories, the duality transformation acts purely in the weak coupling regime and therefore $X_{6,1}^{(\delta=5)}$ also still engineers two theories with gauge groups $U(6)$ (which we termed the horizontal and diagonal one).

Performing a further transformation $\mathcal{F}$ on $X_{6,1}^{(\delta=5)}$, we obtain yet another shifted web $X_{(6,1)}^{(\delta=4)}=\mathcal{F}(X_{(6,1)}^{(\delta=5)})$ whose K\"ahler cone allows again for three separate regions that engineer three different gauge theories. Continuing in this fashion and taking into account that the action of $\mathcal{F}$ on $X_{6,1}$ is of order $6$ (\emph{i.e.} $X_{6,1}^{(\delta)}=\mathcal{F}^{6}(X_{6,1}^{\delta})$), we find a whole orbit of dual Calabi-Yau, each of which engineering three different gauge theories. The gauge groups of the latter are summarised in the following table
\begin{center}
\begin{tabular}{l|c|c|c}
{\bf Calabi-Yau} & $G_{\text{hor}}$ & $G_{\text{vert}}$ & $G_{\text{diag}}$ \\ \hline\hline 
&&&\\[-15pt] 
$X_{6,1}^{(\delta=0)}$ & $[U(1)]^6$ & $U(6)$ & $U(6)$\\[2pt]\hline
&&&\\[-15pt] 
$X_{6,1}^{(\delta=5)}=\mathcal{F}(X_{6,1}^{(\delta=0)})$ & $U(6)$ & $[U(3)]^2$ & $U(6)$\\[2pt]\hline
&&&\\[-15pt] 
$X_{6,1}^{(\delta=4)}=\mathcal{F}^2(X_{6,1}^{(\delta=0)})$ & $[U(3)]^2$ & $[U(2)]^3$ & $U(6)$\\[2pt]\hline
&&&\\[-15pt] 
$X_{6,1}^{(\delta=3)}=\mathcal{F}^3(X_{6,1}^{(\delta=0)})$ & $[U(2)]^3$ & $[U(3)]^2$ & $U(6)$\\[2pt]\hline
&&&\\[-15pt] 
$X_{6,1}^{(\delta=2)}=\mathcal{F}^4(X_{6,1}^{(\delta=0)})$ & $[U(3)]^2$ & $U(6)$ & $U(6)$\\[2pt]\hline
&&&\\[-15pt] 
$X_{6,1}^{(\delta=1)}=\mathcal{F}^5(X_{6,1}^{(\delta=0)})$ & $U(6)$ & $[U(1)]^6$ & $U(6)$\\[2pt]
\end{tabular}
\end{center}
Notice that all gauge groups obtained in this fashion are of the form
\begin{align}
&[U(N')]^{M'}&&\text{with} &&\begin{array}{l}N'M'=6 \text{ and}\\[2pt] \text{gcd}(N',M')=1\,,\end{array}\label{ConstructionShift}
\end{align}
and thus all have the same rank. Moreover, all theories obtained in this way have gauge groups that can also be engineered from unshifted web diagrams that are related to $X_{6,1}$. While the details of these theories might still differ from the ones in (\ref{ConstructionShift}), we leave an in-depth analysis for future work~\cite{ToAppear}. In the following we shall discuss another example, which potentially leads to theories with new gauge groups that are not engineered by unshifted web diagrams.
%%%%%%%%%%%%%%%%%%%%%%%%%%%
%%%%%%%%%%%%%%%%%%%%%%%%%%%
\subsection{Example: $(N,M)=(4,1)$}
\subsubsection{Web Diagram $(4,1)$ Versus $(2,2)$}
\begin{figure}
\begin{center}
\scalebox{0.5}{\parbox{28.75cm}{\begin{tikzpicture}[scale = 1.25]
\draw[ultra thick] (-1,0) -- (0,0) -- (1,1) -- (2,1) -- (3,2) -- (4,2) -- (5,3) -- (6,3) -- (7,4) -- (8,4);
%vertical top
\draw[ultra thick] (1,1) -- (1,2);
\draw[ultra thick] (3,2) -- (3,3);
\draw[ultra thick] (5,3) -- (5,4);
\draw[ultra thick] (7,4) -- (7,5);
%vertical bottom
\draw[ultra thick] (0,0) -- (0,-1);
\draw[ultra thick] (2,1) -- (2,0);
\draw[ultra thick] (4,2) -- (4,1);
\draw[ultra thick] (6,3) -- (6,2);
%top labels
\node at (1,2.25) {\large $\mathbf{1}$};
\node at (3,3.25) {\large $\mathbf{2}$};
\node at (5,4.25) {\large $\mathbf{3}$};
\node at (7,5.25) {\large $\mathbf{4}$};
%bottom labels
\node at (0,-1.25) {\large $\mathbf{1}$};
\node at (2,-0.25) {\large $\mathbf{2}$};
\node at (4,0.75) {\large $\mathbf{3}$};
\node at (6,1.75) {\large $\mathbf{4}$};
%left horizontal
\node at (-1.25,0) {\large {\bf a}};
\node at (8.25,4) {\large {\bf a}};
%stamp
\node at (3.5,-2.5) {\Huge (a) $\delta=0$};
%%%%%
\begin{scope}[xshift=13cm]
\draw[ultra thick] (-1,0) -- (0,0) -- (1,1) -- (2,1) -- (3,2) -- (4,2) -- (5,3) -- (6,3) -- (7,4) -- (8,4);
%vertical top
\draw[ultra thick] (1,1) -- (1,2);
\draw[ultra thick] (3,2) -- (3,3);
\draw[ultra thick] (5,3) -- (5,4);
\draw[ultra thick] (7,4) -- (7,5);
%vertical bottom
\draw[ultra thick] (0,0) -- (0,-1);
\draw[ultra thick] (2,1) -- (2,0);
\draw[ultra thick] (4,2) -- (4,1);
\draw[ultra thick] (6,3) -- (6,2);
%top labels
\node at (1,2.25) {\large $\mathbf{3}$};
\node at (3,3.25) {\large $\mathbf{4}$};
\node at (5,4.25) {\large $\mathbf{1}$};
\node at (7,5.25) {\large $\mathbf{2}$};
%bottom labels
\node at (0,-1.25) {\large $\mathbf{1}$};
\node at (2,-0.25) {\large $\mathbf{2}$};
\node at (4,0.75) {\large $\mathbf{3}$};
\node at (6,1.75) {\large $\mathbf{4}$};
%left horizontal
\node at (-1.25,0) {\large {\bf a}};
\node at (8.25,4) {\large {\bf a}};
%cut
\draw[ultra thick,dashed,red] (3.5,0.5) -- (3.5,3.5);
\node at (3.5,-2.5) {\Huge (b) $\delta=2$};
\end{scope}
\end{tikzpicture}}}
\caption{\sl Toric web diagram for the configuration $(N,M)=(4,1)$ with shift $\delta$.}
\label{Fig:Diagram41Shift2}
\end{center}
\end{figure}
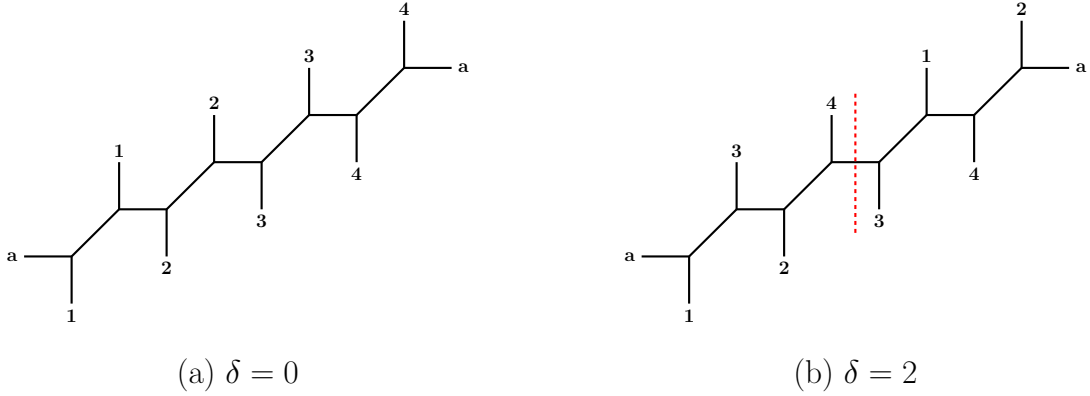

As another example we consider the case $(N,M)=(4,1)$, whose web diagram is shown in \figref{Fig:Diagram41Shift2} (a).
Performing the transformation $\mathcal{F}$ (see appendix~\ref{App:ShiftDuality}) twice, we obtain 
\begin{align}
\mathcal{F}^2(X_{4,1}^{(\delta=0)})=\mathcal{F}^2(X_{4,1})=X_{4,1}^{(\delta=2)}\,,
\end{align}
whose web diagram is shown in \figref{Fig:Diagram41Shift2} (b). Performing repeatedly the transformation $\mathcal{F}$ (recalling that it has order $4$ when acting on $X_{4,1}$) we can construct an orbit of Calabi-Yau manifolds $X_{(4,1)}^{(\delta)}$ for $\delta\in \{0,1,2,3\}$. Analysing again possible parametrisations of the corresponding K\"ahler moduli space, along with suitable decoupling limits to search for areas that engineer weak coupling regions of potential gauge theories, we are lead to the following list of candidate gauge groups
\begin{center}
\begin{tabular}{l|c|c|c}
{\bf Calabi-Yau} & $G_{\text{hor}}$ & $G_{\text{vert}}$ & $G_{\text{diag}}$ \\ \hline\hline 
&&&\\[-15pt] 
$X_{4,1}^{(\delta=0)}$ & $[U(1)]^4$ & $U(4)$ & $U(4)$\\[2pt]\hline
&&&\\[-15pt] 
$X_{4,1}^{(\delta=3)}=\mathcal{F}(X_{4,1}^{(\delta=0)})$ & $U(4)$ & $[U(2)]^2$ & $U(4)$\\[2pt]\hline
&&&\\[-15pt] 
$X_{4,1}^{(\delta=2)}=\mathcal{F}^2(X_{4,1}^{(\delta=0)})$ & $[U(2)]^2$ & $U(4)$ & $U(4)$\\[2pt]\hline
&&&\\[-15pt] 
$X_{4,1}^{(\delta=1)}=\mathcal{F}^3(X_{4,1}^{(\delta=0)})$ & $U(4)$ & $[U(1)]^4$ & $U(4)$\\[2pt]
\end{tabular}
\end{center}

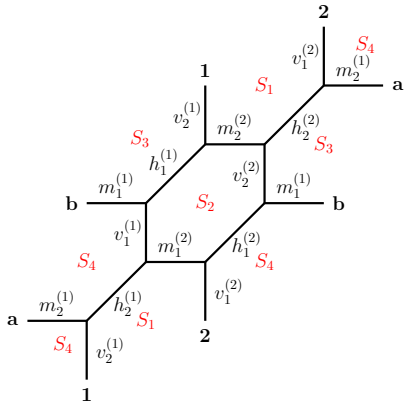
\begin{wrapfigure}{L}{0.33\textwidth}
\centering
\scalebox{0.52}{\parbox{10.5cm}{\begin{tikzpicture}[scale = 1.50]
\draw[ultra thick] (-1,0) -- (0,0) -- (1,1) -- (2,1) -- (3,2) -- (4,2);
\draw[ultra thick] (0,2) -- (1,2) -- (2,3) -- (3,3) -- (4,4) -- (5,4);
\draw[ultra thick] (0,-1) -- (0,0);
\draw[ultra thick] (2,0) -- (2,1);
\draw[ultra thick] (1,1) -- (1,2);
\draw[ultra thick] (3,2) -- (3,3);
\draw[ultra thick] (2,3) -- (2,4);
\draw[ultra thick] (4,4) -- (4,5);
%nodes bottom top
\node at (0,-1.25) {\large  {\bf 1}};
\node at (2,-0.25) {\large  {\bf 2}};
\node at (2,4.25) {\large  {\bf 1}};
\node at (4,5.25) {\large  {\bf 2}};
%nodes left right
\node at (-1.25,0) {\large  {\bf a}};
\node at (-0.25,2) {\large  {\bf b}};
\node at (4.25,2) {\large  {\bf b}};
\node at (5.25,4) {\large  {\bf a}};
%nodes m
\node at (-0.5,0.3) {\large  $m_2^{(1)}$};
\node at (1.5,1.3) {\large  $m_1^{(2)}$};
\node at (3.5,2.3) {\large  $m_1^{(1)}$};
\node at (0.5,2.3) {\large  $m_1^{(1)}$};
\node at (2.5,3.3) {\large  $m_2^{(2)}$};
\node at (4.5,4.3) {\large  $m_2^{(1)}$};
%nodes v
\node at (0.4,-0.5) {\large  $v_2^{(1)}$};
\node at (2.4,0.5) {\large  $v_1^{(2)}$};
\node at (0.7,1.5) {\large  $v_1^{(1)}$};
\node at (2.7,2.5) {\large  $v_2^{(2)}$};
\node at (1.7,3.5) {\large  $v_2^{(1)}$};
\node at (3.7,4.5) {\large  $v_1^{(2)}$};
%nodes v
\node at (0.7,0.3) {\large  $h_2^{(1)}$};
\node at (2.7,1.3) {\large  $h_1^{(2)}$};
\node at (1.3,2.7) {\large  $h_1^{(1)}$};
\node at (3.7,3.3) {\large  $h_2^{(2)}$};
%hexagons
\node[red] at (-0.4,-0.4) {\large  {\bf $S_4$}};
\node[red] at (1,0) {\large  {\bf $S_1$}};
\node[red] at (3,1) {\large  {\bf $S_4$}};
\node[red] at (2,2) {\large  {\bf $S_2$}};
\node[red] at (4,3) {\large  {\bf $S_3$}};
\node[red] at (0,1) {\large  {\bf $S_4$}};
\node[red] at (0.9,3.1) {\large  {\bf $S_3$}};
\node[red] at (3,4) {\large  {\bf $S_1$}};
\node[red] at (4.7,4.7) {\large  {\bf $S_4$}};
\end{tikzpicture}}}
\caption{\sl Toric web diagram and parametrisation of $X_{2,2}^{(1)}$.}
\label{Fig:Diagram22Shift1}
\end{wrapfigure}

\noindent
The appearance of the group $[U(2)]^2$ in this table is rather surprising since it is not of the form $U(N')^{M'}$ with $N'M' =4$ and $\text{gcd}(N',M')=\text{gcd}(4,1)=1$. Thus, if really a quiver gauge theory with this gauge group is engineered from (Calabi-Yau manifolds dual to) $X_{4,1}$, this indicates that the web of possible dual theories is yet even further enhanced. In particular, it would indicate that the condition $\text{gcd}(N,M)=\text{gcd}(N',M')$ could be relaxed in (\ref{DualityCalabiYauFlop}) for the construction of dual gauge theories. However, in the following we will find preliminary indications that $X_{4,1}^{(\delta=2)}$ does not engineer a gauge theory that realises  the gauge group $[U(2)]^2$ in a weak coupling regime (outside the limit of vanishing coupling constant). Rather the appearance of this group seems to be linked to a strong coupling effect in the 6-dimensional description, which may be linked to LSTs.

To discuss this aspect in more detail, in the following we consider $X_{4,1}^{(\delta=2)}$, whose web diagram is shown in \figref{Fig:Diagram41Shift2} (b). Upon cutting the diagram along the dashed red line and re-gluing it along the lines labelled {\bf 3} and {\bf 4}, respectively, the web diagram can be brought into the form of $X_{2,2}^{(\delta=1)}$, whose web diagram is shown in \figref{Fig:Diagram22Shift1} along with a labelling of the K\"ahler parameters. In the following we will present a geometrical presentation of the gauge algebra, by identifying the hexagons $S_{1,2,3,4}$ in the web diagram \figref{Fig:Diagram22Shift1} with the co-roots of the gauge algebra engineered from the web. With this, we shall be able to assign weights to all curves of the toric diagram and the corresponding K\"ahler parameters they represent. This will allow us to analyse in which fashion the algebra $\mathfrak{a}_1\oplus\mathfrak{a}_1$ is realised in the gauge theory engineered from \figref{Fig:Diagram22Shift1}.

%%%%%%%%%%%%%%%%%%%%%%%
\subsubsection{Topological String Partition Function of $X_{N,M}$}\label{Sect:ReviewPartitionFunction}
Our starting point for finding a geometric realisation of the gauge algebra is the expansion of the partition function adapted to a particular gauge theory engineered by $X_{N,M}$ (associated with an unshifted web diagram): as discussed in \cite{Bastian:2017ing}, the refined topological string partition function for $X_{N,M}$ can be expanded in three different fashions. All three can compactly be written in the following form
\begin{align}
\mathcal{Z}_{N,M}&=\sum_{\{\alpha_j^{(i)}\}}\left(\prod_{j=1}^BQ_{g_j}^{\sum_{i=1}^A|\alpha_j^{(i)}|}\right)\,\prod_{i=1}^B W_{\alpha_1^{(i+1)}\ldots \alpha_A^{(i+1)}}^{\alpha_1^{(i)}\ldots \alpha_A^{(i)}}(Q_{r,s}^{(m,n)},\tilde{Q}_{r,s}^{(m,n)};q,t) \,,\label{PartitionGenericFunction}
\end{align} 
where $(A,B)$ are as defined in (\ref{GeneralGroups}). The summation $\alpha_j^{(i)}$ (with $\alpha_j^{(B+1)}=\alpha_j^{(1)}$ for $j=1,\ldots,A$) in (\ref{PartitionGenericFunction}) is over integer partitions of $|\alpha_i^{(j)}|\in\mathbb{N}$ and $Q_{g_i}=e^{-g_i}$ (for $i=1,\ldots,B$) (where $g_i$ has been introduced 
in section~\ref{Sect:BaseChoice}). The $W_{\alpha_1^{(i+1)}\ldots \alpha_A^{(i+1)}}^{\alpha_1^{(i)}\ldots \alpha_A^{(i)}}$ are the fundamental building blocks introduced in  \cite{Bastian:2017ing}, which depend on the $U(1)$ deformation parameters $q=e^{2\pi i\epsilon_1}$ and $t=e^{-2\pi i\epsilon_2}$ (which are related to the refinement of the topological string \cite{Gopakumar:1998ii, Gopakumar:1998jq, Nekrasov:2002qd,Hollowood:2003cv} as well as particular combinations $Q_{r,s}^{(m,n)}$ and $\tilde{Q}_{r,s}^{(m,n)}$ (with $m,n=1,\ldots, B$ and $r,s=1,\ldots,A$) of the remaining K\"ahler parameters of $X_{N,M}$. While the precise definition of $W_{\alpha_1^{(i+1)}\ldots \alpha_A^{(i+1)}}^{\alpha_1^{(i)}\ldots \alpha_A^{(i)}}$ is not important in the following, we remark that the topological string partition function can be written as the following quotient
\begin{align}
\mathcal{Z}_{N,M}&= \sum_{\{\alpha_j^{(i)}\}} \left(\prod_{j=1}^BQ_{g_j}^{\sum_{i=1}^A|\alpha_j^{(i)}|}\right) \prod_{r,s=1}^{A} \prod_{{m,n=1}\atop{|m-n|=1}}^{B} \frac{\vartheta_{\alpha_r^{(m)}\alpha_s^{(n)}}(Q_{r,s}^{(m,n)};\rho)}{\vartheta_{\alpha_r^{(m)}\alpha_s^{(n)}}(\tilde{Q}_{r,s}^{(m,n)};\rho)}\,.\label{TopPartExpansionQuotient}
\end{align}
Here $\vartheta_{\mu\nu}$ is a particular class of theta-functions (see eq.~(\ref{DefCurlTheta}) for the definition and \cite{Haghighat:2013gba,Hohenegger:2013ala} for further information) that implicitly depend on $\epsilon_{1,2}$ and are labelled by two integer partitions. Finally, the (modular) parameter $\rho$ is related to the modular parameter of the base, when viewing the Calabi-Yau threefold $X_{N,M}$ as a particular quotient of $G\cong \mathbb{Z}_N\times \mathbb{Z}_M$. 

As already explained above, the partition function of the gauge theory engineered from $X_{N,M}$ is captured by $\mathcal{Z}_{N,M}$. To interpret the various (independent) K\"ahler parameters of $X_{N,M}$ that appear in the expansions of $\mathcal{Z}_{N,M}$ from a gauge theory perspective, we can compare the combinations of $\vartheta$-functions appearing in the numerator and denominator in (\ref{TopPartExpansionQuotient}) with the Nekrasov subfunctions \cite{Nekrasov:2002qd, Nekrasov:2003rj, Mironov:2009qt} that are reviewed in appendix \ref{NekSub}. In fact, turning this interpretation around, we can associate a weight (with respect to the gauge algebra) to the various K\"ahler parameters of the partition function, and thus also to the curves in the Calabi-Yau manifold $X_{N,M}$ they are represented by. This allows us to give a geometric realisation of the gauge algebra.\footnote{In appendix~\ref{Sect:Intersect22}, we obtain the same result for the particular $(N,M)=(2,2)$ based on an observation concerning the intersection numbers of various curves with (the canonical classes of) certain surfaces in the web diagram. A more abstract and general discussion of the connection between the Calabi-Yau geometry and the gauge algebra of the theories engineered by $X_{N,M}$  will be further discussed in the future~\cite{ToAppear}.}
 
%%%%%%%%%%%%%%%%%%%%%%%%
\label{RootsWeights}\label{Sect:RootsAndWeights}
\subsubsection{Intersection Numbers}
As reviewed above~\cite{Bastian:2017ary}, the area of certain (combinations of) curves in the webdiagram of $X_{N,M}$ are identified with the roots of $A$ copies of the Lie algebra $\widehat{\mathfrak{a}}_{A-1}$ (in the notation of (\ref{GeneralGroups})). The latter is the non-twisted affine extension of $\mathfrak{a}_{A-1}$, which is the Lie algebra of one of the $U(B)$ factors in the gauge group $G$ in (\ref{GeneralGroups}). The correspondence between roots and curves is established by matching the intersection product on the geometry side with the scalar product on the Lie algebra side \cite{Esole:2015xfa}. As we work in a three-dimensional complex geometry, we are taking intersections of curves with compact surfaces $S_i$.\footnote{In appendix~\ref{Sect:Intersect22}, we observe in the case $(N,M)=(2,2)$ that the same relation also applies when replacing the surfaces $S_i$ with their canonical classes $K_{S_i}$. This connection is easier to evaluate technically and will therefore be used later on.}
\begin{figure}[htb]
\begin{center}
\scalebox{0.8}{\parbox{15cm}{\begin{tikzpicture}[scale = 1.5]
%horizontal lines%%%%%%%%%%
%zeroth layer
\draw[ultra thick, dotted] (0,-1) -- (1,-1);
\draw[ultra thick, dotted] (2,0) -- (3,0);
\draw[ultra thick, dotted] (5,0) -- (6,0);
\draw[ultra thick, dotted] (7,1) -- (8,1);
\node at (0,-1.25) {{\footnotesize $\alpha_1^{(i)}$}};
\node at (2.1,-0.25) {{\footnotesize $\alpha_2^{(i)}$}};
\node at (7.1,0.75) {{\footnotesize $\alpha_A^{(i)}$}};
\node at (4,0) {\Large $\cdots$};
%
%first layer
\draw[ultra thick] (-1,0) -- (0,0);
\draw[ultra thick] (1,1) -- (2,1);
\draw[ultra thick] (3,2) -- (4,2);
\node at (0.9,2.25) {{\footnotesize $\alpha_1^{(i+1)}$}};
\node at (2.9,3.25) {{\footnotesize $\alpha_2^{(i+1)}$}};
\node at (7.9,4.25) {{\footnotesize $\alpha_A^{(i+1)}$}};
\node at (5,2) {\Large $\cdots$};
\draw[ultra thick] (6,2) -- (7,2);
\draw[ultra thick] (8,3) -- (9,3);
%second layer
\draw[ultra thick,dotted] (0,2) -- (1,2);
\draw[ultra thick,dotted] (2,3) -- (3,3);
\draw[ultra thick,dotted] (4,4) -- (5,4);
\node at (6,4) {\Large $\cdots$};
\draw[ultra thick,dotted] (7,4) -- (8,4);
%vertical lines%%%%%%%%%%
%first layer
\draw[ultra thick] (0,0) -- (0,-1);
\draw[ultra thick] (2,1) -- (2,0);
\draw[ultra thick,dotted] (4,2) -- (4,1);
\draw[ultra thick,dotted] (5,1) -- (5,0);
\draw[ultra thick] (7,2) -- (7,1);
\draw[ultra thick, dotted] (9,3) -- (9,2);
%second layer
\draw[ultra thick,dotted] (-1,0) -- (-1,1);
\draw[ultra thick] (1,1) -- (1,2);
\draw[ultra thick] (3,2) -- (3,3);
\draw[ultra thick,dotted] (5,3) -- (5,4);
\draw[ultra thick,dotted] (6,2) -- (6,3);
\draw[ultra thick] (8,3) -- (8,4);
%diagonal lines%%%%%%%%%%
%zeroth layer
\draw[ultra thick,dotted] (1,-1) -- (2,0);
\draw[ultra thick,dotted] (3,0) -- (4,1);
\draw[ultra thick,dotted] (6,0) -- (7,1);
\draw[ultra thick,dotted] (8,1) -- (9,2);
%first layer
\draw[ultra thick] (0,0) -- (1,1);
\draw[ultra thick] (2,1) -- (3,2);
\draw[ultra thick,dotted] (4,2) -- (5,3);
\draw[ultra thick,dotted] (5,1) -- (6,2);
\draw[ultra thick] (7,2) -- (8,3);
%second layer
\draw[ultra thick,dotted] (-1,1) -- (0,2);
\draw[ultra thick,dotted] (1,2) -- (2,3);
\draw[ultra thick,dotted] (3,3) -- (4,4);
\draw[ultra thick,dotted] (6,3) -- (7,4);
%%%%%%%%%%%%%%%%%%%%%%
%roots
%upper
\draw[ultra thick,red] (-0.9,0.1) -- (-0.05,0.1) -- (0.85,1);
\draw[ultra thick,red] (1.1,1.1) -- (1.95,1.1) -- (2.85,2);
\draw[ultra thick,red] (3.1,2.1) -- (3.95,2.1) -- (4.85,3);
\draw[ultra thick,red] (6.1,2.1) -- (6.95,2.1) -- (7.85,3);
%lower
\draw[ultra thick,green!75!black] (0.15,0) -- (1.05,0.9) -- (1.95,0.9);
\draw[ultra thick,green!75!black] (2.15,1) -- (3.05,1.9) -- (3.95,1.9);
\draw[ultra thick,green!75!black] (5.15,1) -- (6.05,1.9) -- (6.95,1.9);
\draw[ultra thick,green!75!black] (7.15,2) -- (8.05,2.9) -- (8.95,2.9);
%%%%%%%%%%%%%%%%%%%%%%
%connectors
\node[rotate=90] at (-0.75,0) {$=$}; 
\node at (-0.85,-0.2) {{\footnotesize $1$}};
\node[rotate=90] at (8.75,3) {$=$}; 
\node at (8.85,3.2) {{\footnotesize $1$}};
%%%%%%%%%%%%%%
%surfaces
%upper
\node at (0,1) {{$S_1$}};
\node at (2,2) {{$S_2$}};
\node at (4,3) {{$S_3$}};
\node at (7,3) {{$S_A$}};
%lower
\node at (1,0) {{$S_1'$}};
\node at (3,1) {{$S_2'$}};
\node at (6,1) {{$S_{A-1}'$}};
\node at (8,2) {{$S_A'$}};
\end{tikzpicture}}}
\caption{\sl Strip with two sets of compact surfaces labeled by $S_i$ and $S_j'$. Red (Green) curves correspond to roots with respect to $S_i$ ($S_j'$). The (integer partition) labels $\alpha_{1,\ldots, A}^{(i)}$ and $\alpha_{1,\ldots, A}^{(i+1)}$ indicate how the strip is glued into the web diagram of the Calabi-Yau manifold.}
\label{Fig:strip1}
\end{center}
\end{figure}
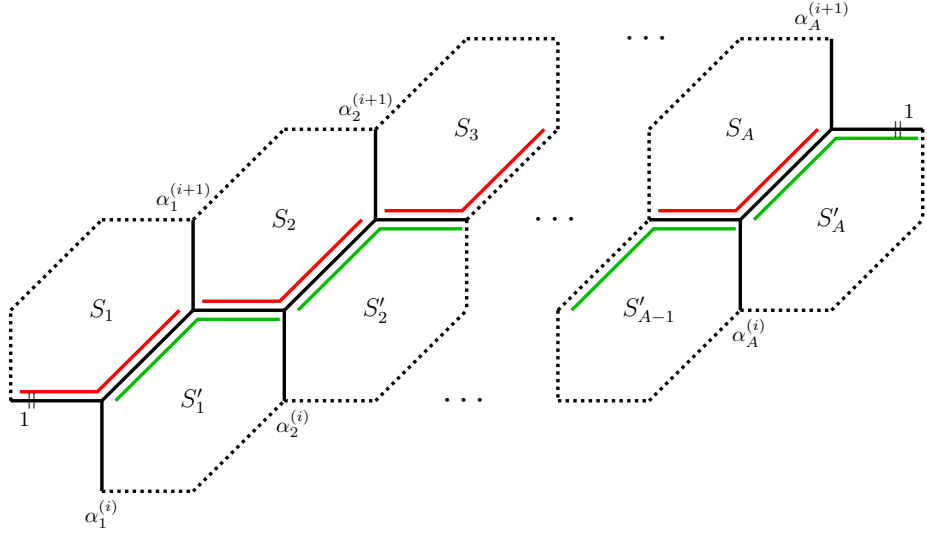
Concretely, we consider a single `strip' of length $A$ in the web diagram of $X_{N,M}$., as shown in \figref{Fig:strip1} and which (from the perspective of the partition function~(\ref{PartitionGenericFunction})) is captured by the building block $W_{\alpha_1^{(i+1)}\ldots \alpha_A^{(i+1)}}^{\alpha_1^{(i)}\ldots \alpha_A^{(i)}}$. In \figref{Fig:strip1} we have indicated two families of compact surfaces $S_{i=1,\ldots,A}$ and $S'_{j=1,\ldots,A}$ that corresponds to hexagons in the web diagram. The basic idea is to interpret these surfaces $S$ and $S'$ as representations of the co-roots of two copies of the affine algebras $\widehat{\mathfrak{a}}_{A-1}$ (see \cite{Esole:2015xfa,Intriligator:1997dh}). Indeed, the matrix of intersection numbers of the red (green) curves with the surfaces $S_i$ ($S_j'$), reproduces (up to a sign) the Cartan matrix of $\widehat{\mathfrak{a}}_{A-1}$.\footnote{A more detailed discussion of this geometric realisation of the gauge algebra shall be given in \cite{ToAppear}. As remarked before, in appendix~\ref{Sect:Intersect22} we argue for the case $(N,M)=(2,2)$ that the role of the co-roots can equally be played by the canonical classes of the surfaces $S_{1,\ldots,A}$ and $S'_{1,\ldots,A}$, which shall be important for the explicit computations in section~\ref{Sect:22ExplicitExample}.} This allows us to identify these curves with the set of positive simple roots of the affine Lie algebra $\widehat{\mathfrak{a}}_{A-1}$. In the following we keep the notation generic (thus treating the horizontal, vertical and diagonal theory in parallel), by calling the roots associated with the red (green) lines collectively $\{\widehat{\beta}_0, \widehat{\beta}_1,\dots, \widehat{\beta}_{A-1}\}$ ($\{\widehat{\beta}'_0, \widehat{\beta}'_1,\dots, \widehat{\beta}'_{A-1}\}$). By convention, $\widehat{\beta}_0$ ($\widehat{\beta}'_0$) denotes the root that extends the root system of $\mathfrak{a}_{A-1}$ to that of the affine Lie algebra $\widehat{\mathfrak{a}}_{A-1}$: The choice of the curve corresponding to $\widehat{\beta}_0$ is fixed up to cyclic permutations, which reflects the rotational symmetry of the affine Dynkin diagram. Although the web diagram realises the structure of an affine Lie algebra, in this work, we mostly focus on aspects pertaining to properties of the finite algebra $\mathfrak{a}_{A-1}$. We shall relegate further discussions of the affine structure to \cite{ToAppear}.

It is important to realise that for $A<NM$ (such that $B>1$) the two sets of roots $\widehat{\beta}_i$ and $\widehat{\beta}'_i$ related to the red and green curves, respectively, are in general distinct and two different (copies of the same) algebras. This can be seen from the intersection numbers\footnote{Here the symbol $\circ$ indicates that the intersection is calculated in the full 6-dimensional Calabi-Yau manifold.}
\begin{align}
&S_i\circ \widehat{\beta}'_j=0=S'_i\circ \widehat{\beta}_j\,,&&\forall i,j\in\{0,\ldots,A-1\}\,,
\end{align} 
\emph{i.e.} the weights of one set of roots with respect to the co-roots of another copy of $\widehat{\mathfrak{a}}_{A-1}$ vanish. In the case $A=NM$ (such that $B=1$) where the whole web diagram in fact only consists of the single strip shown in \figref{Fig:strip1} (with the external legs $\alpha_{1,\ldots,A}^{(1)}$ and $\alpha_{1,\ldots,A}^{(2)}$ being identified, possibly after some cyclic rotation), the sets $S_i$ and $S_j'$ of compact surfaces are identified with each other. Therefore, in this case we in fact only have a single set of roots. Calculating the intersection numbers in this scenario is more intricate and is discussed in \cite{ToAppear}.

Once the co-roots are identified in a geometric fashion, we can assign a (non-affine) weight vector to any curve $\mathcal{C}$ in the web-diagram by calculating the intersections with $S_{i=1,\ldots,A-1}$ and $S'_{i=1,\ldots, A-1}$. Specifically, we define
\begin{align}
w_{\mathcal{C}}:=\big([\lambda_1,\dots,\lambda_{A-1}],[\lambda_1',\dots,\lambda_{A-1}']\big)=\big([\mathcal{C}\circ S_1,\dots,\mathcal{C}\circ S_{A-1}],[\mathcal{C}\circ S'_1,\dots,\mathcal{C}\circ S'_{A-1}]\big) \,, \label{CurveWeight}
\end{align}
and interpret $\lambda_i$ and $\lambda_j'$ as the Dynkin labels of the two copies of $\mathfrak{a}_{A-1}$. In the case $B=1$, the two gauge algebras are identified such that we take the direct sum of the two weight vectors
\begin{align}
w_{\mathcal{C}}=[\lambda_1+\lambda_1',\dots,\lambda_{N-1}+\lambda_{N-1}']\,.
\end{align}
%%%%%%%%%%%%%%%%%%%%%%%%%%%
\subsubsection{The Example of $(N,M)=(2,2)$ with $\delta=0$}
\begin{wrapfigure}{L}{0.33\textwidth}
\centering
\scalebox{0.52}{\parbox{10.5cm}{\begin{tikzpicture}[scale = 1.50]
\draw[ultra thick] (-1,0) -- (0,0) -- (1,1) -- (2,1) -- (3,2) -- (4,2);
\draw[ultra thick] (0,2) -- (1,2) -- (2,3) -- (3,3) -- (4,4) -- (5,4);
\draw[ultra thick] (0,-1) -- (0,0);
\draw[ultra thick] (2,0) -- (2,1);
\draw[ultra thick] (1,1) -- (1,2);
\draw[ultra thick] (3,2) -- (3,3);
\draw[ultra thick] (2,3) -- (2,4);
\draw[ultra thick] (4,4) -- (4,5);
%nodes bottom top
\node at (0,-1.25) {\large  {\bf 1}};
\node at (2,-0.25) {\large  {\bf 2}};
\node at (2,4.25) {\large  {\bf 1}};
\node at (4,5.25) {\large  {\bf 2}};
%nodes left right
\node at (-1.25,0) {\large  {\bf b}};
\node at (-0.25,2) {\large  {\bf a}};
\node at (4.25,2) {\large  {\bf b}};
\node at (5.25,4) {\large  {\bf a}};
%nodes m
\node at (-0.5,0.3) {\large  $h_1$};
\node at (1.5,1.3) {\large  $h_2$};
\node at (0.5,2.3) {\large  $h_1$};
\node at (2.5,3.3) {\large  $h_2$};
%nodes v
\node at (0.7,1.5) {\large  $v_2$};
\node at (2.7,2.5) {\large  $v_2$};
\node at (1.7,3.5) {\large  $v_1$};
\node at (3.7,4.5) {\large  $v_1$};
%nodes v
\node at (0.7,0.3) {\large  $m_1$};
\node at (2.7,1.3) {\large  $m_2$};
\node at (1.3,2.7) {\large  $m_2$};
\node at (3.7,3.3) {\large  $m_1$};
%hexagons
\node[red] at (-0.4,-0.4) {\large  {\bf $S_3$}};
\node[red] at (1,0) {\large  {\bf $S_1$}};
\node[red] at (3,1) {\large  {\bf $S_3$}};
\node[red] at (2,2) {\large  {\bf $S_2$}};
\node[red] at (4,3) {\large  {\bf $S_4$}};
\node[red] at (0,1) {\large  {\bf $S_4$}};
\node[red] at (0.9,3.1) {\large  {\bf $S_3$}};
\node[red] at (3,4) {\large  {\bf $S_1$}};
\node[red] at (4.7,4.7) {\large  {\bf $S_3$}};
\end{tikzpicture}}}
\caption{\sl Toric web diagram and parametrisation of $X_{2,2}$.}
\label{Fig:Diagram22}
\end{wrapfigure}
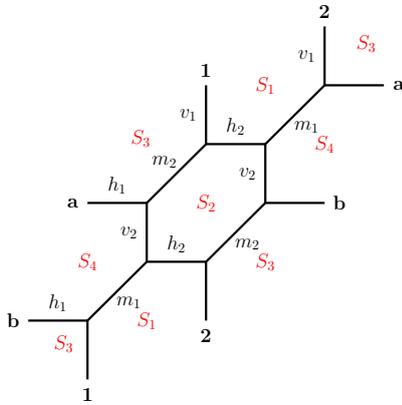

Returning to the example of the configuration $(N,M)=(2,2)$, we first discuss the assignment of weights for $X_{2,2}$ (\emph{i.e.} with $\delta=0$), whose web diagram is show in \figref{Fig:Diagram22}. In this figure, we have also indicated a labelling of the K\"ahler parameters as well as the surfaces $S_i$ introduced in the previous subsection. The former have been chosen in such a manner as to already satisfy the consistency conditions such that $(h_{1,2},v_{1,2},m_{1,2})$ are a set of independent parameters. 

In appendix~\ref{Sect:Intersect22} we have shown that the web diagram~\figref{Fig:Diagram22} is related (through flop transformations) to the geometry shown in~\figref{Fig:22StartingP1P1} (a), corresponding to local geometries of the type $\mathbb{P}^1\times \mathbb{P}^1$. The latter suggests a construction of the canonical classes $K_{S_{1,2,3,4}}$ of the four surfaces $S_{1,2,3,4}$, which is invariant under (certain) flop transformations. We furthermore observe, that a geometric realisation of the gauge algebras can be obtained, where the $K_{S_i}$ play the role of the co-roots and certain combinations of curves of the diagram can be interpreted as the simple positive roots.

%%%%%%%%%%%%%%%%%%%%
It is important to mention that the geometric realisation of the gauge algebra observed in appendix~\ref{Sect:Intersect22}, is very natural from the point of view of assigning weights to the various K\"ahler parameters appearing in the partition function~(\ref{TopPartExpansionQuotient}). Indeed, depending on the gauge theoretic interpretation, we expect the following weights for the individual K\"ahler parameters
\begin{itemize}
\item The coupling constants should be neutral under the gauge group. If the coupling constants transform non-trivially under the gauge group, the latter could not be realised perturbatively 
\item The $\vartheta$-functions in the numerator of~(\ref{TopPartExpansionQuotient}) are the matter contribution, stemming from the hypermultiplet. They should therefore carry the corresponding matter representation
\item The $\vartheta$-functions in the denominator of~(\ref{TopPartExpansionQuotient}) are the contribution from the (gauge) vector multiplet contribution. They should therefore carry the fundamental representation of the gauge group.
\end{itemize}
Specifically for $X_{2,2}$, the partition function takes the form
{\allowdisplaybreaks
\begin{align}
\mathcal{Z}_{2,2}^{(\delta=0)}=&\sum_{\alpha_1^{(1)},\alpha_2^{(1)},\alpha_1^{(2)},\alpha_2^{(2)}}  (Q_{v_1}Q_{m_2})^{|\alpha_1^{(1)}|+|\alpha_2^{(1)}|} (Q_{v_2}Q_{m_1})^{|\alpha_1^{(2)}|+|\alpha_2^{(2)}|}\nonumber\\
&\times \frac{\vartheta_{\alpha_1^{(1)} \alpha_1^{(2)}}(Q_{m_1})\vartheta_{\alpha_1^{(1)} \alpha_2^{(2)}}(Q_{m_2} \widehat{Q}_{1})\vartheta_{\alpha_2^{(1)} \alpha_1^{(2)}}(Q_{m_1}\widehat{Q}_{1}^{-1})\vartheta_{\alpha_2^{(1)} \alpha_2^{(2)}}(Q_{m_2}) }{\vartheta_{\alpha_1^{(1)} \alpha_1^{(1)}}(1)\vartheta_{\alpha_1^{(1)} \alpha_2^{(1)}}(\widehat{Q}_{1}^{-1})\vartheta_{\alpha_2^{(1)} \alpha_1^{(1)}}(\widehat{Q}_{1})\vartheta_{\alpha_2^{(1)} \alpha_2^{(1)}}(1) } \nonumber \\
&\times \frac{\vartheta_{\alpha_1^{(2)} \alpha_1^{(1)}}(Q_{m_2})\vartheta_{\alpha_1^{(2)} \alpha_2^{(1)}}(Q_{m_1} \widehat{Q}_{2})\vartheta_{\alpha_2^{(2)} \alpha_1^{(1)}}(Q_{m_2}\widehat{Q}_{2}^{-1})\vartheta_{\alpha_2^{(2)} \alpha_2^{(1)}}(Q_{m_1})}{\vartheta_{\alpha_1^{(2)} \alpha_1^{(2)}}(1)\vartheta_{\alpha_1^{(2)} \alpha_2^{(2)}}(\widehat{Q}_{2}^{-1})\vartheta_{\alpha_2^{(2)} \alpha_1^{(2)}}(\widehat{Q}_{2})\vartheta_{\alpha_2^{(2)} \alpha_2^{(2)}}(1)}\,, \label{Part22}
\end{align}}
where we introduced the following notation 
\begin{align}
&\widehat{Q}_1= Q_{m_1}\,Q_{h_2}\,,&&\text{and} &&\widehat{Q}_2=Q_{m_2}\,Q_{h_2}\,.
\end{align}
%%%%%%%%%%%%%%%%%%%%
Using the procedure outline in section \ref{RootsWeights}, we can associate weights to the curves whose K\"ahler parameters make up the arguments of the $\vartheta$-functions and to those that are related to the coupling constants (\emph{i.e.} expansion parameters). The geometry $X_{2,2}$, displayed in \figref{Fig:Diagram22}, is known to engineer six-dimensional $U(2) \times U(2)$ quiver gauge theory with 2 bi-fundamentals. We choose the surfaces $S_1$ and $S_2$ to correspond to the finite simple coroots of the respective gauge factors and denote the weight of a curve $\mathcal{C}$ by 
\begin{align}
w_{\mathcal{C}}=\left([\mathcal{C}\cdot K_{S_1}],[\mathcal{C}\cdot K_{S_2}]\right)=([\lambda_1],[\lambda_1'])\,,\label{WeightsKanonical}
\end{align}
which, as explained above, in the current case are computed more efficiently as the intersection numbers of the curve $\mathcal{C}$ with $K_{S_1}$ and $K_{S_2}$.\footnote{Indeed, we have observed in appendix~\ref{Sect:Intersect22} that the canonical classes $K_{S_{1,2,3,4}}$ give rise to a presentation of the gauge algebra. Furthermore, the intersection $\mathcal{C}\cdot K_{S_i}$ is understood to be taken inside the compact surface $S_i$.} Group theoretically, $[\lambda_1]$ and $[\lambda_1']$ are the Dynkin labels corresponding respectively to the first and to the second gauge factor. 

Let us first look at the matter representation, \emph{i.e.} the $\vartheta$-functions in the numerator of (\ref{Part22}). The curves that appear in their arguments have areas $m_{1,2}$, $m_1-\widehat{\alpha}_1$ and $m_2+\widehat{\alpha}_1$. Using (\ref{WeightsKanonical}), their associated weights are in fact those of two bifundamental representations:
\begin{align}
&m_1 \to ([1],[-1])\,, && m_2+\widehat{\alpha}_1 \to ([1],[1]) \nonumber \\
&m_2 \to ([-1],[1]) \,,&& m_1-\widehat{\alpha}_1 \to ([-1],[-1])
\end{align}
These weights correspond to the anticipated product representation. 

For the vector contribution in the denominator we have the following:
\begin{align}
&\pm \widehat{\alpha}_1 \to ([\pm 2],[0]) \,, && \pm \widehat{\alpha}_2 \to ([0],[\pm 2]) \,, && 0 \to ([0],[0])
\end{align}
These give the right product representation for two adjoint representations. \\
Concerning the coupling constants we obtain for the weight factors
\begin{align}
&v_1+m_2 \to ([0],[0])  \,, && v_2+m_1 \to ([0],[0])
\end{align}
As expected they are uncharged under the gauge group.
%%%%%%%%%%%%%%%%%%%%%%%%%%%%%%%%%%%%%%%
%%%%%%%%%%%%%%%%%%%%%%%%%%%%%%%%%%%%%%%
%%%%%%%%%%%%%%%%%%%%%%%%%%%%%%%%%%%%%%%
%%%%%%%%%%%%%%%%%%%%%%%%%%%%%%%%%%%%%%%%%%%%%%
\subsubsection{The Example of $(N,M)=(2,2)$ with $\delta=1$}\label{Sect:22ExplicitExample}

After the example of $X_{2,2}^{(\delta=0)}$ we are finally ready to consider $X_{2,2}^{(\delta=1)}$. The latter can be analysed in a similar fashion as before: The web diagram is shown in \figref{Fig:Diagram22Shift1} where we have also introduced a parametrisation of the K\"ahler parameters by associating an area to each of the curves of the web. As explained in detail in \cite{Hohenegger:2016yuv}, the parameters $\{m_i^{(j)},v_i{^{(j)}},h_i^{(j)}\}$ are not independent, but there are consistency conditions associated with each of the four hexagons $S_{1,2,3,4}$:
\begin{align}
&S_1:&&h_2^{(1)}+m_1^{(2)}=m_2^{(2)}+h_2^{(2)}\,,&&v_2^{(1)}+h_2^{(1)}=v_1^{(2)}+h_2^{(2)}\,,\nonumber\\
&S_2:&&m_1^{(2)}+h_1^{(2)}=h_1^{(1)}+m_2^{(2)}\,,&&v_1^{(1)}+h_1^{(1)}=h_1^{(2)}+v_2^{(2)}\,,\nonumber\\
&S_3:&&m_1^{(1)}+h_1^{(1)}=h_2^{(2)}+m_2^{(1)}\,,&&v_2^{(1)}+h_1^{(1)}=v_2^{(2)}+h_2^{(2)}\,,\nonumber\\
&S_4:&&m_2^{(1)}+h_2^{(1)}=h_1^{(2)}+m_1^{(1)}\,,&&v_1^{(2)}+h_1^{(2)}=h_2^{(1)}+v_1^{(1)}\,.
\end{align}
Furthermore, the canonical classes $K_{S_{1,2,3,4}}$ have been worked out in appendix~\ref{Sect:Intersect22} and have been shown to give a presentation of the gauge algebra. Using the general formula derived in \cite{Bastian:2017ing}, the partition function associated with this web diagram can be written in the form
{\allowdisplaybreaks
\begin{align}
\mathcal{Z}_{2,2}^{(\delta=1)}&=\sum_{\alpha_1^{(1)},\alpha_2^{(1)},\alpha_1^{(2)},\alpha_2^{(2)}} (Q_{m_1^{(1)}}Q_{h_1^{(1)}}Q_{h_2^{(2)}}\widehat{Q}_{1,1})^{|\alpha_1^{(1)}|} (Q_{m_2^{(1)}}Q_{h_1^{(2)}}Q_{h_2^{(1)}}\widehat{Q}_{2,1})^{|\alpha_2^{(1)}|}   \nonumber \\
&\hspace{0.5cm} \times(Q_{m_1^{(2)}}Q_{h_1^{(1)}}Q_{h_1^{(2)}}\widehat{Q}_{1,2})^{|\alpha_1^{(2)}|} (Q_{m_2^{(2)}}Q_{h_2^{(2)}}Q_{h_2^{(1)}}\widehat{Q}_{2,2})^{|\alpha_2^{(2)}|}\, Q_{\rho}^{-\frac{|\alpha_1^{(1)}|+|\alpha_2^{(1)}|+|\alpha_1^{(2)}|+|\alpha_2^{(2)}|}{2} } \nonumber \\
&\hspace{0.5cm} \times \frac{\vartheta_{\alpha_1^{(1)} \alpha_1^{(2)}}(Q_{h_1^{(1)}}\widehat{Q}_{1,2})\vartheta_{\alpha_1^{(1)} \alpha_2^{(2)}}(Q_{h_1^{(1)}})\vartheta_{\alpha_2^{(1)} \alpha_1^{(2)}}(Q_{h_2^{(1)}})\vartheta_{\alpha_2^{(1)} \alpha_2^{(2)}}(Q_{h_2^{(1)}}\widehat{Q}_{2,2})}{\vartheta_{\alpha_1^{(1)} \alpha_1^{(1)}}(1)\vartheta_{\alpha_1^{(1)} \alpha_2^{(1)}}(\widehat{Q}_{2,1})\vartheta_{\alpha_2^{(1)} \alpha_1^{(1)}}(\widehat{Q}_{1,1})\vartheta_{\alpha_2^{(1)} \alpha_2^{(1)}}(1)} \nonumber \\
&\hspace{0.5cm} \times \frac{\vartheta_{\alpha_1^{(2)} \alpha_1^{(1)}}(Q_{h_1^{(2)}})\vartheta_{\alpha_1^{(2)} \alpha_2^{(1)}}(Q_{h_1^{(2)}}\widehat{Q}_{2,1})\vartheta_{\alpha_2^{(2)} \alpha_1^{(1)}}(Q_{h_2^{(2)}}\widehat{Q}_{1,1})\vartheta_{\alpha_2^{(2)} \alpha_2^{(1)}}(Q_{h_2^{(2)}})}{\vartheta_{\alpha_1^{(1)} \alpha_1^{(1)}}(1)\vartheta_{\alpha_1^{(1)} \alpha_2^{(1)}}(\widehat{Q}_{2,2})\vartheta_{\alpha_2^{(1)} \alpha_1^{(1)}}(\widehat{Q}_{1,2})\vartheta_{\alpha_2^{(1)} \alpha_2^{(1)}}(1)}\,, \nonumber 
\end{align}}
where we introduced the notation
\begin{align}
&\widehat{Q}_{1,1}=Q_{v_1^{(1)}}Q_{h_2^{(1)}}\,,&&\widehat{Q}_{2,1}=Q_{v_2^{(1)}}Q_{h_1^{(1)}}\,,&&\widehat{Q}_{1,2}=Q_{v_1^{(2)}}Q_{h_2^{(2)}}\,,&&\widehat{Q}_{2,2}=Q_{v_2^{(2)}}Q_{h_1^{(2)}}\,.
\end{align}
We can analyse the weights of the coupling constants of this expression with respect to the surfaces $\{S_4,S_3,S_1,S_2\}$, as given in appendix~\ref{Sect:Intersect22}. Specifically, we have 
\begin{center}
\begin{tabular}{c||c|c|c|c|c|c}
{\bf coup.} & {\bf curve} $\mathcal{C}$ & {\bf area} & $\mathcal{C}\cdot K_4$ & $\mathcal{C}\cdot K_3$ & $\mathcal{C}\cdot K_1$ & $\mathcal{C}\cdot K_2$\\[2pt]\hline
&&&&&&\\[-12pt]
$g_1^{(1)}$ & \parbox{3cm}{$2 L_1 + L_3 + L_4$\\ 
$ -2 E_1- 2 E_2 + E_4$} & $m_1^{(1)} + h_1^{(1)} + h_2^{(2)} + v_1^{(1)} + h_2^{(1)}$ & $-1$ & $-1$ & $0$ & $2$\\[14pt]\hline
&&&&&&\\[-12pt]
$g_2^{(1)}$ & \parbox{3cm}{$L_1 + L_2 + 2 L_3$\\ 
$ -2 E_1 - 2 E_2 + E_3 $} & $m_2^{(1)} + h_1^{(2)} + h_2^{(1)} + v_2^{(1)} + h_1^{(1)}$ & $-1$ & $-1$ & $2$ & $0$\\[14pt]\hline
&&&&&&\\[-12pt]
$g_1^{(2)}$ & \parbox{3cm}{$L_1 + L_2 + 2 L_3$\\ 
$ -E_1 - 2 E_2 $} & $m_1^{(2)} + h_1^{(1)} + h_1^{(2)} + v_1^{(2)} + h_2^{(2)}$ & $2$ & $0$ & $-1$ & $-1$\\[14pt]\hline
&&&&&&\\[-12pt]
$g_2^{(2)}$ & \parbox{3cm}{$2 L_1 + L_3 + L_4$\\ 
$ -2 E_1 - E_2$} & $m_2^{(2)} + h_2^{(2)} + h_2^{(1)} + v_2^{(2)} + h_1^{(2)}$ & $0$ & $2$ & $-1$ & $-1$\\[14pt]
\end{tabular}
\end{center}
Upon designating $S_4$ and $S_1$ (related to $\widehat{a}_1^{(1)}$ and $\widehat{a}_1^{(2)}$ in \figref{Fig:Diagram22Shift1}) as the co-roots of a (non-affine) $U(2) \times U(2)$ group, we demand that the coupling constants carry no weights under them, in order to have a perturbative realisation of the latter in the form of a gauge group. However, in the present case, we see that it is not possible to realise the full group $U(2) \times U(2)$ at the perturbative level. If at all possible, the latter can therefore only be realised non-perturbatively, thus possibly pertaining to a LST. We leave further analysis of the latter to future work~\cite{ToAppear}.
%%%%%%%%%%%%%%%%%%%%%%%%%%%%%%%%%%%%%%%
%%%%%%%%%%%%%%%%%%%%%%%%%%%%%%%%%%%%%%%
\section{Conclusions}
In this work we have analysed (part of) the web of dualities of a class of gauge theories engineered from toric Calabi-Yau manifolds $X_{N,M}$, whose toric diagram is schematically shown in \figref{Fig:WebDiagramGeneric}. On the one hand side, it has been argued in \cite{Bastian:2017ary} that there are three regions in the K\"ahler cone of the Calabi-Yau manifolds $X_{N,M}$ that correspond to the weak coupling regions of three (in general) different quiver gauge theories. The gauge groups of the latter are given in (\ref{DualGaugeTheories}). On the other hand, it has been argued in \cite{Hohenegger:2016yuv} (and checked explicitly for a large class of cases at the level of the partition function in \cite{Bastian:2017ing}) that $X_{N,M}\sim X_{N',M'}$ if $NM=N'M'$ and $\text{gcd}(N,M)=\text{gcd}(N',M')$. Combining these two facts implies the existence of a large number of dual quiver gauge theories in 6 dimensions and below. For each of these theories, the instanton partition function can be computed explicitly as a specific expansion of the topological string partition function $Z_{N,M}$ of the Calabi-Yau manifold $X_{N,M}$. 

Furthermore, besides the dual manifolds $X_{N',M'}$ as mentioned above, the extended K\"ahler moduli space of $X_{N,M}$ contains yet other regions which represent new types of manifolds. Among these, there are some Calabi-Yau threefolds whose toric diagrams look very similar to those of $X_{N,M}$, except that (some of) their external legs are glued together after a cyclic shift $\delta$, as is schematically shown in \figref{Fig:WebDiagramGenericShift}. In this paper we have undertaken a first step towards analysing whether also these manifolds $X_{N,M}^{(\delta)}$ engineer (weakly coupled) supersymmetric gauge theories. Upon discussing in some detail the cases $(N,M)=(6,1)$ and $(N,M)=(4,1)$, we have found evidence that this is indeed the case. However, as showcased in the case of $X_{4,1}^{(\delta=2)}=X_{2,2}^{(\delta=1)}$, for some of these theories, part of the expected gauge group might not be realised perturbatively, but rather only appears in the strong coupling regime. The latter may be a  Little String Theory.

To analyse the particular cases $X_{N,M}^{(\delta)}$ mentioned above in more detail, requires to develop a geometric realisation of the gauge algebra, at the level of the web diagram. In this paper, we have used an observation specific to the web diagrams of $X_{2,2}^{(\delta=0)}$ and $X_{2,2}^{(\delta=1)}$. In the upcoming work~\cite{ToAppear} we will further extend this discussion and apply it to more general cases. Using these tools, a question which will be interesting to address in the future is to obtain a complete picture of the web of dualities for all low-energy gauge theories engineered by (manifolds dual to) $X_{N,M}$. It will also be interesting to analyse their strong coupling counterparts.

%%%%%%%%%%%%%%%%%%%%%%%%%%%%%%%%%%%%%%%
\section*{Acknowledgements}
We would like to thank J.~ Heckman, G.~Moore and C.~Koz\c{c}az for many useful discussions. S.H. would like to thank the organisers of the workshop `Superconformal Field Theories in 6 and Lower Dimensions' at the Tsinghua Sanya International Mathematics Forum (TSIMF) from January 15-19, 2018 for creating a stimulating atmosphere, during which part of this work was done. A.I. would like to thank the hospitality of the Simons Center for Geometry and Physics during the 2017 Summer Workshop in Mathematics and Physics. S.J.R. would like to thank the hospitality of the University of Pennsylvania during this work. 

\appendix
%%%%%%%%%%%%%%%%%%%%%%%%%%%%%%%%%%%%%%%
%%%%%%%%%%%%%%%%%%%%%%%%%%%%%%%%%%%%%%%
%%%%%%%%%%%%%%%%%%%%%%%%%%%%%%%%%%%%%%%
%%%%%%%%%%%%%%%%%%%%%%%%%%%%%%%%%%%%%%%
%%%%%%%%%%%%%%%%%%%%%%%%%%
%%%%%%%%%%%%%%%%%%%%%%%%%%
\section{Intersection Numbers for $(N,M)=(2,2)$}\label{Sect:Intersect22}
Many algebraic properties of the six-dimensional quiver gauge theories (and their five-dimensional limits \cite{Morrison:1996xf,Intriligator:1997pq}) discussed in the main body of this paper can directly be read off from the underlying web diagram. Indeed, as explained in section~\ref{Sect:RootsAndWeights}, for a given toric diagram, geometrically the roots $\widehat{a}_i$ (with $i=1,\ldots,MN$) of the (affine) algebra $\widehat{\mathfrak{g}}$ that is realised in the six-dimensional quiver gauge theory can be identified with certain curves of the geometry. Furthermore, the co-roots are given by compact surfaces $S_i$ that are associated with the various hexagons appearing in the web (see \cite{Esole:2015xfa}): The intersection numbers $S_i\circ \widehat{a}_j$ reproduce (up to a sign) the Cartan matrix of the (affine) algebra $\widehat{\mathfrak{g}}$. 

In this appendix, we observe that for $(N,M)=(2,2)$ (with $\delta=0$ or $\delta=1$), the co-roots can also be realised in terms of the canonical classes associated with the $S_i$. We remark that all intersections discussed in this appendix shall be intersections of curves within a compact divisor. This simplifies the computation of the weights associated with different curves in the toric web diagram (and thus the corresponding contributions in the partition functions of the associated quiver gauge theories) and is exploited in section~\ref{Sect:22ExplicitExample}.

\begin{figure}
\begin{center}
\scalebox{0.55}{\parbox{23.5cm}{\begin{tikzpicture}[scale = 1.50]
\node at (-1.2,-1.2) {\large  {\bf a}};
\node at (4.2,4.2) {\large  {\bf a}};
\draw[ultra thick] (-1,-1) -- (0,0);
\node at (-0.8,-0.2) {\large  {\bf $-E_2$}};
\draw[ultra thick] (0,0) -- (1,0) -- (1,1) -- (0,1) -- (0,0);
\node at (0.5,-0.3) {\large  {\bf $L_1$}};
\node at (0.5,1.3) {\large  {\bf $L_1$}};
\node at (1.3,0.5) {\large  {\bf $L_2$}};
\node at (-0.3,0.5) {\large  {\bf $L_2$}};
\draw[ultra thick] (1,0) -- (2,-1);
\draw[ultra thick] (0,1) -- (-1,2);
\node at (1.2,1.8) {\large  {\bf $-E_1$}};
\draw[ultra thick] (1,1) -- (2,2);
\draw[ultra thick] (2,2) -- (3,2) -- (3,3) -- (2,3) -- (2,2);
\draw[ultra thick] (3,2) -- (4,1);
\node at (2.5,1.7) {\large  {\bf $L_3$}};
\node at (2.5,3.3) {\large  {\bf $L_3$}};
\node at (3.3,2.5) {\large  {\bf $L_4$}};
\node at (1.7,2.5) {\large  {\bf $L_4$}};
\draw[ultra thick] (2,3) -- (1,4);
\draw[ultra thick] (3,3) -- (4,4);
\node at (3.2,3.8) {\large  {\bf $-E_2$}};
%%
%empties right
\node at (2.2,-1.2) {\large  {\bf 1}};
\node at (1.7,-0.3) {\large  {\bf $E_3$}};
\node at (4.2,0.8) {\large  {\bf 2}};
\node at (3.7,1.7) {\large  {\bf $E_4$}};
%empties left
\node at (-1.2,2.2) {\large  {\bf 2}};
\node at (-0.3,1.7) {\large  {\bf $E_4$}};
\node at (0.8,4.2) {\large  {\bf 1}};
\node at (1.7,3.7) {\large  {\bf $E_3$}};
%stamp
\node at (1.5,-2.5) {\Large  {\bf (a) $\delta=0$}};
%%%%%%%%%
%%%%%%%%%
%%%%%%%%%
\begin{scope}[xshift=10cm]
\node at (-1.2,-1.2) {\large  {\bf a}};
\node at (4.2,4.2) {\large  {\bf a}};
\draw[ultra thick] (-1,-1) -- (0,0);
\node at (-0.8,-0.2) {\large  {\bf $-E_2$}};
\draw[ultra thick] (0,0) -- (1,0) -- (1,1) -- (0,1) -- (0,0);
\node at (0.5,-0.3) {\large  {\bf $L_1$}};
\node at (0.5,1.3) {\large  {\bf $L_1$}};
\node at (1.3,0.5) {\large  {\bf $L_2$}};
\node at (-0.3,0.5) {\large  {\bf $L_2$}};
\draw[ultra thick] (1,0) -- (2,-1);
\draw[ultra thick] (0,1) -- (-1,2);
\node at (1.2,1.8) {\large  {\bf $-E_1$}};
\draw[ultra thick] (1,1) -- (2,2);
\draw[ultra thick] (2,2) -- (3,2) -- (3,3) -- (2,3) -- (2,2);
\draw[ultra thick] (3,2) -- (4,1);
\node at (2.5,1.7) {\large  {\bf $L_3$}};
\node at (2.5,3.3) {\large  {\bf $L_3$}};
\node at (3.3,2.5) {\large  {\bf $L_4$}};
\node at (1.7,2.5) {\large  {\bf $L_4$}};
\draw[ultra thick] (2,3) -- (1,4);
\draw[ultra thick] (3,3) -- (4,4);
\node at (3.2,3.8) {\large  {\bf $-E_2$}};
%%
%empties right
\node at (2.2,-1.2) {\large  {\bf 1}};
\node at (1.7,-0.3) {\large  {\bf $E_3$}};
\node at (4.2,0.8) {\large  {\bf 2}};
\node at (3.7,1.7) {\large  {\bf $E_4$}};
%empties left
\node at (-1.2,2.2) {\large  {\bf 1}};
\node at (-0.3,1.7) {\large  {\bf $E_3$}};
\node at (0.8,4.2) {\large  {\bf 2}};
\node at (1.7,3.7) {\large  {\bf $E_4$}};
%stamp
\node at (1.5,-2.5) {\Large  {\bf (b) $\delta=1$}};
\end{scope}
\end{tikzpicture}}}
\caption{\sl Gluing two local geometries of the type $\mathbb{P}^1\times \mathbb{P}^1$ related to the web diagram $(N,M)=(2,2)$ with $\delta=0$ (diagram (a)) and $\delta=1$ (diagram (b)).}
\label{Fig:22StartingP1P1}
\end{center}
\end{figure}
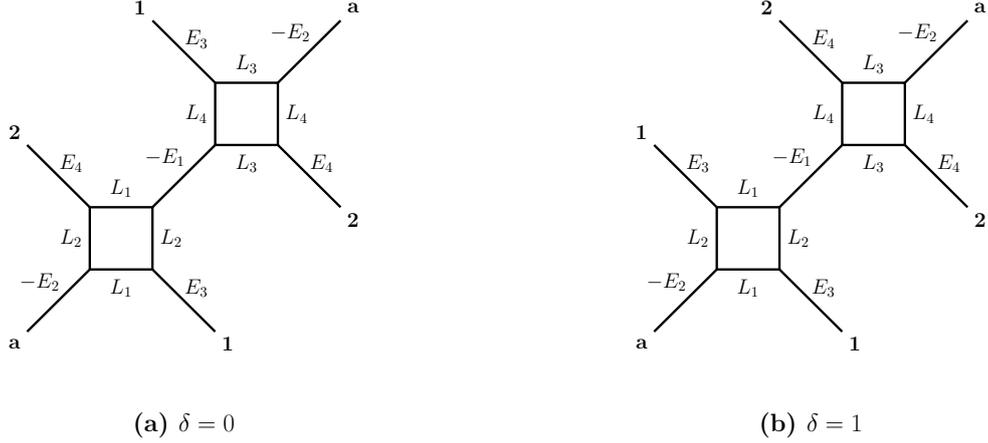

\begin{figure}
\begin{center}
\scalebox{0.55}{\parbox{23.5cm}{\begin{tikzpicture}[scale = 1.50]
\node at (-1.2,-1.2) {\large  {\bf a}};
\node at (-0.8,-0.2) {\large  {\bf $-E_2$}};
\node at (3.2,3.8) {\large  {\bf $-E_2$}};
\node at (4.2,4.2) {\large  {\bf a}};
\draw[ultra thick] (-1,-1) -- (0,0);
\draw[ultra thick] (0,0) -- (2,0) -- (2,1) -- (1,2) -- (0,2) -- (0,0);
\draw[ultra thick] (1,2) -- (1,3) -- (3,3) -- (3,1) -- (2,1);
\draw[ultra thick] (3,3) -- (4,4);
%
%empties right
\draw[ultra thick] (2,0) -- (3,-1);
\node at (3.2,-1.2) {\large  {\bf 1}};
\node at (2.4,-0.8) {\large  {\bf $E_3$}};
\draw[ultra thick] (3,1) -- (4,0);
\node at (4.2,-0.2) {\large  {\bf 2}};
\node at (3.8,0.6) {\large  {\bf $E_4$}};
%empties left
\draw[ultra thick] (0,2) -- (-1,3);
\node at (-1.2,3.2) {\large  {\bf 2}};
\node at (-0.6,2.2) {\large  {\bf $E_4$}};
\draw[ultra thick] (1,3) -- (0,4);
\node at (-0.2,4.2) {\large  {\bf 1}};
\node at (0.8,3.6) {\large  {\bf $E_3$}};
\node at (1,-0.3) {\large  {\bf $L_1$}};
\node at (-0.3,1) {\large  {\bf $L_2$}};
\node at (1.3,0.5) {\large  {\bf $L_2-E_1$}};
\node[rotate=270] at (0.5,1.3) {\large  {\bf $L_1-E_1$}};
\node at (1.3,1.3) {\large  {\bf $E_1$}};
\node[rotate=315] at (3,0.4) {\large  {\bf $L_3-E_1$}};
\node at (3.3,2) {\large  {\bf $L_4$}};
\node at (1.7,2.5) {\large  {\bf $L_4-E_1$}};
\node at (2,3.3) {\large  {\bf $L_3$}};
%cut
\draw[dashed,red, ultra thick] (2.4,0.5) -- (2.4,3.5);
%stamp
\node at (1.5,-2.5) {\Large  {\bf (a) $\delta=0$}};
%%%%%%%%
%%%%%%%%
%%%%%%%%
\begin{scope}[xshift=10cm]
\node at (-1.2,-1.2) {\large  {\bf a}};
\node at (-0.8,-0.2) {\large  {\bf $-E_2$}};
\node at (3.2,3.8) {\large  {\bf $-E_2$}};
\node at (4.2,4.2) {\large  {\bf a}};
\draw[ultra thick] (-1,-1) -- (0,0);
\draw[ultra thick] (0,0) -- (2,0) -- (2,1) -- (1,2) -- (0,2) -- (0,0);
\draw[ultra thick] (1,2) -- (1,3) -- (3,3) -- (3,1) -- (2,1);
\draw[ultra thick] (3,3) -- (4,4);
%
%empties right
\draw[ultra thick] (2,0) -- (3,-1);
\node at (3.2,-1.2) {\large  {\bf 1}};
\node at (2.4,-0.8) {\large  {\bf $E_3$}};
\draw[ultra thick] (3,1) -- (4,0);
\node at (4.2,-0.2) {\large  {\bf 2}};
\node at (3.8,0.6) {\large  {\bf $E_4$}};
%empties left
\draw[ultra thick] (0,2) -- (-1,3);
\node at (-1.2,3.2) {\large  {\bf 1}};
\node at (-0.6,2.2) {\large  {\bf $E_3$}};
\draw[ultra thick] (1,3) -- (0,4);
\node at (-0.2,4.2) {\large  {\bf 2}};
\node at (0.8,3.6) {\large  {\bf $E_4$}};
\node at (1,-0.3) {\large  {\bf $L_1$}};
\node at (-0.3,1) {\large  {\bf $L_2$}};
\node at (1.3,0.5) {\large  {\bf $L_2-E_1$}};
\node[rotate=270] at (0.5,1.3) {\large  {\bf $L_1-E_1$}};
\node at (1.3,1.3) {\large  {\bf $E_1$}};
\node[rotate=315] at (3,0.4) {\large  {\bf $L_3-E_1$}};
\node at (3.3,2) {\large  {\bf $L_4$}};
\node at (1.7,2.5) {\large  {\bf $L_4-E_1$}};
\node at (2,3.3) {\large  {\bf $L_3$}};
%cut
\draw[dashed,red, ultra thick] (2.4,0.5) -- (2.4,3.5);
%stamp
\node at (1.5,-2.5) {\Large  {\bf (b) $\delta=1$}};
\end{scope}
%
%%%%%%%%%
%%%%%%%%%
%%%%%%%%%
\end{tikzpicture}}}
\caption{\sl Geometry \figref{Fig:22StartingP1P1} after a flop transition of the curve $-E_1$ for $\delta=0$ (diagram (a)) and $\delta=1$ (diagram (b)).}
\label{Fig:22StartingP1P1Flop1}
\end{center}
\end{figure}
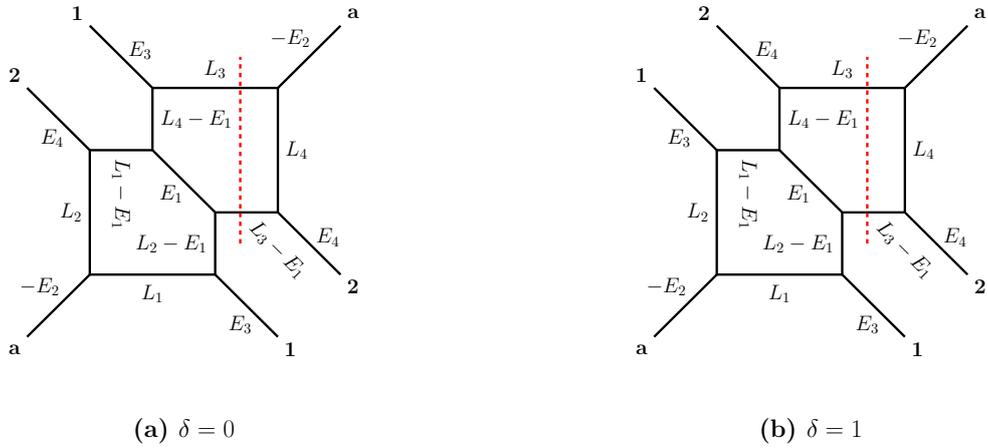

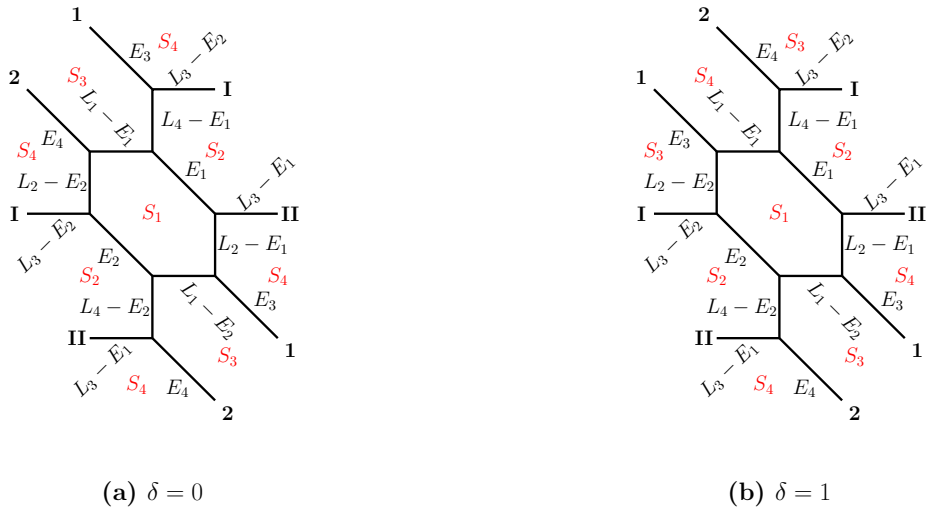
\begin{figure}
\begin{center}
\scalebox{0.55}{\parbox{22.5cm}{\begin{tikzpicture}[scale = 1.50]
\draw[ultra thick] (1,0) -- (1,-1) -- (2,-2);
\draw[ultra thick] (0,-1) -- (1,-1);
\draw[ultra thick] (-1,1) -- (0,1);
\draw[ultra thick] (1,0) -- (2,0) -- (2,1) -- (1,2) -- (0,2) -- (0,1) -- (1,0);
\draw[ultra thick] (1,2) -- (1,3) -- (2,3);
\draw[ultra thick] (2,1) -- (3,1);
%horizontal nodes
%
%
%empties right
\draw[ultra thick] (2,0) -- (3,-1);
\node at (3.2,-1.2) {\large   {\bf 1}};
\node at (2.2,-2.2) {\large   {\bf 2}};
%empties left
\draw[ultra thick] (0,2) -- (-1,3);
\node at (-1.2,3.2) {\large  {\bf 2}};
\draw[ultra thick] (1,3) -- (0,4);
\node at (-0.2,4.2) {\large  {\bf 1}};
\node at (0.3,0.3) {\large  {\bf $E_2$}};
\node[rotate=315] at (1.9,-0.55) {\large  {\bf $L_1-E_2$}};
\node at (0.4,-0.5) {\large  {\bf $L_{4}-E_2$}};
\node at (-0.6,1.5) {\large  {\bf $L_2-E_2$}};
\node[rotate=45] at (-0.7,0.5) {\large  {\bf $L_{3}-E_2$}};
\node at (2.6,0.5) {\large  {\bf $L_2-E_1$}};
\node[rotate=315] at (0.3,2.55) {\large  {\bf $L_1-E_1$}};
\node at (1.7,1.7) {\large  {\bf $E_1$}};
\node[rotate=45] at (0.2,-1.5) {\large  {\bf $L_{3}-E_{1}$}};
\node[rotate=45] at (2.8,1.5) {\large  {\bf $L_3-E_1$}};
\node at (1.7,2.5) {\large  {\bf $L_4-E_1$}};
\node[rotate=45] at (1.7,3.5) {\large  {\bf $L_3-E_2$}};
%horizontal labels
\node at (-1.2,1) {\large  {\bf I}};
\node at (-0.2,-1) {\large  {\bf II}};
\node at (2.2,3) {\large  {\bf I}};
\node at (3.2,1) {\large  {\bf II}};
%diagonal curves
\node at (2.8,-0.4) {\large  {\bf $E_3$}};
\node at (1.4,-1.8) {\large  {\bf $E_4$}};
\node at (0.8,3.6) {\large  {\bf $E_3$}};
\node at (-0.6,2.2) {\large  {\bf $E_4$}};
%hexagons
\node[red] at (0,0) {\large  {\bf $S_2$}};
\node[red] at (1,1) {\large  {\bf $S_1$}};
\node[red] at (2,2) {\large  {\bf $S_2$}};
\node[red] at (-0.2,3.2) {\large  {\bf $S_3$}};
\node[red] at (-1,2) {\large  {\bf $S_4$}};
\node[red] at (2.2,-1.3) {\large  {\bf $S_3$}};
\node[red] at (3,-0) {\large  {\bf $S_4$}};
\node[red] at (1.25,3.75) {\large  {\bf $S_4$}};
\node[red] at (0.75,-1.75) {\large  {\bf $S_4$}};
%stamp
\node at (1,-3.5) {\Large  {\bf (a) $\delta=0$}};
%%%%%%
%%%%%%
%%%%%%
\begin{scope}[xshift=10cm]
\draw[ultra thick] (1,0) -- (1,-1) -- (2,-2);
\draw[ultra thick] (0,-1) -- (1,-1);
\draw[ultra thick] (-1,1) -- (0,1);
\draw[ultra thick] (1,0) -- (2,0) -- (2,1) -- (1,2) -- (0,2) -- (0,1) -- (1,0);
\draw[ultra thick] (1,2) -- (1,3) -- (2,3);
\draw[ultra thick] (2,1) -- (3,1);
%horizontal nodes
%
%
%empties right
\draw[ultra thick] (2,0) -- (3,-1);
\node at (3.2,-1.2) {\large   {\bf 1}};
\node at (2.2,-2.2) {\large   {\bf 2}};
%empties left
\draw[ultra thick] (0,2) -- (-1,3);
\node at (-1.2,3.2) {\large  {\bf 1}};
\draw[ultra thick] (1,3) -- (0,4);
\node at (-0.2,4.2) {\large  {\bf 2}};
\node at (0.3,0.3) {\large  {\bf $E_2$}};
\node[rotate=315] at (1.9,-0.55) {\large  {\bf $L_1-E_2$}};
\node at (0.4,-0.5) {\large  {\bf $L_{4}-E_2$}};
\node at (-0.6,1.5) {\large  {\bf $L_2-E_2$}};
\node[rotate=45] at (-0.7,0.5) {\large  {\bf $L_{3}-E_2$}};
\node at (2.6,0.5) {\large  {\bf $L_2-E_1$}};
\node[rotate=315] at (0.3,2.55) {\large  {\bf $L_1-E_1$}};
\node at (1.7,1.7) {\large  {\bf $E_1$}};
\node[rotate=45] at (0.2,-1.5) {\large  {\bf $L_{3}-E_{1}$}};
\node[rotate=45] at (2.8,1.5) {\large  {\bf $L_3-E_1$}};
\node at (1.7,2.5) {\large  {\bf $L_4-E_1$}};
\node[rotate=45] at (1.7,3.5) {\large  {\bf $L_3-E_2$}};
%horizontal labels
\node at (-1.2,1) {\large  {\bf I}};
\node at (-0.2,-1) {\large  {\bf II}};
\node at (2.2,3) {\large  {\bf I}};
\node at (3.2,1) {\large  {\bf II}};
%diagonal curves
\node at (2.8,-0.4) {\large  {\bf $E_3$}};
\node at (1.4,-1.8) {\large  {\bf $E_4$}};
\node at (0.8,3.6) {\large  {\bf $E_4$}};
\node at (-0.6,2.2) {\large  {\bf $E_3$}};
%hexagons
\node[red] at (0,0) {\large  {\bf $S_2$}};
\node[red] at (1,1) {\large  {\bf $S_1$}};
\node[red] at (2,2) {\large  {\bf $S_2$}};
\node[red] at (-0.2,3.2) {\large  {\bf $S_4$}};
\node[red] at (-1,2) {\large  {\bf $S_3$}};
\node[red] at (2.2,-1.3) {\large  {\bf $S_3$}};
\node[red] at (3,-0) {\large  {\bf $S_4$}};
\node[red] at (1.25,3.75) {\large  {\bf $S_3$}};
\node[red] at (0.75,-1.75) {\large  {\bf $S_4$}};
%stamp
\node at (1,-3.5) {\Large  {\bf (b) $\delta=1$}};
\end{scope}
\end{tikzpicture}}}
\caption{\sl The geometry shown in \figref{Fig:22StartingP1P1Flop1} after a flop transition of the curve $-E_{1}$ for $\delta=0$ (diagram (a)) and $\delta=1$ (diagram (b)). The curves $S_{1,2,3,4}$ are explained in the text.}
\label{Fig:22StartingP1P1Flop2}
\end{center}
\end{figure}

\begin{figure}
\begin{center}
\scalebox{0.55}{\parbox{25.5cm}{\begin{tikzpicture}[scale = 1.50]
\draw[ultra thick] (-1,0) -- (0,0) -- (1,1) -- (2,1) -- (3,2) -- (4,2);
\draw[ultra thick] (0,2) -- (1,2) -- (2,3) -- (3,3) -- (4,4) -- (5,4);
\draw[ultra thick] (0,-1) -- (0,0);
\draw[ultra thick] (2,0) -- (2,1);
\draw[ultra thick] (1,1) -- (1,2);
\draw[ultra thick] (3,2) -- (3,3);
\draw[ultra thick] (2,3) -- (2,4);
\draw[ultra thick] (4,4) -- (4,5);
%nodes bottom top
\node at (0,-1.2) {\large  {\bf a}};
\node at (2,-0.2) {\large  {\bf b}};
\node at (2,4.2) {\large  {\bf a}};
\node at (4,5.2) {\large  {\bf b}};
%nodes left right
\node at (-1.2,0) {\large  {\bf 2}};
\node at (-0.2,2) {\large  {\bf 1}};
\node at (4.2,2) {\large  {\bf 2}};
\node at (5.2,4) {\large  {\bf 1}};
%horizontal curves
\node at (-0.5,0.3) {\large  {\bf $E_4$}};
\node at (1.5,1.3) {\large  {\bf $E_1$}};
\node at (3.5,2.3) {\large  {\bf $E_4$}};
\node at (0.5,2.3) {\large  {\bf $E_3$}};
\node at (2.5,3.3) {\large  {\bf $E_2$}};
\node at (4.5,4.3) {\large  {\bf $E_3$}};
%diagonal curves
\node[rotate=45] at (0.6,0.3) {\large  {\bf $L_1-E_1$}};
\node[rotate=45] at (2.6,1.3) {\large  {\bf $L_3-E_1$}};
\node[rotate=45] at (1.6,2.3) {\large  {\bf $L_3-E_2$}};
\node[rotate=45] at (3.6,3.3) {\large  {\bf $L_1-E_2$}};
%vertical curves
\node at (-0.65,-0.8) {\large  {\bf $L_2-E_2$}};
\node at (2.65,0.3) {\large  {\bf $L_2-E_1$}};
\node at (0.35,1.5) {\large  {\bf $L_4-E_1$}};
\node[rotate=45] at (2.45,2.2) {\large  {\bf $L_4-E_2$}};
\node at (1.35,3.5) {\large  {\bf $L_2-E_2$}};
\node[rotate=45] at (3.45,4.3) {\large  {\bf $L_2-E_1$}};
%hexagons
\node[red] at (-0.4,-0.4) {\large  {\bf $S_4$}};
\node[red] at (1,0) {\large  {\bf $S_1$}};
\node[red] at (3,1) {\large  {\bf $S_4$}};
\node[red] at (1.8,1.8) {\large  {\bf $S_2$}};
\node[red] at (4,3) {\large  {\bf $S_3$}};
\node[red] at (0,1) {\large  {\bf $S_3$}};
\node[red] at (1,3) {\large  {\bf $S_4$}};
\node[red] at (2.8,4) {\large  {\bf $S_1$}};
\node[red] at (4.7,4.7) {\large  {\bf $S_4$}};
%stamp
\node at (2,-2.5) {\Large  {\bf (a) $\delta=0$}};
%%%%%%
%%%%%%
%%%%%%
\begin{scope}[xshift=10cm]
\draw[ultra thick] (-1,0) -- (0,0) -- (1,1) -- (2,1) -- (3,2) -- (4,2);
\draw[ultra thick] (0,2) -- (1,2) -- (2,3) -- (3,3) -- (4,4) -- (5,4);
\draw[ultra thick] (0,-1) -- (0,0);
\draw[ultra thick] (2,0) -- (2,1);
\draw[ultra thick] (1,1) -- (1,2);
\draw[ultra thick] (3,2) -- (3,3);
\draw[ultra thick] (2,3) -- (2,4);
\draw[ultra thick] (4,4) -- (4,5);
%nodes bottom top
\node at (0,-1.2) {\large  {\bf a}};
\node at (2,-0.2) {\large  {\bf b}};
\node at (2,4.2) {\large  {\bf a}};
\node at (4,5.2) {\large  {\bf b}};
%nodes left right
\node at (-1.2,0) {\large  {\bf 1}};
\node at (-0.2,2) {\large  {\bf 2}};
\node at (4.2,2) {\large  {\bf 2}};
\node at (5.2,4) {\large  {\bf 1}};
%horizontal curves
\node at (-0.5,0.3) {\large  {\bf $E_3$}};
\node at (1.5,1.3) {\large  {\bf $E_1$}};
\node at (3.5,2.3) {\large  {\bf $E_4$}};
\node at (0.5,2.3) {\large  {\bf $E_4$}};
\node at (2.5,3.3) {\large  {\bf $E_2$}};
\node at (4.5,4.3) {\large  {\bf $E_3$}};
%diagonal curves
\node[rotate=45] at (0.6,0.3) {\large  {\bf $L_1-E_1$}};
\node[rotate=45] at (2.6,1.3) {\large  {\bf $L_3-E_1$}};
\node[rotate=45] at (1.6,2.3) {\large  {\bf $L_3-E_2$}};
\node[rotate=45] at (3.6,3.3) {\large  {\bf $L_1-E_2$}};
%vertical curves
\node at (-0.65,-0.8) {\large  {\bf $L_2-E_2$}};
\node at (2.65,0.3) {\large  {\bf $L_2-E_1$}};
\node at (0.35,1.5) {\large  {\bf $L_4-E_1$}};
\node[rotate=45] at (2.45,2.2) {\large  {\bf $L_4-E_2$}};
\node at (1.35,3.5) {\large  {\bf $L_2-E_2$}};
\node[rotate=45] at (3.45,4.3) {\large  {\bf $L_2-E_1$}};
%hexagons
\node[red] at (-0.4,-0.4) {\large  {\bf $S_4$}};
\node[red] at (1,0) {\large  {\bf $S_1$}};
\node[red] at (3,1) {\large  {\bf $S_4$}};
\node[red] at (1.8,1.8) {\large  {\bf $S_2$}};
\node[red] at (4,3) {\large  {\bf $S_3$}};
\node[red] at (0,1) {\large  {\bf $S_4$}};
\node[red] at (1,3) {\large  {\bf $S_3$}};
\node[red] at (2.8,4) {\large  {\bf $S_1$}};
\node[red] at (4.7,4.7) {\large  {\bf $S_4$}};
%stamp
\node at (2,-2.5) {\Large  {\bf (b) $\delta=1$}};
\end{scope}
\end{tikzpicture}}}
\caption{\sl Alternative presentation of the geometry in \figref{Fig:22StartingP1P1Flop2} in the form of a (twisted) $(2,2)$ web with $\delta=0$ (diagram (a)) and $\delta=1$ (diagram (b)). }
\label{Fig:22StartingP1P1Flop3}
\end{center}
\end{figure}
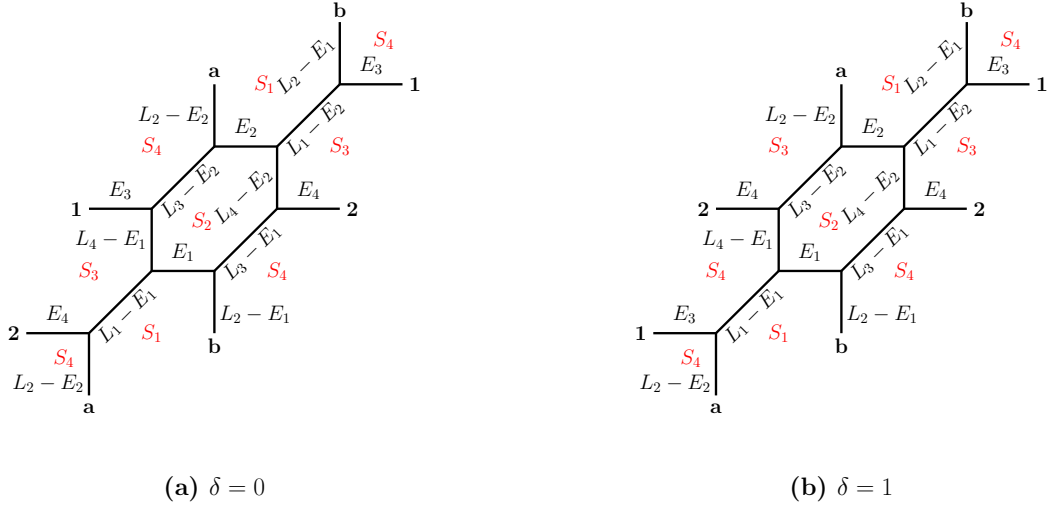

The starting point is the geometry shown in~\figref{Fig:22StartingP1P1} (see \cite{Shabbir:2017kuv} for a five-dimensional limit), where we presented the cases $\delta=0$ and $\delta=1$ in parallel. The labelling of the curves has been chosen in such a way as to reflect the local geometry of $\mathbb{P}^1\times \mathbb{P}^1$, glued together along the exceptional cuves $-E_{1,2}$ and $E_{3,4}$, respectively. Notice that the difference in $\delta$ is entirely related to the latter. 
To related the geometry~\figref{Fig:22StartingP1P1} to the usual toric diagram with $(N,M)=(2,2)$ as shown in \figref{Fig:WebDiagramGenericShift}, we first perform a flop transformation on the curve $-E_1$, to obtain the diagrams shown in~\figref{Fig:22StartingP1P1Flop1}.
Next, we cut both diagrams along the red line and re-glue them along the curve labelled $-E_2$. After a flop transition of the latter, we obtain the geometry shown in \figref{Fig:22StartingP1P1Flop2}.
Finally, after an $SL(2,\mathbb{Z})$ transformation, we can bring both geometries into the form of a $(2,2)$ web with shifts $\delta=0$ and $\delta=1$, respectively, as shown in \figref{Fig:22StartingP1P1Flop3}.
Each of the two geometries contains four divisors (hexagons in the toric web) $S_{1,2,3,4}$ with canonical classes $K_{S_{1,2,3,4}}$. The area of canonical class of $S_i$ is given by $A_i=\int_{K_{S_i}} \omega$, where $\omega$ is the K\"ahler form. The areas $A_i$ are not all independent of one another but satisfy
\begin{align}
\sum_{i=1}^4 A_i=\sum_{i=1}^4 \int_{K_{S_i}} \omega =0\,.
\end{align}
Furthermore, within every $S_i$ we can identify three different curves $\widehat{a}_i$, $\widehat{b}_i$ and $\widehat{c}_i$ (for $i=1,2,3,4$) as shown schematically in \figref{Fig:22Roots} for a generic hexagon of a toric web diagram. From the perspective of the quiver gauge theories (or the Little String Theories) engineered by the web diagram, these curves are related to the positive simple roots of the (affine extensions of the) underlying gauge algebra: To be precise, the curves $\widehat{a}_i$ are associated with the roots of $\widehat{\mathfrak{g}}_{\text{hor}}$, the curves $\widehat{b}_i$ are associated with the roots of $\widehat{\mathfrak{g}}_{\text{vert}}$ and the curves $\widehat{c}_i$ are associated with the roots of $\widehat{\mathfrak{g}}_{\text{diag}}$, corresponding to the horizontal, vertical and diagonal gauge theories, respectively.

Concretely, the explicit expressions (for $\delta=0$ and $\delta=1$) for $K_{S_i}$ as well as $\widehat{a}_i$, $\widehat{b}_i$ and $\widehat{c}_i$ can be summarised in the following table
\begin{center}
\begin{tabular}{c|c}
$\mathbf{\delta=0}$ & $\mathbf{\delta=1}$\\[4pt]\hline\hline
&\\[-32pt]
\parbox{7cm}{\begin{align}K_{S_1}&=-(2L_1+2L_2-E_1-E_2+E_3+E_4)\,,\nonumber\\
K_{S_2}&=-(2L_3+2L_4-E_1-E_2+E_3+E_4)\,,\nonumber\\
K_{S_3}&=2L_1+2L_4-E_1-E_2+E_3+E_4\,,\nonumber\\
K_{S_4}&=2L_2+2L_3-E_1-E_2+E_3+E_4\,.\nonumber\end{align}} 
&
\parbox{7cm}{\begin{align}K_{S_1}&=-(2L_1+2L_2-E_1-E_2+2E_3)\,,\nonumber\\
K_{S_2}&=-(2L_3+2L_4-E_1-E_2+2E_4)\,,\nonumber\\
K_{S_3}&=L_1+L_2+L_3+L_4-2E_2+E_3+E_4\,,\nonumber\\
K_{S_4}&=L_1+L_2+L_3+L_4-2E_1+E_3+E_4\,.\nonumber\end{align}}
\\[-8pt]\hline
%%%%%%%%
&\\[-32pt]
\parbox{7cm}{\begin{align}
&\widehat{a}_1=L_1\,,&&\widehat{a}_2=L_3\,,\nonumber\\
&\widehat{a}_3=L_1-E_1+E_4\,,&&\widehat{a}_4=L_3-E_1+E_4\,.\nonumber
\end{align}}
&
\parbox{7cm}{\begin{align}
&\widehat{a}_1=L_1\,,&&\widehat{a}_2=L_3\,,\nonumber\\
&\widehat{a}_3=L_3-E_1+E_4\,,&&\widehat{a}_4=L_4-E_1+E_4\,.\nonumber
\end{align}}
\\[-8pt]\hline
%%%%%%%%
&\\[-32pt]
\parbox{7cm}{\begin{align}
&\widehat{b}_1=L_1+L_2-E_1-E_2\,,\nonumber\\
&\widehat{b}_2=L_3+L_4-E_1-E_2\,,\nonumber\\
&\widehat{b}_3=L_1+L_4-2E_2\,,\nonumber\\
&\widehat{b}_4=L_2+L_3-2E_1\,.\nonumber
\end{align}}
&
\parbox{7cm}{\begin{align}
&\widehat{b}_1=L_1+L_2-E_1-E_2\,,\nonumber\\
&\widehat{b}_2=L_3+L_4-E_1-E_2\,,\nonumber\\
&\widehat{b}_3=L_1+L_4-2E_1\,,\nonumber\\
&\widehat{b}_4=L_2+L_3-2E_2\,.\nonumber
\end{align}}
\\[-8pt]\hline
%%%%%%%%
&\\[-32pt]
\parbox{7cm}{\begin{align}
&\widehat{c}_1=L_2\,,&&\widehat{c}_2=L_4\,,\nonumber\\
&\widehat{c}_3=L_4-E_2+E_4\,,&&\widehat{c}_4=L_2-E_1+E_3\,.\nonumber
\end{align}}
&
\parbox{7cm}{\begin{align}
&\widehat{c}_1=L_2\,,&&\widehat{c}_2=L_4\,,\nonumber\\
&\widehat{c}_3=L_2-E_2+E_3\,,&&\widehat{c}_4=L_4-E_1+E_4\,.\nonumber
\end{align}}
\end{tabular}
\end{center}
Notice here that the canonical classes $K_{S_i}$ are chosen in such a manner that the areas $A(S_i)$ are invariant under flop transformations of the curves $E_{1,2}$. Using the intersection numbers\footnote{The intersections in eq.~(\ref{InterSectP1P1}) are calculated in $\mathbb{P}^1\times \mathbb{P}^1$ blown up at two points.}
{\allowdisplaybreaks
\begin{align}
&L_1\cdot L_1=L_2\cdot L_2=L_3\cdot L_3=L_4\cdot L_4\,,\nonumber\\
&L_1\cdot L_2=1\,,\hspace{1cm} L_3\cdot L_4=1\,,\hspace{1cm}L_1\cdot L_3=L_1\cdot L_4=L_2\cdot L_3=L_2\cdot L_4\,,\nonumber \\
&E_i\cdot E_j=-\delta_{ij}\,,\hspace{1cm} L_i\cdot E_j=0\hspace{2cm}\forall i,j=1,2,3,4\,,\label{InterSectP1P1}
\end{align}}
we can recover (up to an overall sign) the Cartan matrices of Lie algebras of the horizontal, vertical and diagonal gauge theory, respectively. This can be summarised in the following table 
   \begin{center}
\begin{tabular}{c|c||c|c}
\multicolumn{2}{c||}{$\mathbf{\delta=0}$} & \multicolumn{2}{c}{$\mathbf{\delta=1}$} \\\hline\hline
{\bf intersections}  & {\bf Lie algebra} & {\bf intersections} & {\bf Lie algebra} \\\hline\hline
&\\[-28pt]
\parbox{3cm}{\begin{align}&\left(\begin{array}{c}\widehat{a}_1 \\ \widehat{a}_2 \\ \widehat{a}_3 \\ \widehat{a}_4\end{array}\right)\cdot \left(K_{S_1} \, K_{S_4} \, K_{S_2} \, K_{S_3}\right)\nonumber\\
&=\left(\begin{array}{cccc}-2 & 2 & 0 & 0 \\  2 & -2 & 0 & 0 \\ 0 & 0 & -2 & 2 \\ 0 & 0 & 2 & -2\end{array}\right)\nonumber\end{align}} 
& $\widehat{\mathfrak{a}}_1\oplus \widehat{\mathfrak{a}}_1$ 
&
\parbox{3cm}{\begin{align}&\left(\begin{array}{c}\widehat{a}_1 \\ \widehat{a}_2 \\ \widehat{a}_3 \\ \widehat{a}_4\end{array}\right)\cdot \left(K_{S_1} \, K_{S_4} \, K_{S_2} \, K_{S_3}\right)\nonumber\\
&=\left(\begin{array}{cccc}-2 & 1 & 0 & 1 \\  1 & -2 & 1 & 0 \\ 0 & 1 & -2 & 1 \\ 1 & 0 & 1 & -2\end{array}\right)\nonumber\end{align}} 
& $\widehat{\mathfrak{a}}_3$
\\[-8pt]\hline
%%%%%%%%
&\\[-28pt]
\parbox{3cm}{\begin{align}&\left(\begin{array}{c}\widehat{b}_1 \\ \widehat{b}_2 \\ \widehat{b}_4 \\ \widehat{b}_3\end{array}\right)\cdot \left(K_{S_1} \, K_{S_2} \, K_{S_4} \, K_{S_3}\right)\nonumber\\
&=\left(\begin{array}{cccc}-2 & 2 & 0 & 0 \\  2 & -2 & 0 & 0 \\ 0 & 0 & -2 & 2 \\ 0 & 0 & 2 & -2\end{array}\right)\nonumber\end{align}} 
& $\widehat{\mathfrak{a}}_1\oplus \widehat{\mathfrak{a}}_1$ 
&
\parbox{3cm}{\begin{align}&\left(\begin{array}{c}\widehat{b}_1 \\ \widehat{b}_2 \\ \widehat{b}_4 \\ \widehat{b}_3\end{array}\right)\cdot \left(K_{S_1} \, K_{S_2} \, K_{S_4} \, K_{S_3}\right)\nonumber\\
&=\left(\begin{array}{cccc}-2 & 2 & 0 & 0 \\  2 & -2 & 0 & 0 \\ 0 & 0 & -2 & 2 \\ 0 & 0 & 2 & -2\end{array}\right)\nonumber\end{align}} 
& $\widehat{\mathfrak{a}}_1\oplus \widehat{\mathfrak{a}}_1$
\\[-8pt]\hline
%%%%%%%%
&\\[-28pt]
\parbox{3cm}{\begin{align}&\left(\begin{array}{c}\widehat{c}_1 \\ \widehat{c}_3 \\ \widehat{c}_2 \\ \widehat{c}_4\end{array}\right)\cdot \left(K_{S_1} \, K_{S_3} \, K_{S_2} \, K_{S_4}\right)\nonumber\\
&=\left(\begin{array}{cccc}-2 & 2 & 0 & 0 \\  2 & -2 & 0 & 0 \\ 0 & 0 & -2 & 2 \\ 0 & 0 & 2 & -2\end{array}\right)\nonumber\end{align}} 
& $\widehat{\mathfrak{a}}_1\oplus \widehat{\mathfrak{a}}_1$ 
&
\parbox{3cm}{\begin{align}&\left(\begin{array}{c}\widehat{c}_1 \\ \widehat{c}_3 \\ \widehat{c}_2 \\ \widehat{c}_4\end{array}\right)\cdot \left(K_{S_1} \, K_{S_3} \, K_{S_2} \, K_{S_4}\right)\nonumber\\
&=\left(\begin{array}{cccc}-2 & 1 & 0 & 1 \\  1 & -2 & 1 & 0 \\ 0 & 1 & -2 & 1 \\ 1 & 0 & 1 & -2\end{array}\right)\nonumber\end{align}} 
& $\widehat{\mathfrak{a}}_3$
\end{tabular}
\end{center}
Notice that the same result can also be obtained from computing the intersection numbers with the surfaces $S_i$. However, using the canonical classes for the computation is in practice easier and will be used for all computations in the main body of the paper.
%%%%%%%%%%%%%%%%%%%%%%%%%%
%%%%%%%%%%%%%%%%%%%%%%%%%%%%%%%%%%%%%%%%%%%%%%%%%%%%
\section{Nekrasov Subfunctions} 
\label{NekSub}
The Nekrasov subfunctions capture the contribution of the various multiplets to the instanton partition function of the four-dimensional \cite{Nekrasov:2002qd}, five-dimensional \cite{Nekrasov:2003rj} or the six-dimensional gauge theories \cite{Iqbal:2015fvd,Nieri:2015dts,Hayling:2017}. For a review of these see \cite{Mironov:2009qt, Tachikawa:2014dja}. The six-dimensional instanton partition function we can identify the different contribution from the tensor or matter multiplets. This correspondence relates the tensor branch parameters and hypermultiplet masses from the gauge theory side to linear combination of K\"ahler parameters on the geometry side. The first contribution that will be of interest to us comes from the vector multiplet 
\begin{align} 
&z^{\text{vector}}(\vec{a},\vec{\alpha})=\prod_{i,j=1}^N \vartheta_{\alpha_i \alpha_j}^{-1}(e^{a_i-a_j+\frac{1}{2}\epsilon_{+}};\rho)\,,&&\text{with} &&\epsilon_+=\epsilon_1+\epsilon_2\,,
\label{vect}
\end{align}
where $\vec{\alpha}=(\alpha_1,\dots,\alpha_N)$ corresponds to a vector of integer partitions, $\vec{a}=(a_1,\dots,a_N)$ are the vacuum expectation values of the scalar field in the five-dimensional vector multiplet and
\bea\nonumber
\vartheta_{\alpha_{a}\alpha_{b}}(x;\rho)&=&\prod_{(i,j)\in\alpha_{a}}\vartheta\Big(x\,q^{\alpha_{a,i}-j+\frac{1}{2}}\,t^{\alpha_{b,j}^{t}-i+\frac{1}{2}};\rho\Big)
\prod_{(i,j)\in \alpha_{b}}\vartheta(x\,q^{-\alpha_{b,i}+j-\frac{1}{2}}\,t^{-\alpha_{a,j}^{t}+i-\frac{1}{2}};\rho\Big)\,,\\
\vartheta(x;\rho)&=&\Big(x^{\frac{1}{2}}-x^{-\frac{1}{2}}\Big)\prod_{k=1}^{\infty}\Big(1-x\,e^{2\pi i k\,\rho}\Big)\Big(1-x^{-1}e^{2\pi i k\rho}\Big)\,.\label{DefCurlTheta}
\eea   Another contribution that appears in our partition functions comes from bi-fundamental matter 
\begin{align}
z^{\text{bifund}}(\vec{a},\vec{\alpha},\vec{b},\vec{\beta},m)=\prod_{i,j=1}^N \vartheta_{\alpha_i \beta_j}(e^{a_i-b_j-m+\frac{1}{2}\epsilon_+};\rho) \,.  \label{bifund} 
\end{align}
Here, $a_i$ and $b_j$ are the Coulomb branch parameters corresponding to the two gauge groups the bi-fundamental matter is coupled to. The $m$ corresponds to the mass parameter. The adjoint matter contribution corresponds to a special case of (\ref{bifund})
\begin{align}
z^{\text{adj}}(\vec{a},\vec{\alpha},\vec{a},\vec{\alpha},m)=\prod_{i,j=1}^N \vartheta_{\alpha_i \alpha_j}(e^{a_i-a_j-m+\frac{1}{2}\epsilon_+};\rho) \,.
\label{adj}
\end{align}
The interpretation of (\ref{vect}), (\ref{bifund}) and (\ref{adj}) as specific gauge theory contributions is justified by the fact that when we take the five-dimensional limit ($\rho \to i \infty$), we recover the well known respective contributions obtained from a localization calculation on the gauge theory side.
%%%%%%%%%%%%%%%%%%%%
\section{Duality Transformations}\label{App:ShiftDuality}
In this appendix, we discuss a series of flop- and symmetry transformations which relate the toric Calabi-Yau manifolds associated to the web-diagrams in \figref{Fig:WebDiagramGenericShift}  with different values of $\delta$. For concreteness, we shall denote this transformation $\mathcal{F}$, which maps a shifted Calabi-Yau manifold $X_{N,M}^{(\delta)}$ to a similar Calabi-Yau manifold that only differs in its shift parameter 
\begin{align}
\mathcal{F}:\,X_{N,M}^{(\delta)}\longrightarrow X_{N,M}^{(\delta+s)}\,,&&\text{with} &&s=M-N\text{ mod }N\,.\label{FlopShift}
\end{align}
Rather than discussing the most general case (for which the corresponding web diagrams are cumbersome to draw), we exemplify the transformation for $(N,M)=(6,4)$ (but generic $\delta$) and indicate how the results can be generalised. 

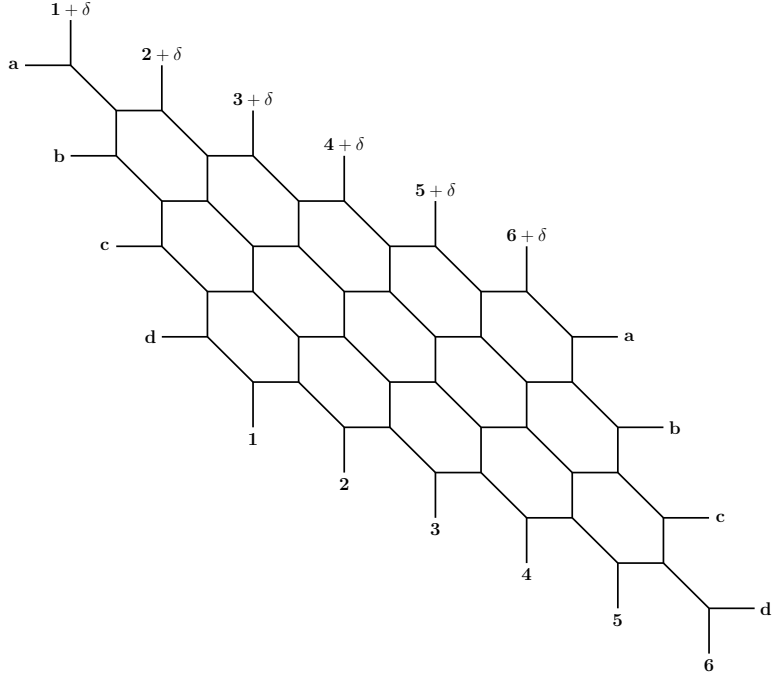
\begin{figure}
\begin{center}
\scalebox{0.4}{\parbox{25.5cm}{\begin{tikzpicture}[scale = 1.50]
\draw[ultra thick] (-1,0) -- (0,0) -- (1,-1) -- (2,-1) -- (3,-2) -- (4,-2) -- (5,-3) -- (6,-3) -- (7,-4) -- (8,-4) -- (9,-5) -- (10,-5) -- (11,-6) -- (12,-6);
\draw[ultra thick] (0,-2) -- (1,-2) -- (2,-3) -- (3,-3) -- (4,-4) -- (5,-4) -- (6,-5) -- (7,-5) -- (8,-6) -- (9,-6) -- (10,-7) -- (11,-7) -- (12,-8) -- (13,-8);
\draw[ultra thick] (1,-4) -- (2,-4) -- (3,-5) -- (4,-5) -- (5,-6) -- (6,-6) -- (7,-7) -- (8,-7) -- (9,-8) -- (10,-8) -- (11,-9) -- (12,-9) -- (13,-10) -- (14,-10);
\draw[ultra thick] (2,-6) -- (3,-6) -- (4,-7) -- (5,-7) -- (6,-8) -- (7,-8) -- (8,-9) -- (9,-9) -- (10,-10) -- (11,-10) -- (12,-11) -- (13,-11) -- (14,-12) -- (15,-12);
%top line
\draw[ultra thick] (0,1) -- (0,0);
\draw[ultra thick] (2,-1) -- (2,0);
\draw[ultra thick] (4,-2) -- (4,-1);
\draw[ultra thick] (6,-3) -- (6,-2);
\draw[ultra thick] (8,-4) -- (8,-3);
\draw[ultra thick] (10,-5) -- (10,-4);
%first line
\draw[ultra thick] (1,-1) -- (1,-2);
\draw[ultra thick] (3,-2) -- (3,-3);
\draw[ultra thick] (5,-3) -- (5,-4);
\draw[ultra thick] (7,-4) -- (7,-5);
\draw[ultra thick] (9,-5) -- (9,-6);
\draw[ultra thick] (11,-6) -- (11,-7);
%second line
\draw[ultra thick] (2,-3) -- (2,-4);
\draw[ultra thick] (4,-4) -- (4,-5);
\draw[ultra thick] (6,-5) -- (6,-6);
\draw[ultra thick] (8,-6) -- (8,-7);
\draw[ultra thick] (10,-7) -- (10,-8);
\draw[ultra thick] (12,-8) -- (12,-9);
%third line
\draw[ultra thick] (3,-5) -- (3,-6);
\draw[ultra thick] (5,-6) -- (5,-7);
\draw[ultra thick] (7,-7) -- (7,-8);
\draw[ultra thick] (9,-8) -- (9,-9);
\draw[ultra thick] (11,-9) -- (11,-10);
\draw[ultra thick] (13,-10) -- (13,-11);
%bottom line
\draw[ultra thick] (4,-7) -- (4,-8);
\draw[ultra thick] (6,-8) -- (6,-9);
\draw[ultra thick] (8,-9) -- (8,-10);
\draw[ultra thick] (10,-10) -- (10,-11);
\draw[ultra thick] (12,-11) -- (12,-12);
\draw[ultra thick] (14,-12) -- (14,-13);
%top labels
\node at (0,1.25) {\Large $\mathbf{1+\delta}$};
\node at (2,0.25) {\Large $\mathbf{2+\delta}$};
\node at (4,-0.75) {\Large $\mathbf{3+\delta}$};
\node at (6,-1.75) {\Large $\mathbf{4+\delta}$};
\node at (8,-2.75) {\Large $\mathbf{5+\delta}$};
\node at (10,-3.75) {\Large $\mathbf{6+\delta}$};
%bottom labels
\node at (4,-8.25) { \Large {\bf 1}};
\node at (6,-9.25) {\Large {\bf 2}};
\node at (8,-10.25) {\Large {\bf 3}};
\node at (10,-11.25) { \Large {\bf 4}};
\node at (12,-12.25) {\Large {\bf 5}};
\node at (14,-13.25) {\Large {\bf 6}};
%left labels
\node at (-1.25,0) {\Large {\bf a}};
\node at (-0.25,-2) {\Large {\bf b}};
\node at (0.75,-4) {\Large {\bf c}};
\node at (1.75,-6) {\Large {\bf d}};
%right labels
\node at (12.25,-6) {\Large {\bf a}};
\node at (13.25,-8) {\Large {\bf b}};
\node at (14.25,-10) {\Large {\bf c}};
\node at (15.25,-12) {\Large  {\bf d}};
\end{tikzpicture}}}
\caption{\sl Toric web diagram with $(N,M)=(6,4)$ and shift $\delta$.}
\label{Fig:Web64Diag}
\end{center}
\end{figure}

\begin{figure}
\begin{center}
\scalebox{0.53}{\parbox{25.5cm}{\begin{tikzpicture}[scale = 1.50]
%first line
\draw[ultra thick] (-1,-1) -- (0,0) -- (1,0);
\draw[ultra thick, red] (1,0) -- (2,1);
\draw[ultra thick] (2,1) -- (3,1);
\draw[ultra thick, red] (3,1) -- (4,2);
\draw[ultra thick] (4,2) -- (5,2);
\draw[ultra thick, red] (5,2) -- (6,3);
\draw[ultra thick] (6,3) -- (7,3);
\draw[ultra thick, red] (7,3) -- (8,4);
\draw[ultra thick] (8,4) -- (9,4);
\draw[ultra thick, red] (9,4) -- (10,5);
\draw[ultra thick] (10,5) -- (11,5) -- (12,6);
%second line
\draw[ultra thick] (0,-2) -- (1,-1) -- (2,-1);
\draw[ultra thick, red] (2,-1) -- (3,0);
\draw[ultra thick] (3,0) -- (4,0);
\draw[ultra thick, red] (4,0) -- (5,1);
\draw[ultra thick] (5,1) -- (6,1);
\draw[ultra thick, red] (6,1) -- (7,2);
\draw[ultra thick] (7,2) -- (8,2);
\draw[ultra thick, red] (8,2) -- (9,3);
\draw[ultra thick] (9,3) -- (10,3);
\draw[ultra thick, red](10,3) -- (11,4);
\draw[ultra thick] (11,4) -- (12,4) -- (13,5);
%third line
\draw[ultra thick] (1,-3) -- (2,-2) -- (3,-2);
\draw[ultra thick, red] (3,-2) -- (4,-1);
\draw[ultra thick] (4,-1) -- (5,-1);
\draw[ultra thick, red] (5,-1) -- (6,0);
\draw[ultra thick] (6,0) -- (7,0);
\draw[ultra thick, red] (7,0) -- (8,1);
\draw[ultra thick] (8,1) -- (9,1);
\draw[ultra thick, red] (9,1) -- (10,2);
\draw[ultra thick] (10,2) -- (11,2);
\draw[ultra thick, red] (11,2) -- (12,3);
\draw[ultra thick] (12,3) -- (13,3) -- (14,4);
%fourth line
\draw[ultra thick] (2,-4) -- (3,-3) -- (4,-3);
\draw[ultra thick, red] (4,-3) -- (5,-2);
\draw[ultra thick] (5,-2) -- (6,-2);
\draw[ultra thick, red] (6,-2) -- (7,-1);
\draw[ultra thick] (7,-1) -- (8,-1);
\draw[ultra thick, red] (8,-1) -- (9,0);
\draw[ultra thick] (9,0) -- (10,0);
\draw[ultra thick, red] (10,0) -- (11,1);
\draw[ultra thick] (11,1) -- (12,1);
\draw[ultra thick, red] (12,1) -- (13,2);
\draw[ultra thick] (13,2) -- (14,2) -- (15,3);
%top line
\draw[ultra thick] (0,1) -- (0,0);
\draw[ultra thick] (2,2) -- (2,1);
\draw[ultra thick] (4,3) -- (4,2);
\draw[ultra thick] (6,4) -- (6,3);
\draw[ultra thick] (8,5) -- (8,4);
\draw[ultra thick] (10,6) -- (10,5);
%second line
\draw[ultra thick] (1,0) -- (1,-1);
\draw[ultra thick] (3,1) -- (3,0);
\draw[ultra thick] (5,2) -- (5,1);
\draw[ultra thick] (7,3) -- (7,2);
\draw[ultra thick] (9,4) -- (9,3);
\draw[ultra thick] (11,5) -- (11,4);
%third line
\draw[ultra thick] (2,-2) -- (2,-1);
\draw[ultra thick] (4,-1) -- (4,0);
\draw[ultra thick] (6,0) -- (6,1);
\draw[ultra thick] (8,1) -- (8,2);
\draw[ultra thick] (10,2) -- (10,3);
\draw[ultra thick] (12,3) -- (12,4);
%fourth line
\draw[ultra thick] (3,-3) -- (3,-2);
\draw[ultra thick] (5,-2) -- (5,-1);
\draw[ultra thick] (7,-1) -- (7,0);
\draw[ultra thick] (9,0) -- (9,1);
\draw[ultra thick] (11,1) -- (11,2);
\draw[ultra thick] (13,2) -- (13,3);
%bottom line
\draw[ultra thick] (4,-3) -- (4,-4);
\draw[ultra thick] (6,-2) -- (6,-3);
\draw[ultra thick] (8,-1) -- (8,-2);
\draw[ultra thick] (10,0) -- (10,-1);
\draw[ultra thick] (12,1) -- (12,0);
\draw[ultra thick] (14,2) -- (14,1);
%top labels
\node at (0,1.25) {\large $\mathbf{1+\delta}$};
\node at (2,2.25) {\large $\mathbf{2+\delta}$};
\node at (4,3.25) {\large $\mathbf{3+\delta}$};
\node at (6,4.25) {\large $\mathbf{4+\delta}$};
\node at (8,5.25) {\large $\mathbf{5+\delta}$};
\node at (10,6.25) {\large $\mathbf{6+\delta}$};
%bottom labels
\node at (4,-4.25) {\large $\mathbf{1}$};
\node at (6,-3.25) {\large $\mathbf{2}$};
\node at (8,-2.25) {\large $\mathbf{3}$};
\node at (10,-1.25) {\large $\mathbf{4}$};
\node at (12,-0.25) {\large $\mathbf{5}$};
\node at (14,0.75) {\large $\mathbf{6}$};
%left labels
\node at (-1.2,-1.2) {\large $\mathbf{a}$};
\node at (-0.2,-2.2) {\large $\mathbf{b}$};
\node at (0.8,-3.2) {\large $\mathbf{c}$};
\node at (1.8,-4.2) {\large $\mathbf{d}$};
%right labels
\node at (12.2,6.2) {\large $\mathbf{a}$};
\node at (13.2,5.2) {\large $\mathbf{b}$};
\node at (14.2,4.2) {\large $\mathbf{c}$};
\node at (15.2,3.2) {\large $\mathbf{d}$};
\end{tikzpicture}}}
\caption{\sl $(6,4)$ web with shift $\delta$ after $SL(2,\mathbb{Z})$ transformation.}
\label{Fig:Web64DiagSL2}
\end{center}
\end{figure}
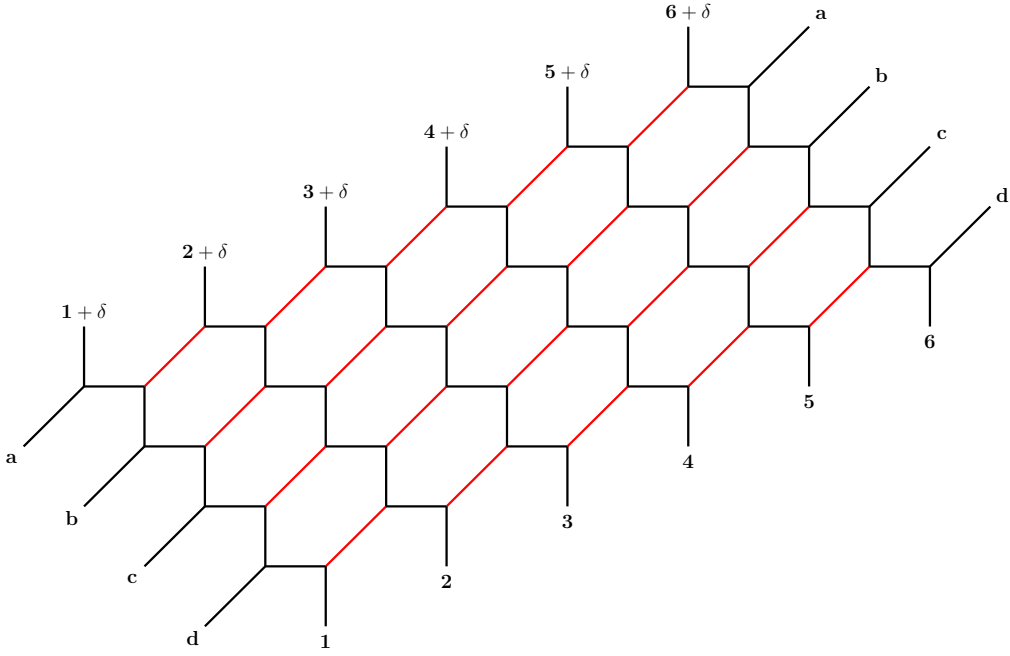

Upon flopping all the diagonal lines in \figref{Fig:WebDiagramGenericShift}, the (shifted) $(N,M)=(6,4)$ web takes the form given in \figref{Fig:Web64Diag} and after an $SL(2,\mathbb{Z})$-transformation as in \figref{Fig:Web64DiagSL2}. 
%%%%%%%%%%%%%%

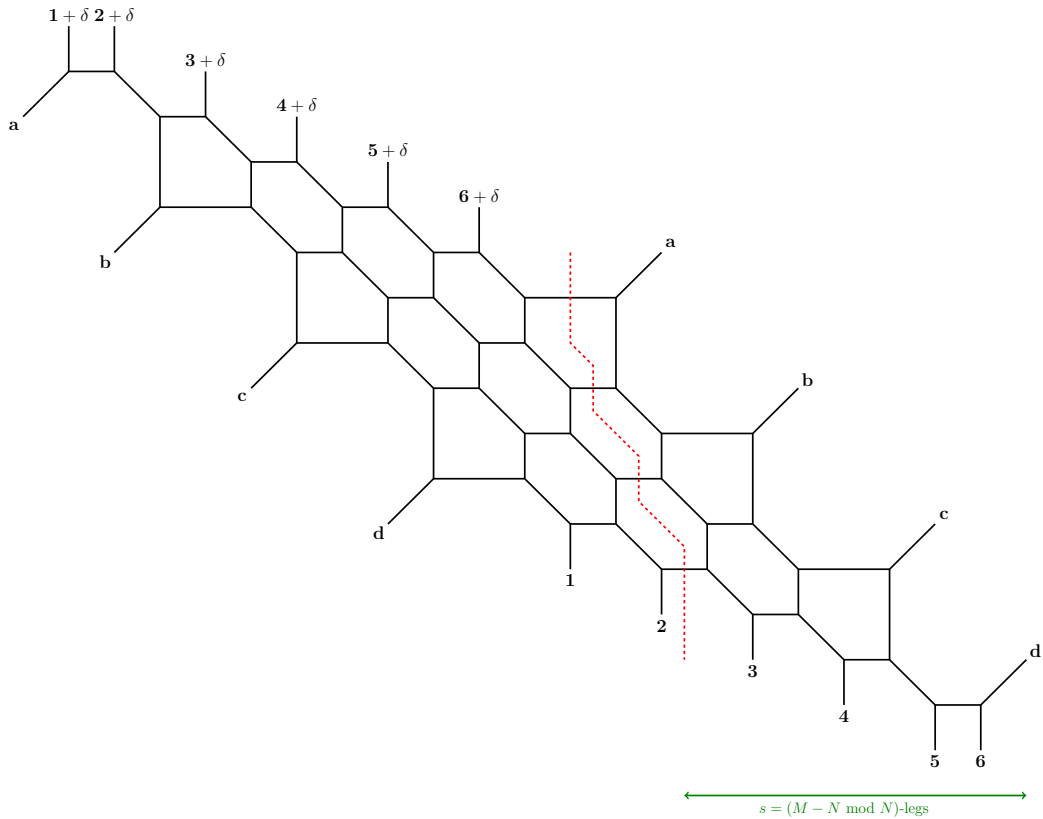
\begin{figure}
\begin{center}
\scalebox{0.4}{\parbox{33.5cm}{\begin{tikzpicture}[scale = 1.50]
\draw[ultra thick] (-1,-1) -- (0,0) -- (1,0) -- (2,-1) -- (3,-1) -- (4,-2) -- (5,-2) -- (6,-3) -- (7,-3) -- (8,-4) -- (9,-4) -- (10,-5) -- (12,-5) -- (13,-4);
\draw[ultra thick] (1,-4) -- (2,-3) -- (4,-3) -- (5,-4) -- (6,-4) -- (7,-5) -- (8,-5) -- (9,-6) -- (10,-6) -- (11,-7) -- (12,-7) -- (13,-8) -- (15,-8) -- (16,-7);
\draw[ultra thick] (4,-7) -- (5,-6) -- (7,-6) -- (8,-7) -- (9,-7) -- (10,-8)-- (11,-8) -- (12,-9) -- (13,-9) -- (14,-10) -- (15,-10) -- (16,-11) -- (18,-11) -- (19,-10);
\draw[ultra thick] (7,-10) -- (8,-9) -- (10,-9) -- (11,-10) -- (12,-10) -- (13,-11) -- (14,-11) -- (15,-12) -- (16,-12) -- (17,-13) -- (18,-13) -- (19,-14) -- (20,-14) -- (21,-13);
%top row
\draw[ultra thick] (0,1) -- (0,0);
\draw[ultra thick] (1,1) -- (1,0);
\draw[ultra thick] (3,0) -- (3,-1);
\draw[ultra thick] (5,-1) -- (5,-2);
\draw[ultra thick] (7,-2) -- (7,-3);
\draw[ultra thick] (9,-3) -- (9,-4);
%first row
\draw[ultra thick] (2,-1) -- (2,-3);
\draw[ultra thick] (4,-2) -- (4,-3);
\draw[ultra thick] (6,-3) -- (6,-4);
\draw[ultra thick] (8,-4) -- (8,-5);
\draw[ultra thick] (10,-5) -- (10,-6);
\draw[ultra thick] (12,-5) -- (12,-7);
%second row
\draw[ultra thick] (5,-4) -- (5,-6);
\draw[ultra thick] (7,-5) -- (7,-6);
\draw[ultra thick] (9,-6) -- (9,-7);
\draw[ultra thick] (11,-7) -- (11,-8);
\draw[ultra thick] (13,-8) -- (13,-9);
\draw[ultra thick] (15,-8) -- (15,-10);
%third row
\draw[ultra thick] (8,-7) -- (8,-9);
\draw[ultra thick] (10,-8) -- (10,-9);
\draw[ultra thick] (12,-9) -- (12,-10);
\draw[ultra thick] (14,-10) -- (14,-11);
\draw[ultra thick] (16,-11) -- (16,-12);
\draw[ultra thick] (18,-11) -- (18,-13);
%third row
\draw[ultra thick] (11,-10) -- (11,-11);
\draw[ultra thick] (13,-11) -- (13,-12);
\draw[ultra thick] (15,-12) -- (15,-13);
\draw[ultra thick] (17,-13) -- (17,-14);
\draw[ultra thick] (19,-14) -- (19,-15);
\draw[ultra thick] (20,-14) -- (20,-15);
%top labels
\node at (0,1.25) {\Large $\mathbf{1+\delta}$};
\node at (1,1.25) {\Large $\mathbf{2+\delta}$};
\node at (3,0.25) {\Large $\mathbf{3+\delta}$};
\node at (5,-0.75) {\Large $\mathbf{4+\delta}$};
\node at (7,-1.75) {\Large $\mathbf{5+\delta}$};
\node at (9,-2.75) {\Large $\mathbf{6+\delta}$};
%bottom labels
\node at (11,-11.25) {\Large $\mathbf{1}$};
\node at (13,-12.25) {\Large $\mathbf{2}$};
\node at (15,-13.25) {\Large $\mathbf{3}$};
\node at (17,-14.25) {\Large $\mathbf{4}$};
\node at (19,-15.25) {\Large $\mathbf{5}$};
\node at (20,-15.25) {\Large $\mathbf{6}$};
%left labels
\node at (-1.2,-1.2) {\Large $\mathbf{a}$};
\node at (0.8,-4.2) {\Large $\mathbf{b}$};
\node at (3.8,-7.2) {\Large $\mathbf{c}$};
\node at (6.8,-10.2) {\Large $\mathbf{d}$};
%right labels
\node at (13.2,-3.8) {\Large $\mathbf{a}$};
\node at (16.2,-6.8) {\Large $\mathbf{b}$};
\node at (19.2,-9.8) {\Large $\mathbf{c}$};
\node at (21.2,-12.8) {\Large $\mathbf{d}$};
%cut
\draw[red, ultra thick, dashed] (11,-4) -- (11,-6) -- (11.5,-6.5) -- (11.5,-7.5) -- (12.5,-8.5) -- (12.5,-9.5) -- (13.5,-10.5) -- (13.5,-13);
%length
\draw[<->, ultra thick, green!50!black] (13.5,-16) -- (21,-16); 
\node[green!50!black] at (17,-16.3) {\large $s=(M-N$ mod $N)$-legs};
\end{tikzpicture}}}
\caption{\sl Web diagram after a flop transformation of the diagonal lines in \figref{Fig:Web64DiagSL2}.}
\label{Fig:Web64DiagCut}
\end{center}
\end{figure}

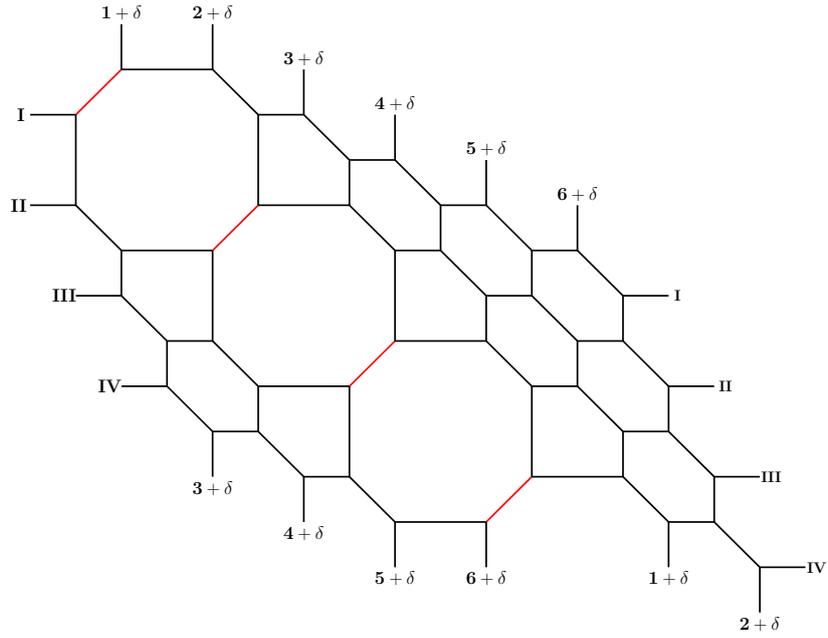
\begin{figure}
\begin{center}
\scalebox{0.4}{\parbox{27cm}{\begin{tikzpicture}[scale = 1.50]
\draw[ultra thick] (-1,0) -- (0,0);
\draw[ultra thick,red] (0,0) -- (1,1);
\draw[ultra thick] (1,1) -- (3,1) -- (4,0) -- (5,0) -- (6,-1) -- (7,-1) -- (8,-2) -- (9,-2) -- (10,-3) -- (11,-3) -- (12,-4) -- (13,-4);
%%%
\draw[ultra thick] (-1,-2) -- (0,-2) -- (1,-3) -- (3,-3);
\draw[ultra thick, red] (3,-3) -- (4,-2);
\draw[ultra thick] (4,-2) -- (6,-2) -- (7,-3) -- (8,-3) -- (9,-4) -- (10,-4) -- (11,-5) -- (12,-5) -- (13,-6) -- (14,-6);
%%%
\draw[ultra thick] (0,-4) -- (1,-4) -- (2,-5) -- (3,-5) -- (4,-6) -- (6,-6);
\draw[ultra thick,red] (6,-6) -- (7,-5);
\draw[ultra thick](7,-5) -- (9,-5) -- (10,-6) -- (11,-6) -- (12,-7) -- (13,-7) -- (14,-8) -- (15,-8);
%%%
\draw[ultra thick] (1,-6) -- (2,-6) -- (3,-7) -- (4,-7) -- (5,-8) -- (6,-8) -- (7,-9) -- (9,-9);
\draw[ultra thick,red] (9,-9) -- (10,-8);
\draw[ultra thick] (10,-8) -- (12,-8) -- (13,-9) -- (14,-9) -- (15,-10) -- (16,-10);
%
%top row
\draw[ultra thick] (1,2) -- (1,1);
\draw[ultra thick] (3,2) -- (3,1);
\draw[ultra thick] (5,1) -- (5,0);
\draw[ultra thick] (7,0) -- (7,-1);
\draw[ultra thick] (9,-1) -- (9,-2);
\draw[ultra thick] (11,-2) -- (11,-3);
%first row
\draw[ultra thick] (0,0) -- (0,-2);
\draw[ultra thick] (4,0) -- (4,-2);
\draw[ultra thick] (6,-1) -- (6,-2);
\draw[ultra thick] (8,-2) -- (8,-3);
\draw[ultra thick] (10,-3) -- (10,-4);
\draw[ultra thick] (12,-4) -- (12,-5);
%second row
\draw[ultra thick] (1,-3) -- (1,-4);
\draw[ultra thick] (3,-3) -- (3,-5);
\draw[ultra thick] (7,-3) -- (7,-5);
\draw[ultra thick] (9,-4) -- (9,-5);
\draw[ultra thick] (11,-5) -- (11,-6);
\draw[ultra thick] (13,-6) -- (13,-7);
%third row
\draw[ultra thick] (2,-5) -- (2,-6);
\draw[ultra thick] (4,-6) -- (4,-7);
\draw[ultra thick] (6,-6) -- (6,-8);
\draw[ultra thick] (10,-6) -- (10,-8);
\draw[ultra thick] (12,-7) -- (12,-8);
\draw[ultra thick] (14,-8) -- (14,-9);
%bottom row
\draw[ultra thick] (3,-7) -- (3,-8);
\draw[ultra thick] (5,-8) -- (5,-9);
\draw[ultra thick] (7,-9) -- (7,-10);
\draw[ultra thick] (9,-9) -- (9,-10);
\draw[ultra thick] (13,-9) -- (13,-10);
\draw[ultra thick] (15,-10) -- (15,-11);
%top labels
\node at (1,2.25) {\Large $\mathbf{1+\delta}$};
\node at (3,2.25) {\Large $\mathbf{2+\delta}$};
\node at (5,1.25) {\Large $\mathbf{3+\delta}$};
\node at (7,0.25) {\Large $\mathbf{4+\delta}$};
\node at (9,-0.75) {\Large $\mathbf{5+\delta}$};
\node at (11,-1.75) {\Large $\mathbf{6+\delta}$};
%bottom labels
\node at (3,-8.25) {\Large $\mathbf{3+\delta}$};
\node at (5,-9.25) {\Large $\mathbf{4+\delta}$};
\node at (7,-10.25) {\Large $\mathbf{5+\delta}$};
\node at (9,-10.25) {\Large $\mathbf{6+\delta}$};
\node at (13,-10.25) {\Large $\mathbf{1+\delta}$};
\node at (15,-11.25) {\Large $\mathbf{2+\delta}$};
%left labels
\node at (-1.2,0) {\Large {\bf I}};
\node at (-1.25,-2) {\Large {\bf II}};
\node at (-0.25,-4) {\Large {\bf III}};
\node at (0.75,-6) {\Large {\bf IV}};
%right labels
\node at (13.2,-4) {\large {\bf I}};
\node at (14.25,-6) {\large {\bf II}};
\node at (15.25,-8) {\large {\bf III}};
\node at (16.25,-10) {\large {\bf IV}};
\end{tikzpicture}}}
\caption{\sl Cutting and re-gluing of the web diagram in \figref{Fig:Web64DiagCut}.}
\label{Fig:Web64DiagPreFlop}
\end{center}
\end{figure}

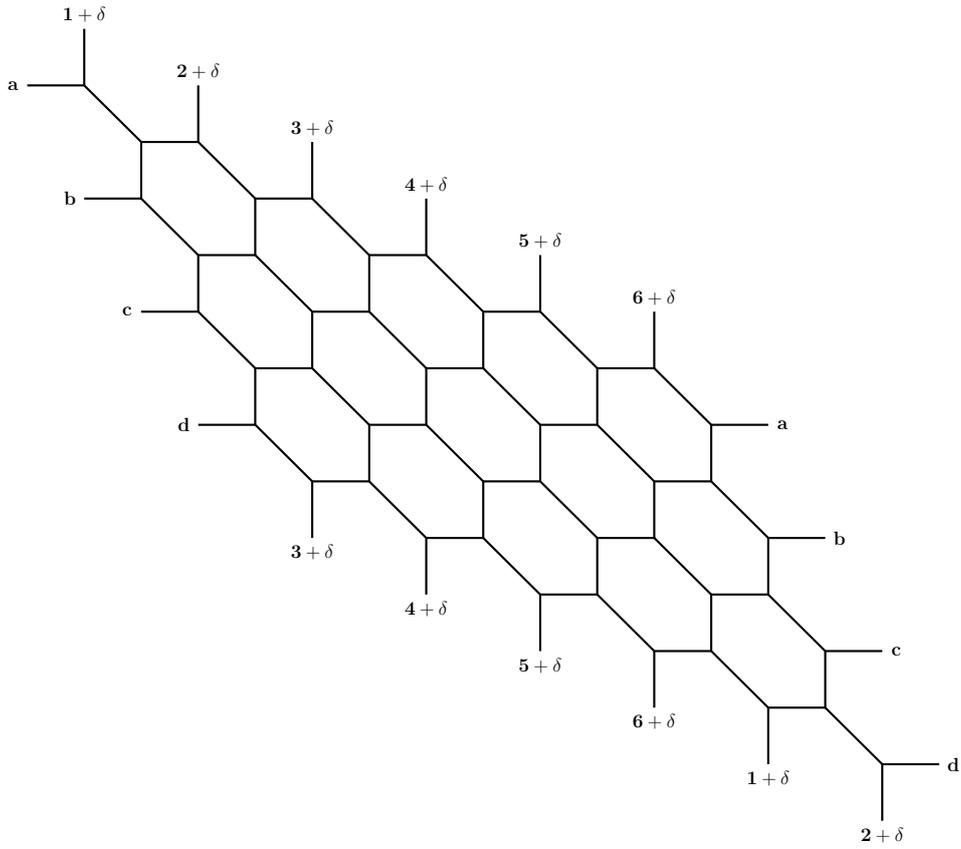
\begin{figure}
\begin{center}
\scalebox{0.5}{\parbox{25.5cm}{\begin{tikzpicture}[scale = 1.50]
\draw[ultra thick] (-1,0) -- (0,0) -- (1,-1) -- (2,-1) -- (3,-2) -- (4,-2) -- (5,-3) -- (6,-3) -- (7,-4) -- (8,-4) -- (9,-5) -- (10,-5) -- (11,-6) -- (12,-6);
\draw[ultra thick] (0,-2) -- (1,-2) -- (2,-3) -- (3,-3) -- (4,-4) -- (5,-4) -- (6,-5) -- (7,-5) -- (8,-6) -- (9,-6) -- (10,-7) -- (11,-7) -- (12,-8) -- (13,-8);
\draw[ultra thick] (1,-4) -- (2,-4) -- (3,-5) -- (4,-5) -- (5,-6) -- (6,-6) -- (7,-7) -- (8,-7) -- (9,-8) -- (10,-8) -- (11,-9) -- (12,-9) -- (13,-10) -- (14,-10);
\draw[ultra thick] (2,-6) -- (3,-6) -- (4,-7) -- (5,-7) -- (6,-8) -- (7,-8) -- (8,-9) -- (9,-9) -- (10,-10) -- (11,-10) -- (12,-11) -- (13,-11) -- (14,-12) -- (15,-12);
%top line
\draw[ultra thick] (0,1) -- (0,0);
\draw[ultra thick] (2,-1) -- (2,0);
\draw[ultra thick] (4,-2) -- (4,-1);
\draw[ultra thick] (6,-3) -- (6,-2);
\draw[ultra thick] (8,-4) -- (8,-3);
\draw[ultra thick] (10,-5) -- (10,-4);
%first line
\draw[ultra thick] (1,-1) -- (1,-2);
\draw[ultra thick] (3,-2) -- (3,-3);
\draw[ultra thick] (5,-3) -- (5,-4);
\draw[ultra thick] (7,-4) -- (7,-5);
\draw[ultra thick] (9,-5) -- (9,-6);
\draw[ultra thick] (11,-6) -- (11,-7);
%second line
\draw[ultra thick] (2,-3) -- (2,-4);
\draw[ultra thick] (4,-4) -- (4,-5);
\draw[ultra thick] (6,-5) -- (6,-6);
\draw[ultra thick] (8,-6) -- (8,-7);
\draw[ultra thick] (10,-7) -- (10,-8);
\draw[ultra thick] (12,-8) -- (12,-9);
%third line
\draw[ultra thick] (3,-5) -- (3,-6);
\draw[ultra thick] (5,-6) -- (5,-7);
\draw[ultra thick] (7,-7) -- (7,-8);
\draw[ultra thick] (9,-8) -- (9,-9);
\draw[ultra thick] (11,-9) -- (11,-10);
\draw[ultra thick] (13,-10) -- (13,-11);
%bottom line
\draw[ultra thick] (4,-7) -- (4,-8);
\draw[ultra thick] (6,-8) -- (6,-9);
\draw[ultra thick] (8,-9) -- (8,-10);
\draw[ultra thick] (10,-10) -- (10,-11);
\draw[ultra thick] (12,-11) -- (12,-12);
\draw[ultra thick] (14,-12) -- (14,-13);
%top labels
\node at (0,1.25) {\large $\mathbf{1+\delta}$};
\node at (2,0.25) {\large $\mathbf{2+\delta}$};
\node at (4,-0.75) {\large $\mathbf{3+\delta}$};
\node at (6,-1.75) {\large $\mathbf{4+\delta}$};
\node at (8,-2.75) {\large $\mathbf{5+\delta}$};
\node at (10,-3.75) {\large $\mathbf{6+\delta}$};
%bottom labels
\node at (4,-8.25) {\large $\mathbf{3+\delta}$};
\node at (6,-9.25) {\large $\mathbf{4+\delta}$};
\node at (8,-10.25) {\large $\mathbf{5+\delta}$};
\node at (10,-11.25) {\large $\mathbf{6+\delta}$};
\node at (12,-12.25) {\large $\mathbf{1+\delta}$};
\node at (14,-13.25) {\large $\mathbf{2+\delta}$};
%left labels
\node at (-1.25,0) {\large {\bf a}};
\node at (-0.25,-2) {\large {\bf b}};
\node at (0.75,-4) {\large {\bf c}};
\node at (1.75,-6) {\large {\bf d}};
%right labels
\node at (12.25,-6) {\large {\bf a}};
\node at (13.25,-8) {\large {\bf b}};
\node at (14.25,-10) {\large {\bf c}};
\node at (15.25,-12) {\large  {\bf d}};
\end{tikzpicture}}}
\caption{\sl Toric web diagram with $(N,M)=(6,4)$ and shift $\delta-2$.}
\label{Fig:Web64DiagShiftTwist}
\end{center}
\end{figure}

\noindent
In the latter presentation, we perform a flop transformation on all diagonal lines drawn in red. This leads to a new web diagram in the extended K\"ahler moduli space of $X_{N,M}$, as shown in \figref{Fig:Web64DiagCut}.
Next, cutting the diagram along the red line and re-gluing the diagonal lines labelled {\bf a}, {\bf b}, {\bf c} and {\bf d} we obtain an equivalent presentation as shown in \figref{Fig:Web64DiagPreFlop}. 
Finally, performing a flop transformation on the diagonal lines shown in red, we obtain the web diagram shown in \figref{Fig:Web64DiagShiftTwist}. 
The latter corresponds to a web diagram in which the shift has been changed to $\delta-s$, where  $s=M-N\text{ mod } N$.

%%%%%%%%%%%%%%%%%%%%%%%%%%%%%%%%%%%
%%%%%%%%%%%%%%%%%%%%%%%%%%%%%%%%%%%
%%%%%%%%%%%%%%%%%%%%%%%%%%%%%%%%%%%
%%%%%%%%%%%%%%%%%%%%%%%%%%%%%%%%%%%
%%%%%%%%%%%%%%%%%%%%%%%%%%%%%%%%%%%
%%%%%%%%%%%%%%%%%%%%%%%%%%%%%%%%%%%

\end{document}